\documentclass[11pt,a4paper]{article}
\usepackage{jheppub}
\usepackage{hyphenat}

\usepackage{amsfonts}
\usepackage{amssymb}
\usepackage{booktabs}
\usepackage{tikz-cd,tikz}
\usetikzlibrary{angles,quotes}
\usepackage{physics}
\usepackage{float}
\usepackage{listings}
\usepackage{calc}
\usepackage{mathbbol}

\usepackage{tikz}
\usetikzlibrary{arrows.meta,calc}
\usepackage{subcaption}

\DeclareFixedFont{\ttb}{T1}{txtt}{bx}{n}{12}
\DeclareFixedFont{\ttm}{T1}{txtt}{m}{n}{12}
\DeclareFixedFont{\ttms}{T1}{txtt}{m}{n}{9}
\DeclareFixedFont{\ttmss}{T1}{txtt}{m}{n}{7}

\definecolor{deepblue}{rgb}{0.0, 0.0, 0.55}
\definecolor{shadecolor}{rgb}{0.96,0.96,0.91}
\definecolor{shadecolor2}{rgb}{1,1,0.99}

\newcommand{\mcomment}[1]{}
\newcommand\mathstyle{\lstset{
language=Mathematica,
basicstyle={\scriptsize\def\fvm@Scale{.5}\fontfamily{fvm}\selectfont},
otherkeywords={self},
keywordstyle=\ttb\scriptsize\color{deepblue},
emph={MyClass,__init__},
emphstyle=\ttb\color{deepred},
backgroundcolor=\color{pink!20!white},
stringstyle=\color{deepgreen},
commentstyle=\color{SkyBlue3!70!PaleGreen4},
frame=tb,
showstringspaces=false
}}

\newwrite\todofile
\immediate\openout\todofile=\jobname.tdo

\newcounter{todocounter}

\newcommand{\printtodos}{
        \section*{To-Do List}
        \immediate\closeout\todofile
        \input{\jobname.tdo}
}
\lstnewenvironment{mathematica}[1][]
{
\mathstyle
\lstset{#1}
}
{}

\newcommand{\la}[1]{\label{#1}}
\newcommand{\eq}[1]{(\ref{#1})}
\newcommand{\nn}{\nonumber}
\newcommand{\stot}{s_{\text{tot}}}

    \newcommand{\beq}{\begin{equation}}
    \newcommand{\eeq}{\end{equation}}
    \newcommand\beqa{\begin{eqnarray}}
    \newcommand\eeqa{\end{eqnarray}}
\newcommand\bea{\begin{array}}
\newcommand\eea{\end{array}}

\newcommand{\da}{\dot{a}}

\newcommand{\ii}{i}

\newlength{\widthLOne}
\newlength{\widthLTwo}
\newcommand{\lM}{\mathcal{M}}

\newcommand{\lO}{\mathcal{O}}

\newcommand{\algsl}{\mathfrak{sl}}
\newcommand{\algso}{\mathfrak{so}}

\newcommand{\betheQ}{\mathbb{Q}}

\def\[{\left[}
\def\]{\right]}
\def\({\left(}
\def\){\right)}
\def\<{\left<}
\def\>{\right>}
\def\d{\partial}

\title{Integrability and the spectrum of two-dimensional fishnet CFT}
\author[1,2]{Simon Ekhammar,}
\author[1]{Nikolay Gromov,}
\author[3]{Fedor Levkovich-Maslyuk,}
\author[4]{Paul Ryan}

\affiliation[1]{
Department of Mathematics, King's College London, The Strand, London WC2R 2LS, UK}
\affiliation[2]{
Department of Physics and Astronomy,
     Uppsala University,
     Box 516,
     SE-751\,20 Uppsala,
     Sweden}
\affiliation[3]{Centre for Mathematical Science, City St George’s, University of London, Northampton Square, EC1V 0HB,
London, UK}
\affiliation[4]{Deutsches Elektronen-Synchrotron DESY, Notkestr. 85, 22607 Hamburg, Germany}

\emailAdd{simon.ekhammar@kcl.ac.uk}
\emailAdd{nikolay.gromov@kcl.ac.uk}
\emailAdd{fedor.levkovich@gmail.com}
\emailAdd{paul.ryan@desy.de}

\abstract{
We formulate a closed set of equations for the spectrum of two-dimensional bi-scalar fishnet conformal field theory, comprising Baxter equations and quantisation conditions, which we derive operatorially from the underlying $\algsl(2)$ spin chain. These equations are reminiscent of the Quantum Spectral Curve framework found in other holographic conformal field theories and are expected to provide a complete non-perturbative description of the spectrum at arbitrary coupling.
We solve our equations numerically at finite coupling and uncover a rich analytic structure, including state collisions and complex energy levels.
Analytically, we introduce a new method to derive the Asymptotic Bethe Ansatz equations, which control the spectrum up to wrapping order and incorporate spinning states. We further extend our results to the twisted case, which may be particularly useful for future separation of variables analyses of correlation functions in this theory.
}

\date{April 2025}

\begin{document}

\maketitle

\section{Introduction}

The full solution of a non-trivial interacting conformal field theory is a notoriously difficult problem. However, in special cases such as 4D $\mathcal{N}=4$ super Yang-Mills (SYM) or 3D ABJM theory in the planar large-$N$ limit, the presence of integrability gives hope that such a momentous feat will one day be accomplished \cite{Beisert:2010jr}. A major obstacle towards this goal is that the origin of integrability in the above theories remains largely obscure, except at a few loops at weak coupling or infinitely strong coupling. It was realised in the seminal paper \cite{Gurdogan:2015csr} that after $\gamma$-deforming the theory \cite{Lunin:2005jy,Frolov:2005dj,Beisert:2005if,Zoubos:2010kh} and subsequently taking the strong deformation limit, it is possible to find much simpler theories in which integrability is manifest at the level of Feynman diagrams. This also connects with Zamolodchikov's remarkable diagrammatic models \cite{Zamolodchikov:1980mb}. The simplest options lead to the 4D bi-scalar fishnet theory, which was generalised to arbitrary dimension $D$ in \cite{Kazakov:2018qbr}. The Lagrangian for these theories includes two $N\times N$ matrix scalar fields $\phi_1,\phi_2$ and a quartic interaction with coupling constant $\xi$,
\begin{equation}\label{eq:LagrangianBiscalar}
    \mathcal{L}= N\, {\rm tr}\[\phi_1^\dagger(-\partial_\mu \partial^\mu)^{\frac{D}{4}}\phi_1 +\phi_2^\dagger(-\partial_\mu \partial^\mu)^{\frac{D}{4}}\phi_2+ (4 \pi)^{\tfrac{D}{2}}\, \xi^2 \phi_1^\dagger\phi_2^\dagger \phi_1 \phi_2\] \ .
\end{equation}
Strictly speaking, it should also be complemented with a set of double-trace counter-terms in order for the theory to remain conformal in the planar limit \cite{Sieg:2016vap,Grabner:2017pgm}.  While the obtained theory is no longer unitary, it is a highly non-trivial interacting CFT with a rich structure. More general versions are also known that include
fermions \cite{Caetano:2016ydc,Kazakov:2018gcy,Pittelli:2019ceq,Kade:2023xet,Kade:2024ucz,Kade:2024lkc}, more scalar fields \cite{Kazakov:2022dbd,Alfimov:2023vev}, or even gauge fields \cite{Ferrando:2023ogg}. Exploration of various facets of integrability for these theories has led to a wide range of remarkable results regarding their spectrum \cite{Gromov:2017cja,Grabner:2017pgm,Basso:2018agi,Basso:2019xay,Ahn:2020zly,Ahn:2021emp,Ahn:2022snr,NietoGarcia:2021kgh}, correlators \cite{Gromov:2018hut,Basso:2018cvy,Korchemsky:2018hnb,Basso:2021omx,Derkachov:2018rot,Derkachov:2019tzo,Derkachov:2021rrf,Olivucci:2021pss,Olivucci:2021cfy,Derkachov:2020zvv,Kostov:2022vup,Olivucci:2023tnw}, generalisations to the massive case \cite{Loebbert:2020tje,Loebbert:2020hxk}, Yangian symmetries and Feynman graphs \cite{Chicherin:2017cns,Chicherin:2022nqq,Loebbert:2019vcj,Loebbert:2020glj,Gurdogan:2020ppd,Kazakov:2023nyu,Levkovich-Maslyuk:2024zdy,Ferrando:2025duw,Beisert:2025mtn}, and even the formulation of string-like dual {\it fishchain} models \cite{Gromov:2019aku,Gromov:2019bsj,Gromov:2019jfh,Gromov:2021ahm}.

Currently, the 4D bi-scalar fishnet theory remains the most well-understood example. This is so because one can export integrability-based tools from the parent $\mathcal{N}=4$ SYM theory, in particular the Quantum Spectral Curve (QSC) \cite{Gromov:2013pga}.  This powerful framework can be extended to the $\gamma$-deformed case as well \cite{Kazakov:2015efa}\footnote{See \cite{Marboe:2019wyc,Levkovich-Maslyuk:2020rlp} for some of its applications in deformed models.}, and after taking the strong deformation limit, it can be applied to the 4D fishnet theory, leading to a wide selection of results for its spectrum \cite{Gromov:2017cja}. Yet obtaining the QSC (and thus efficient access to the spectrum) for fishnets in other dimensions has so far remained out of reach.

Let us note that, besides the 4D and 3D cases, there are no known parent theories that give rise to the fishnet models in other dimensions, and there are currently no proposals for QSCs for fishnet theories away from 4D. Although progress towards computing the spectrum for some operators for general $D$ has been made by using the TBA and Bethe ansatz approaches in \cite{Basso:2019xay}, the QSC formulation should provide access to the full spectrum.

\paragraph{QSC for 2D bi-scalar fishnet theory.}

In this paper, we formulate the QSC framework for 2D bi-scalar fishnet theory, and use it to explore many features of the model's spectrum.
There are several reasons to focus on the particular case of 2D. The main motivation for us is that it serves as an ideal playground for developing new methods for computing correlation functions in the separation of variables (SoV) approach. In the SoV framework, the wave functions factorise into Q-functions captured by the QSC. SoV was pioneered by Sklyanin in the late 1980's \cite{Sklyanin:1984sb,Sklyanin:1987ih,Sklyanin:1991ss,Sklyanin:1992sm,Sklyanin:1995bm} and the approach has been the subject of active exploration since then.  In particular, it was recently extended to higher-rank spin chains (see e.g. \cite{Gromov:2016itr,Maillet:2018bim,Maillet:2018czd,Cavaglia:2019pow,Gromov:2019wmz,Maillet:2020ykb} and the recent review \cite{Levkovich-Maslyuk:2025ipl}), leading to many compact results for correlators \cite{Gromov:2020fwh,Gromov:2022waj,Ekhammar:2023iph}. In a range of cases, SoV has also been shown to provide significant simplifications for SYM correlators \cite{Cavaglia:2018lxi,Giombi:2018qox,Basso:2022nny}, and it is a highly promising direction for exploration (see \cite{Bargheer:2025kli,Basso:2025mca} for some very recent findings). Yet, in order to push this programme further and understand the origin of newly emerging structures, it is important to first explore simpler examples than SYM itself. In line with this, in \cite{Cavaglia:2021mft} a part of this programme was established for the 4D fishnet theory, yet this setup is still rather complicated. At the same time, the fishnet theory in 2D provides another excellent arena for developing these ideas, as SoV is especially simple for $\algsl(2)$, and in 2D the conformal group is precisely made out of two copies of $\algsl(2)$. Furthermore, for $\algsl(2)$ the SoV construction can be realised at the operatorial level with full control over all components while higher-rank cases remain noticeably more involved. and feature additional complications. While the 1D case is potentially even simpler, the 2D case has the important advantage of having the possibility to twist the spin chain via a spacetime rotation (similar to cusp angle for Wilson lines \cite{Cavaglia:2018lxi} and colour twist from \cite{Cavaglia:2020hdb}) which is an extremely useful technical tool in SoV, as was emphasised for the fishnet case in \cite{Cavaglia:2021mft}. Here we make the first crucial step to the full SoV realisation by uncovering the QSC of the 2D theory and understanding its key components on the operatorial level.

Among other key reasons for interest in the 2D case is its relation to integrability in BFKL high-energy scattering in QCD \cite{Lipatov:1993qn,Faddeev:1994zg} (see \cite{Alfimov:2020obh} for a review) where the same spin chains arise \cite{Kazakov:2018qbr,Alfimov:2023vev}. Furthermore, the BFKL regime of $\mathcal{N}=4$ was recently described using a mixture of algebraic and functional equations \cite{Ekhammar:2024neh,Ekhammar:2025vig}, a structure highly reminiscent of a supersymmetric version of the 2D bi-scalar QSC we will present in this paper. Let us also mention the beautiful link of fishnet Feynman graphs, specifically in 2D, to Calabi-Yau geometry \cite{Duhr:2022pch,Duhr:2023eld}. We will see that the QSC offers a window into the non-perturbative spectrum, which should lead to non-trivial connections with this approach as well.

Finally, we expect 2D fishnet theory to play a role in the context of AdS$_3$/CFT$_2$ integrability, as it may eventually be possible to obtain it as a strongly twisted limit of some parent CFT$_2$, similarly to the $\gamma$-deformed 4D case. Thus the 2D model can provide insights for AdS$_3$ integrability and may offer a non-trivial testing ground for the proposed QSC \cite{Ekhammar:2021pys,Cavaglia:2021eqr,Chernikov:2025jko,Cavaglia:2025icd} and TBA \cite{Frolov:2021bwp}. Indeed, as we show below, the 2D fishnet QSC that we will derive already shows structural similarities to its AdS$_3$/CFT$_2$ counterpart.

\paragraph{Structure of the paper.} This paper is structured as follows. In section~\ref{sec:DefinitionAndQSC} we provide background on 2D bi-scalar fishnet theory and present the main outcome of this work, a QSC formulation of the model. In section~\ref{sec:Operatorial} we discuss the derivation of these equations from the underlying $\algsl(2)$ spin chain. In section~\ref{sec:ExactL2Solution} we present the exact solution of the QSC for length-two operators. In section~\ref{sec:FiniteCoupling}, we present the finite-coupling spectrum and perturbative expansion. In section~\ref{sec:ABA} we show how the QSC simplifies to a set of Algebraic Bethe Ansatz equations at weak coupling. In section~\ref{sec:Twisting} we discuss how to generalise our results to the case of a twisted chain, setting the stage for the SoV approach. Finally, we conclude in section~\ref{sec:Conclusions}.
The main text is complemented by a number of appendices with technical details.

\section{Quantum Spectral Curve for 2D bi-scalar fishnet theory}\label{sec:DefinitionAndQSC}
In this section, we introduce 2D bi-scalar conformal fishnet theory and present the Baxter equations and the quantisation conditions, which together constitute an analogue of a QSC for this theory.

\subsection{Quick recap of 2D bi-scalar fishnet model}
The 2D bi-scalar fishnet theory was initially introduced in~\cite{Kazakov:2018qbr}. Below, we introduce the definitions and conventions used throughout this paper.

\paragraph{Lagrangian.}

Following the conventions of \cite{Kazakov:2018qbr}, the Lagrangian for the isotropic two-dimensional bi-scalar fishnet model is given by
\eqref{eq:LagrangianBiscalar}, which we repeat here for convenience
\begin{equation}
      \mathcal{L}= N\, {\rm tr}\[\phi_1^\dagger\Box^{\frac{1}{2}}\phi_1 +\phi_2^\dagger \Box^{\frac{1}{2}}\phi_2+4 \pi\, \xi^2 \phi_1^\dagger\phi_2^\dagger \phi_1 \phi_2\],\quad \Box = - \partial_\mu \partial^\mu\,.
\end{equation}
As one can see, the two elementary fields $\phi_1$ and $\phi_2$ have engineering dimensions given by
\begin{equation}
    \Delta_{\phi_1}\big|_{\xi = 0}=\Delta_{\phi_2}\big|_{\xi = 0} = \frac{1}{2}\,.
\end{equation}
The d'Alembert operator $\partial_\mu \partial^{\mu}$ appears in the Lagrangian with a fractional power, we define the action of this operator on functions as the following integral operator:
\begin{equation}
    (\partial_\mu \partial^\mu)^\beta f(x) \equiv \frac{(-4)^\beta \, \Gamma\left(1 + \beta\right)}{\pi \Gamma(-\beta)} \int \frac{d^2 y \, f(y)}{|x - y|^{2 + 2\beta}}\;.
\end{equation}
The propagator of scalar fields
is given by the Green's function $D(x)$ defined as
\begin{equation}
(-\partial_\mu \partial^\mu)^\beta D(x) = \delta^{(2)}(x)\, \ , \ \  \quad
D(x - y) = \frac{\Gamma\left(1 - \beta\right)}{4^\beta \pi \Gamma(\beta) |x - y|^{2- 2\beta}}\,.
\end{equation}
For what follows, it is very useful to introduce complex coordinates $z$ and $\bar{z}$
\begin{equation}
    z=x_1+i\, x_2,\quad \bar{z}=x_1-i \,x_2\,.
\end{equation}
Note that they are complex conjugates to each other.

\paragraph{Local operators.}
In this work, we will scrutinise the spectrum of non-protected local operators. These operators are schematically of the form (up to a permutation of the fields)
\begin{equation}\label{eq:LocalOperatorsDef}
    \lO=\partial^{K} \bar{\partial}^{\dot{K}}\tr \left(\phi^J_1 \phi^{M}_2\left(\phi_1\phi^{\dagger}_1\right)^n\left(\phi_2^{\dagger}\phi_2\right)^{m} \right)
\end{equation}
where $\partial,\bar{\partial}$ are the (anti-)holomorphic derivatives and all fields sit at the same point. Note that some combinations of the fields do not change the quantum numbers and only affect the bare dimensions of the operator. Due to the $\mathbb{Z}_4$ symmetry $\{\phi_1,\phi_2,\phi_1^{\dagger},\phi_2^{\dagger}\} \mapsto \{\phi_2,\phi_1^{\dagger},\phi_2^{\dagger},\phi_1\}$ that leaves the spectrum invariant, we will restrict ourselves to states with $J\geq M$. In the spin chain language, this means that we pick $\tr \phi^{J}_1$ as our vacuum.

\paragraph{Symmetries.}
The $2D$ conformal algebra is not simple, rather we have $\algso(2,2) \simeq \algsl(2) \oplus \algsl(2)$. We will refer to these two $\algsl(2)$ algebras as undotted and dotted, respectively.  The undotted copy is generated by $ \hat S^z,  \hat S^\pm$ satisfying
\begin{equation}\label{eqn:sl2gens}
    [ \hat S^z, \hat S^\pm]= \pm  \hat S^\pm,\quad [ \hat S^+, \hat S^-]=2 \hat S^z\,.
\end{equation}
Similarly, the dotted copy is generated by $ \hat {\dot{S}}^z, \hat {\dot{S}}^\pm$, subject to the same commutation relations \eqref{eqn:sl2gens}.

It will be useful to have an explicit realisation of the $\algsl(2)$ generators as differential operators
\begin{equation}\label{eqn:sl2gensop}
\begin{split}
    & \hat  S^z = -z\partial -s,\quad  \hat S^+ = \partial,\quad  \hat S^- = -z^2 \partial -2sz\,, 
    \\
    & \hat { \dot{S}}^z = -\bar{z}\bar{\partial} - \dot{s},\quad  \hat {\dot{S}}^+ = \bar{\partial},\quad  \hat {\dot{S}}^- = -\bar{z}^2 \bar{\partial} -2\dot{s}\bar{z}\,.
\end{split}
\end{equation}
The parameters $s$, $\dot{s}$ label the spin of the representations of each copy of $\algsl(2)$. We see that the two copies of the algebra act on functions of $z$ and $\bar{z}$, respectively. We will denote these spaces of functions as $V_{s}$ and $\bar{V}_{\dot{s}}$ for the holomorphic and anti-holomorphic parts. Notice also that we have the Casimir operator acting as a scalar,
\beq
\label{scas}
    ( \hat S^z)^2+\frac{1}{2}\left( \hat S^+ \hat S^-+ \hat S^- \hat S^+\right)=s(s-1)
\eeq
and similarly for the dotted sector.

At this point, we need to stress that, while the dotted and undotted copies of $\algsl(2)$ act on functions of $\bar{z}$ and $z$, respectively, there is no complex conjugation relation between the parameters $s,\dot{s}$ (or between eigenvalues of $S^z$ and $\dot{S}^z$) which are, in principle, independent. This is in distinction from the unitary principal series representations, where these parameters \textit{are} related by a relation involving complex conjugation, and the symmetry is denoted in that case as $SL(2,{\mathbb C})$; see, for example, \cite{Derkachov:2001yn}. Even though our methods should apply in general, in this paper, we limit ourselves to the case where $\dot{s}=s$.

For simplicity, we write a function $f(z,\bar{z})$ simply as $f(z)$, omitting the second argument, with the understanding that the anti-holomorphic dependence is implicit.

One useful combination of the coordinates that will appear repeatedly is the complex propagator, for which we introduce a shorthand notation
\begin{equation}\label{eqn:complexpropagator}
    [z]^{\{\alpha,\dot \alpha\}} :=   |z|^{\alpha+\dot\alpha}e^{i(\alpha-\dot\alpha)\arg z} =
    (z)^\alpha (\bar{z})^{\dot{\alpha}}
    \,,
\end{equation}
To ensure the single-valuedness of the above function on the plane, we require $\alpha-\dot\alpha\in{\mathbb Z}$.
Note that the second expression has to be defined by continuity by adding a small imaginary part, which can be of any sign.

\paragraph{Integrability and graph-building operator.}
As emphasised in the
Introduction, integrability emerges in 2D bi-scalar fishnet theory at the level of Feynman diagrams, as elucidated in \cite{Kazakov:2018qbr}, building on the 4D construction. The key object is the CFT wave function \cite{Gromov:2019jfh}
\begin{equation}\label{eq:CFTWaveFunction}
    \varphi_{\mathcal{O}}(z_1,\dots,z_J) = \expval{\mathcal{O}(0)  {\rm tr}(\chi_{\mathbf{I}_1}(z_1)\dots \chi_{\mathbf{I}_J}(z_J))}
\end{equation}
 where we have denoted particular combinations of the fields by
\begin{equation}
    \chi_0 = \phi_1^\dagger,\quad \chi_{+1} = \phi^\dagger_2 \phi_1^\dagger, \quad \chi_{-1} = \phi_1^\dagger \phi_2,\quad \chi_{\bar{0}}=\phi_2^\dagger \phi_1^\dagger \phi_2 \ .
\end{equation}
We refer to $\chi_{+1}$ fields as magnons, $\chi_{-1}$ fields as anti-magnons, and $\chi_{\bar{0}}$ fields as magnon-anti-magnon pairs. The CFT wave function is only non-zero if the total number of magnons minus the number of anti-magnons matches the charge $M$ in \eq{eq:LocalOperatorsDef}.

We will view the CFT wave function as a state of a spin chain with $J$ sites where at each site $k$ we have a representation of ${\frak{sl}}(2)$ with parameter $s_k$, defined by the operators \eq{eqn:sl2gensop} acting on $z_k, \bar{z}_k$. We denote these operators by adding a subscript $k$, and generators of the global conformal symmetry algebra are obtained by summing over the sites, for instance
\beqa
\label{bbs1}
    \hat{\mathbb{S}}^z = \hat S^z_1 + \dots +\hat S^z_J\,,
    \quad
    \hat{\dot{\mathbb{S}}}^z = \hat {\dot{ S}}^z_1 + \dots +\hat{\dot {S}}^z_J\ .
\eeqa
We denote the eigenvalues of these two operators on CFT wave functions as $h$ and $\dot{h}$,
\beq\la{eigenh}
\hat{\mathbb{S}}^z\varphi_\lO(z) = h\,\varphi_\lO(z)\;\;,\;\;
\hat{\dot{\mathbb{S}}}^z\varphi_\lO(z) = \dot h\,\varphi_\lO(z)\;,
\eeq
and they encode the conformal dimension and the spin of the operator $\lO$ via
\beq
\label{dshh}
    \Delta=h+\dot{h}\,,
    \quad
    S=\dot{h}-h\,.
\eeq
Knowledge of the non-vanishing CFT wave function is equivalent to knowing the operator $\lO$, and it uplifts the perturbative notion of the operator wave function (as a scheme-dependent combination of the elementary single traces  \eq{eq:LocalOperatorsDef}) to a scheme-independent finite-coupling notion. The Feynman diagrams contributing to the correlation function \eqref{eq:CFTWaveFunction} possess a simple iterative structure through the successive application of a graph-building operator $\hat{B}$. The CFT wave functions are then the stationary states under the action of $\hat{B}$
\begin{equation}
    \hat{B}\circ\varphi_{\mathcal{O}}(z_1,\dots,z_J) = \varphi_{\mathcal{O}}(z_1,\dots,z_J) \,.
\end{equation}

In the simplest case of no magnons, anti-magnons, or magnon-anti-magnon pairs, in complex coordinates and using the complex propagator \eqref{eqn:complexpropagator}, the graph-building operator is given explicitly by
\begin{equation}
    \hat{B}\circ f(z_1,\dots,z_J) = \frac{\xi^{2J}}{\pi^J}\displaystyle \int \prod_{j=1}^J \frac{{\rm d^2}w_j}{[z_j-w_j]^{1/2}[w_j-w_{j-1}]^{1/2}}f(w_1,\dots,w_J)
\end{equation}
with the integration measure is ${\rm d}^2w:={\rm d}w {\rm d}\bar{w}/2$. $\hat{B}$ pictorially acts by adding a wheel to our correlator (see figure~\ref{fig:GraphBuilding}) with periodic boundary conditions $w_0 = w_J$ assumed. The complex propagators, defined in general by \eq{eqn:complexpropagator}, can be explicitly  written as
\begin{equation}
    [z-w]^{1/2}=(z-w)^{1/2}(\bar{z}-\bar{w})^{1/2}\,.
    \la{convensions}
\end{equation}
In terms of real coordinates, this graph-building operator coincides with that of \cite{Kazakov:2018qbr}.

Similarly, one can define the graph-building operator for the case with magnons, $M\neq 0$, as shown in \cite{Gromov:2019jfh} with
\begin{equation}\label{eqn:generalgraphB}
    \hat{B}\circ f(z_1,\dots z_J) = \frac{\xi^{2J}}{\pi^J}\int \prod_{j=1}^J {\rm d}^2 w_j \, b_{I_j} (z_j,w_j,w_{j-1}) f(w_1,\dots,w_J)
\end{equation}
where
\begin{equation}
    b_I(z,w,\tilde{w}) = \frac{1}{[z-w]^{1/2}} \times \left\{
      \begin{array}{ll}
        \dfrac{1}{[w-\tilde w]^{1/2}}, & \ \ I = 0 \quad\text{(no magnon)}, \\
        \dfrac{1}{[z-\tilde w]^{1/2}}, & \ \ I = 1 \quad\text{(magnon)}.
      \end{array}
    \right.
\end{equation}

\newcommand{\MagnonPic}[4]{
\begin{tikzpicture}[baseline={(current bounding box.center)}]
  \def\rmin{2}
  \def\rstep{10}
  \def\q{1.41}
  \def\rmax{#1}
  \def\a0{#2}
  \def\da{30}
  \def\magnon{#3}
  \def\highlightlast{#4}

  \foreach \a in {0,30,...,150}{
    \draw[black,very thick] ({-(#1+\rstep pt)*cos(\a)},{-(#1+\rstep pt)*sin(\a)})--({(#1+\rstep pt)*cos(\a)},{(#1+\rstep pt)*sin(\a)});
  }

  \pgfmathsetmacro{\Nreal}{ln(\rmax/\rmin)/ln(\q)}
  \pgfmathtruncatemacro{\N}{floor(\Nreal)}

  \foreach \k in {0,...,\N}{
    \pgfmathsetmacro{\r}{\rmin+\k*\rstep}
    \pgfmathsetmacro{\r}{\rmin*pow(\q,\k)}
    \pgfmathsetmacro{\rnext}{\rmin*pow(\q,\k+1)}

    \ifnum\magnon=1

      \ifnum\k<\N
        \draw[red,very thick]
        ({\a0+2*\da}:\r pt) arc[start angle=\a0+2*\da,end angle=\a0+360,radius=\r pt];
        \draw[red,very thick] ({\a0+\da}:\r pt) -- ({\a0}:\rnext pt);
        \draw[red,very thick] ({\a0+2*\da}:\r pt) -- ({\a0+\da}:\rnext pt);
      \fi
      \ifnum\k=\N
        \ifnum\highlightlast=1

              \foreach \a in {0,30,...,330}{
    \draw[gray,very thick] ({(\r pt)*cos(\a)},{(\r pt)*sin(\a)})--({(\rnext pt)*cos(\a)},{(\rnext pt)*sin(\a)});
  }

            \draw[red!40,very thick]
            ({\a0+2*\da}:\r pt) arc[start angle=\a0+2*\da,end angle=\a0+360,radius=\r pt];
            \draw[red!40,very thick] ({\a0+\da}:\r pt) -- ({\a0}:\rnext pt);
            \draw[red!40,very thick] ({\a0+2*\da}:\r pt) -- ({\a0+\da}:\rnext pt);

        \else
            \draw[red,very thick]
            ({\a0+2*\da}:\r pt) arc[start angle=\a0+2*\da,end angle=\a0+360,radius=\r pt];
            \draw[red,very thick] ({\a0+\da}:\r pt) -- ({\a0}:\rnext pt);
            \draw[red,very thick] ({\a0+2*\da}:\r pt) -- ({\a0+\da}:\rnext pt);
        \fi
      \fi
    \else
      \draw[red,very thick] (0,0) circle[radius=\r pt];
      \ifnum\k=\N
        \ifnum\highlightlast=1
              \foreach \a in {0,30,...,330}{
    \draw[gray,very thick] ({(\r pt)*cos(\a)},{(\r pt)*sin(\a)})--({(\rnext pt)*cos(\a)},{(\rnext pt)*sin(\a)});
  }

            \draw[red!40,very thick] (0,0) circle[radius=\r pt];
      \fi
      \fi
    \fi
  }
\end{tikzpicture}
}

\begin{figure}[t]
\centering

\subcaptionbox{No magnons: $J=12,\;M=0$.\label{fig:GB-a}}{
  \begin{tikzpicture}
    \node at (-3,0) {\MagnonPic{34}{0}{0}{0}};
    \node at ( 3,0) {\MagnonPic{51}{0}{0}{1}};
    \draw[thick,black,->] (-0.8,0) -- (0.2,0) node[midway,below] {$\hat{B}$};
  \end{tikzpicture}
}

\vspace{0.8em}

\subcaptionbox{Two adjacent magnons: $J=12,\;M=2$.\label{fig:GB-b}}{
  \begin{tikzpicture}
    \node at (-3,0) {\MagnonPic{34}{0}{1}{0}};
    \node at ( 3,0) {\MagnonPic{51}{0}{1}{1}};
    \draw[thick,black,->] (-0.8,0) -- (0.2,0) node[midway,below] {$\hat{B}$};
  \end{tikzpicture}
}

\caption{Graphical illustration of the action of the graph-building operator $\hat{B}$ on the fishnet graphs.
Panel~(a) shows the vacuum configuration (no magnons), while panel~(b) shows the case of two adjacent magnons.}
\label{fig:GraphBuilding}
\end{figure}

Integrability of 2D bi-scalar fishnet theory arises from the fact that the graph-building operator is part of a large family of commuting operators that can be simultaneously diagonalised, which ultimately allows the spectral problem to be recast as a set of Quantum Spectral Curve equations, as we show in the next section. Before that, we present the final equations explicitly in the next subsection.

\subsection{The 2D bi-scalar Quantum Spectral Curve}\label{sec:2DQSCDef}

The main result of this work is a compact formulation of the spectrum of operators of the form \eqref{eq:LocalOperatorsDef}, circumventing any operatorial definitions. We will refer to this fully non-perturbative description as the 2D bi-scalar Quantum Spectral Curve. Here, we give a concise formulation of our the main results.  The rest of the paper is dedicated to the derivation of these equations and the study of their consequences. Several parts of the construction come from generalising the approach of \cite{Derkachov:2001yn,Derkachov:2002wz,Derkachov:2002pb} for solving the $SL(2,{\mathbb C})$ spin chain. All the key changes are discussed in more detail in section \ref{sec:Operatorial}.

As we will see, the operatorial construction depends on the choice of CFT wave function used to probe the local operator, namely on the ordering of magnons and anti-magnons along the chain.  In general, the CFT wave function remains non-zero when magnons and anti-magnons are added or removed in pairs, so that the overall quantum numbers are preserved, and the way one constructs conserved quantities is sensitive to these details as well. However, in the four-dimensional case, it was shown that different operatorial constructions are related by a similarity transformation, dubbed a {\it discrete reparametrization symmetry} in \cite{Gromov:2019jfh}. As we will show in section~\ref{sec:Operatorial} the same is true in 2D and we can thus restrict ourselves to configurations containing only magnons without loss of generality.

As we derive below, the QSC for 2D bi-scalar fishnet consists of 4 Q-functions $q_{i},\dot{q}_{i},i=1,2$ with power-like asymptotics at large $u$,
\begin{equation}\label{eqn:qasymptotics}
    \{ q_{1}, q_{2}, \dot{q}_{1}, \dot{q}_{2}\} \simeq \{u^{h-\stot},u^{1-h-\stot},u^{\dot{h}-\stot},u^{1-\dot{h}-\stot}\}  \,,
    \quad
    \stot = \frac{J+M}{4}\,,
\end{equation}
which satisfy the Baxter equations
\beqa
\label{eq:Baxter1ABA}
    (u-\tfrac{i}{4})^{J-M}(u-\tfrac{3i}{4})^M q_i^{--} - t\, q_i+(u+\tfrac{i}{4})^{J} q_i^{++}&=&0\;,\\
\label{eq:Baxter2ABA}
    (u-\tfrac{i}{4})^{J-M}(u-\tfrac{3i}{4})^M \dot q_i^{--} - \dot t \, \dot q_i+ (u+\tfrac{i}{4})^{J} \dot q_i^{++}&=&0\;.
\eeqa
Here $t,\dot{t}$ are degree $J$ polynomials in the spectral parameter $u$, and we use the shorthand notation
\beq
    f^{\pm}=f(u\pm i/2)\,,
    \quad
    f^{\pm\pm}=f(u\pm i)\,,
    \quad
    f^{[\pm n]}=f(u\pm in/2)\,.
\eeq
The parameters $h$ and $\dot{h}$ in the asymptotics \eq{eqn:qasymptotics} encode the scaling dimension and spin of the state via \eq{dshh}.

When solving the Baxter equations \eqref{eq:Baxter1ABA} and \eqref{eq:Baxter2ABA}, we can either demand that $q_{i}$ are upper-half-plane analytic, in which case we write $q_i^{\downarrow}$, or lower-half-plane analytic, in which case we write $q_{i}^{\uparrow}$.
As there are only two linearly independent solutions to a second order finite-difference equation,
these two different options are related by an $\ii$-periodic matrix, that is
\begin{equation}\label{eq:Omega2DQSC}
    q^{\uparrow}_i = \Omega_{i}{}^{j}\, q^{\downarrow}_{j}\,,
    \quad
    \dot{q}^{\uparrow}_i = \dot{\Omega}_{i}{}^{j} \, \dot{q}^{\downarrow}_j\,,
    \quad
    \Omega=\Omega^{[2]}\,, \,\,\dot{\Omega} = \dot{\Omega}^{[2]}\,.
\end{equation}

The Baxter equations need to be supplemented with quantisation conditions that determine the discrete admissible set of the coefficients in the polynomials $t$ and $\dot t$ (each typically corresponding to a local operator\footnote{Since the Lagrangian density of this theory is not, strictly speaking, a local combination of the elementary fields, some care is required in defining local operators. It may be necessary to extend the usual notion of locality so as to associate each solution of the Baxter equations with an operator at weak coupling. We return to this issue when deriving the Asymptotic Bethe Ansatz equations below.}). These conditions take the form
\begin{equation}\label{eq:quantizationOmega}
    \Gamma^{ik}\Omega_{k}{}^{j}=\Gamma^{jk}\dot{\Omega}_k{}^{i}\,,
    \quad
     \Gamma=\begin{pmatrix}
        1 & 0 \\
        0 & c
    \end{pmatrix}\,,
\end{equation}
where $c$ is a complex number and we have set $\Gamma^{11}=1$ since \eqref{eq:quantizationOmega} can be freely rescaled.
Writing out \eqref{eq:quantizationOmega} on the level of components we find
\begin{equation}\label{eq:OmegaQuantisation}
    \Omega_{1}{}{}^1 = \dot{\Omega}_1{}^{1}\,,
    \quad
    \Omega_{2}{}{}^2 = \dot{\Omega}_2{}^{2}\,,
    \quad
    \Omega_{1}{}{}^2 = c\,\dot{\Omega}_2{}^{1}\,,
     \quad
    \dot{\Omega}_{1}{}{}^2 = c\,\Omega_2{}^{1}\,.
\end{equation}

\paragraph{The coupling and cyclicity.}

At this stage, we still have to introduce the coupling into our system of equations. Furthermore, we found that some of the solutions to the above equations do not correspond to CFT wave functions realized in the bi-scalar fishnet CFT, because they do not satisfy additional cyclicity properties inherit from the single-trace operator origin of the CFT wave-function (similarly to the situation in ${\cal N}=4$). To restrict to physical states and introduce the coupling, we define the following combination of $q$-functions
\beq
\label{qgq}
    {\mathbb Q}_+(u,\dot{u})
    \propto \Gamma^{ij}\, q_i^\downarrow(u)\, \dot{q}_j^\uparrow(\dot{u})\
\eeq
which we will see can be viewed as a certain Q-operator in our model.
We will use the notation  ${\mathbb Q}_{+}(u)$ for the case when $u$ and $\dot{u}$ in \eq{qgq} are equal. To introduce the coupling constant $\xi$ into our system of equations we impose\footnote{Similar relations were also established in \cite{Derkachov:2001yn,Derkachov:2002wz}.}
\begin{equation}\label{IntroquantisationXi}
    \lim_{\epsilon\rightarrow 0}\epsilon^J \frac{\mathbb{Q}_+\left(\frac{3i}{4}-i\epsilon \right)}{\mathbb{Q}_+\left(\frac{i}{4}\right)}=\xi^{2J}\, \ .
\end{equation}
To impose cyclicity, we will construct an explicit operator that moves magnons, and the eigenvalue of this operator is given, in the case $M<J$, by 
\beq\la{introShiftMnJ}
U=
\frac{(-1)^J}{ (4\xi^2)^M}
\lim_{\epsilon\to 0}
\frac{1}{\epsilon^M}
\frac{{\mathbb Q}_{+}\left(-\frac{3\ii}{4}-i\epsilon\right)}
{{\mathbb Q}_{+}\left(+\frac{3\ii}{4}-i\epsilon\right)}\;.
\eeq
For the case $M=J$ we find a different expression,
\beq\la{introShiftMJ}
U_{M=J}=
\(-1\)^J
\lim_{\epsilon\to0}
\frac{{\mathbb Q}_+\(-\tfrac{i}{4}-i\epsilon\)}{
{\mathbb Q}_+\(+\tfrac{3i}{4}-i\epsilon\)
}\;.
\eeq
As the selection rule that imposes cyclicity, we require that this eigenvalue (\eq{introShiftMnJ} or \eq{introShiftMJ}) is equal to 1.

This completes the full non-perturbative formulation of the spectrum of 2D bi-scalar fishnet theory.

\subsection{Asymptotic Bethe Ansatz}\label{subsec:IntroducingABA}
At weak coupling, the 2D bi-scalar QSC simplifies significantly, and we find that the spectrum of the theory is encoded in a set of algebraic Bethe equations which we will call the Asymptotic Bethe Ansatz (ABA). These equations generalise the results of \cite{Caetano:2016ydc,Basso:2019xay} to the full spectrum of 2D bi-scalar fishnets. We found that for the generic state with magnons the weak coupling expansion of the dimension $\Delta$ goes in powers of $\xi^2$ and the ABA equations are valid at least up to order $\xi^{2J}$.

The ABA consists of $3$ sets of equations written for $3$ sets of roots that we denote by $\{u_{i}\}_{i=1}^{M},\{v_{i}\}_{i=1}^{K},\{\dot{v}_i\}_{i=1}^{\dot{K}}$.
To write them in a manageable form, we introduce notation
%\begin{align}\label{eq:LambdaNuDefEq}
%    \lambda(u) &= \frac{\Gamma[-\ii u+\frac{1}{4}]\Gamma[\ii u+\frac{1}{4}]}{\Gamma[-\ii u+\frac{3}{4}]\Gamma[\ii u+\frac{3}{4}]}\,,
%    \quad
%    \nu(u) = \frac{\Gamma[\frac{1}{4}+\ii u]\Gamma[\frac{3}{4}+\ii u]}{\Gamma[\frac{1}{4}-\ii u]\Gamma[\frac{3}{4}-\ii u]}\,,
%    \quad
%    \nu_{f} = \prod_{i=1}^{M}\nu(u_{i})\,,\\
%    {\cal Q}(u) &= \prod_{k=1}^{K}(u-v_k)\,,
%    \quad
%    \dot{\cal Q}(u) = \prod_{k=1}^{\dot{K}}(u-\dot{v}_k)\,.
%\end{align}
\begin{align}\label{eq:LambdaNuDefEq}
    \lambda(u) &= \frac{\Gamma[-\ii u+\frac{1}{4}]\Gamma[\ii u+\frac{1}{4}]}{\Gamma[-\ii u+\frac{3}{4}]\Gamma[\ii u+\frac{3}{4}]}\,,
    \quad
    \nu(u) = \frac{\Gamma[\frac{1}{4}+\ii u]\Gamma[\frac{3}{4}+\ii u]}{\Gamma[\frac{1}{4}-\ii u]\Gamma[\frac{3}{4}-\ii u]}\,,
    \quad
    \nu_{f} = \prod_{i=1}^{M}\nu(u_{i})\,.
\end{align}
With this notation, the Bethe equations take the form
%\begin{align}\label{eq:BetheAux}
%    -\left(\frac{v_i-\frac{\ii}{4}}{v_i+\frac{\ii}{4}}\right)^{J} &= \prod_{j=1}^{K} \frac{v_i-v_j+\ii}{v_i-v_j-\ii}\,\prod_{j=1}^{M} \frac{(v_i+\frac{\ii}{4})(v_i-\frac{\ii}{4})}{(v_i-u_j+\frac{\ii}{2})(v_i-u_j-\frac{\ii}{2})}\;\;,\;\;i=1,\dots,K\;,
%    \\\label{eq:BetheMiddle}
%     \xi^{2J}\lambda^{J}(u_{i}) \nu^{M}(u_{i}) &= \frac{{\cal Q}(u_{i}+\frac{\ii}{2}){\cal Q}(u_{i}-\frac{\ii}{2})\dot{{\cal Q}}(u_{i}+\frac{\ii}{2})\dot{{\cal Q}}(u_{i}-\frac{\ii}{2})}{{\cal Q}(\frac{\ii}{4}){\cal Q}(-\frac{\ii}{4})\dot{{\cal Q}}(\frac{\ii}{4})\dot{{\cal Q}}(-\frac{\ii}{4})}\\
%     &\times \nu_{f}\prod_{\substack{j=1\\j\neq i}}^{M}\frac{\Gamma[1+\ii(u_{i}-u_{j})]}{\Gamma[1-\ii(u_{i}-u_{j})]}\frac{\Gamma[\ii(u_{i}-u_{j})]}{\Gamma[-\ii(u_{i}-u_{j})]}\;\;,\;\;i=1,\dots,M \nonumber
%     \;,
%    \\
%     -\left(\frac{\dot{v}_i-\frac{\ii}{4}}{\dot{v}_i+\frac{\ii}{4}}\right)^{J} &= \prod_{j=1}^{\dot{K}} \frac{\dot{v}_i-\dot{v}_j+\ii}{\dot{v}_i-\dot{v}_j-\ii}\,\prod_{j=1}^{M} \frac{(\dot{v}_i+\frac{\ii}{4})(\dot{v}_i-\frac{\ii}{4})}{(\dot{v}_i-u_j+\frac{\ii}{2})(\dot{v}_i-u_j-\frac{\ii}{2})}\;\;,\;\;i=1,\dots,\dot K\;.\label{eq:BetheAuxDot}
%\end{align}
\begin{align}\label{eq:BetheAux}
    -\left(\frac{v_i-\frac{\ii}{4}}{v_i+\frac{\ii}{4}}\right)^{J} &= \prod_{j=1}^{K} \frac{v_i-v_j+\ii}{v_i-v_j-\ii}\,\prod_{j=1}^{M} \frac{(v_i+\frac{\ii}{4})(v_i-\frac{\ii}{4})}{(v_i-u_j+\frac{\ii}{2})(v_i-u_j-\frac{\ii}{2})}\;\;,\;\;i=1,\dots,K\;,
    \\\label{eq:BetheMiddle}
     \xi^{2J}\lambda^{J}(u_{i}) \nu^{M}(u_{i}) &= \prod_{j=1}^{K}\frac{(u_i-v_j+\tfrac{\ii}{2})(u_i-v_j-\tfrac{\ii}{2})}{(v_{i}+\frac{\ii}{4})(v_{i}-\frac{\ii}{4})}\prod_{j=1}^{\dot{K}}\frac{(u_i-\dot{v}_j+\tfrac{\ii}{2})(u_i-\dot{v}_j-\tfrac{\ii}{2})}{(\dot{v}_{i}+\frac{\ii}{4})(\dot{v}_{i}-\frac{\ii}{4})}\\
     &\times \nu_{f}\prod_{\substack{j=1\\j\neq i}}^{M}\frac{\Gamma[1+\ii(u_{i}-u_{j})]}{\Gamma[1-\ii(u_{i}-u_{j})]}\frac{\Gamma[\ii(u_{i}-u_{j})]}{\Gamma[-\ii(u_{i}-u_{j})]}\;\;,\;\;i=1,\dots,M \nonumber
     \;,
    \\
     -\left(\frac{\dot{v}_i-\frac{\ii}{4}}{\dot{v}_i+\frac{\ii}{4}}\right)^{J} &= \prod_{j=1}^{\dot{K}} \frac{\dot{v}_i-\dot{v}_j+\ii}{\dot{v}_i-\dot{v}_j-\ii}\,\prod_{j=1}^{M} \frac{(\dot{v}_i+\frac{\ii}{4})(\dot{v}_i-\frac{\ii}{4})}{(\dot{v}_i-u_j+\frac{\ii}{2})(\dot{v}_i-u_j-\frac{\ii}{2})}\;\;,\;\;i=1,\dots,\dot K\;.\label{eq:BetheAuxDot}
\end{align}

Here we assume that for the generic states at weak coupling $u_{k} = \frac{\ii}{4}+\mathcal{O}(\xi^2)$. The solutions of these equations are classified by an integer $n=0,\dots , J-1$ defined by the equation
%\begin{equation}
%    \frac{{\cal Q}(-\frac{\ii}{4})}{{\cal Q}(\frac{\ii}{4})}\frac{\dot{\cal Q}(-\frac{\ii}{4})}{\dot{\cal Q}(\frac{\ii}{4})}\prod_{i=1}^{M}
%    \frac{1}{\xi^{2}\lambda(u_i)} = e^{\frac{2\pi \ii}{J}n}\,.
%\end{equation}
\begin{equation}
    \prod_{i=1}^{K}\frac{v_{i}+\tfrac{\ii}{4}}{v_{i}-\tfrac{\ii}{4}}\prod_{i=1}^{\dot{K}}\frac{\dot{v}_{i}+\tfrac{\ii}{4}}{\dot{v}_{i}-\tfrac{\ii}{4}}\prod_{i=1}^{M}
    \frac{1}{\xi^{2}\lambda(u_i)} = e^{\frac{2\pi \ii}{J}n}\,.
\end{equation}
For cyclically invariant states we must have $n=0$, but as we comment above our equations are valid for a more general spin chain.

Upon solving the above equations, the scaling dimension $\Delta$ and the spin $S$ are obtained as
\begin{equation}
    \Delta = \frac{J+M}{2}+K+\dot{K} + \gamma + \mathcal{O}(\xi^{2J+2})\,,
    \quad
    S = \dot{K}-K\,,
    \quad
    \gamma = \sum_{i=1}^{M}(-2\ii u_k-\frac{1}{2})\,.
\end{equation}
We note that \eqref{eq:BetheMiddle} with $K=\dot{K}=0$ was derived in \cite{Basso:2019xay} \footnote{See \cite{Ipsen:2018fmu} for inclusions of derivatives at $1$-loop in 4D.}. 

\section{Operatorial derivation of the QSC}\label{sec:Operatorial}

In this section, we map the problem of computing the CFT wave function (defined as a correlation function in CFT) into a problem of diagonalising a family of mutually commuting operators. Let us begin by introducing a key operator in this endeavour: the Q-operator $\hat{\mathbb{Q}}_+$ defined~\cite{Derkachov:2001yn} by its action on functions $\Phi(z_1,\dots,z_J)$ as
\begin{equation}\label{eq:QOperatorIntro}
    [\hat{\mathbb{Q}}_+(u,\dot u)\Phi](z_1,\dots,z_J) = \int {\rm d}^{2J} w\, {\mathcal{Q}}_+(u, \dot u)(z_1,\dots,z_J|w_1,\dots,w_J)\,\Phi(w_1,\dots,w_J)\,,
\end{equation}
where we have used the shorthand notation ${\rm d}^{2J} w:={\rm d}^2 w_1 \dots {\rm d}^2 w_J$ and the kernel ${\mathcal{Q}}_+$ is given by
\begin{equation}\label{eqn:Qopkernelsin}
   {\mathcal{Q}}_+(u,\dot u)(z|w)=\displaystyle \prod_{k=1}^J  [w_{k}-z_k]^{\{\alpha^+_k,\dot\alpha^+_k\}}[w_{k-1}-z_k]^{\{\beta_k^+,\dot\beta_k^+\}} [w_{k}-w_{k-1}]^{\{\gamma^+_k,\dot\gamma^+_k\}}\,.
\end{equation}
We used the notation \eq{eqn:complexpropagator} for the spectral-parameter dependent propagators. Finally, the exponents are given as
\begin{equation}\label{eqn:Qopexponentsin}
    \alpha_k^+ = i(u-\theta_k)-s_k,
    \quad
    \beta^+_k = -i(u-\theta_k)-s_k\,,
    \quad
    \gamma_k^+ = -3/4+s_k-\ii \theta_k\,,
\end{equation}
and the same for the dotted exponentials with $u$ replaced by $\dot u$. Here $s_k$ are precisely the spin variables attached to each site introduced in \eqref{eqn:sl2gensop} while $\theta_k$ is a new parameter attached to each site, referred to as an inhomogeneity.

A crucial property of $\hat{\mathbb{Q}}_{+}$ is that it contains the graph-building operator, namely we have
\begin{equation}\la{QPtoB1}
    \hat{\mathbb{Q}}_+\left(\frac{i}{4}\right)= \frac{\pi^J}{\xi^{2J}}\hat{B}\,
\end{equation}
if we also tune the inhomogeneities and local spin variables appropriately, see Table~\ref{tab:configs} for the explicit values corresponding to different choices of CFT wave function. It is easy to see by a direct computation in the case of only magnons that \eqref{QPtoB1} holds, with the other set-ups obtained by applying the discrete reparametrization transforms.
\begin{table}
\begin{equation}\nn
\begin{array}{c|c|c|c}
{I}_n & \text{fields} & s_n & \theta_n \\
\midrule
0 & \phi_1^\dagger(x_n) & \frac{1}{4} & 0 \\
+1 &\phi_1^\dagger(x_n) \phi_2^\dagger(x_n) & \frac{1}{2} & +\frac{i}{4} \\
-1 & \phi_1^\dagger(x_n)\phi_2(x_n) & \frac{1}{2} & -\frac{i}{4} \\
\bar{0} & \phi_2^\dagger(x_n)\phi_1^\dagger(x_n)\phi_2(x_n) & \frac{3}{4} & 0
\end{array}\,.
\end{equation}
\caption{
\label{tab:configs}
The table represents the map between the fields probing the non-trivial operator in the CFT wave functions and parameters sitting in the corresponding sites of the spin-chain.}
\end{table}

Thus, if we can diagonalize $\hat{\mathbb{Q}}_+$ we have also succeeded in diagonalizing $\hat{B}$. At first, this does not seem very encouraging since $\hat{\mathbb{Q}}_+$ certainly appears more complicated than $\hat{B}$. The saving grace is integrability, and the remainder of this section will be devoted to showing that $\hat{\mathbb{Q}}_+$ is a member of a family of commuting operators whose spectrum can be obtained from a simple functional equation, namely the Baxter equation that we already introduced in section~\ref{sec:2DQSCDef}.

As the remainder of this section contains many technical steps we believe it is helpful to here outline our approach. In section~\ref{sec:FiniteDim} we introduce a family of commuting operators from finite-dimensional transfer matrices, then we turn to infinite-dimensional transfer matrices in section~\ref{sec:InfiniteDim}, and in section~\ref{sec:QOperator} we find that the Q-operator \eqref{eq:QOperatorIntro} can be obtained from these infinite-dimensional transfer matrices. We study properties of $\hat{\mathbb{Q}}_+$ in section~\ref{sec:PropertiesOfQoperators}, the Baxter equation in \ref{sec:BaxterEquation} and finally in section~\ref{sec:OperatorialQSC} we show how to obtain the QSC of section~\ref{sec:2DQSCDef} from our operatorial considerations.

\subsection{Finite-dimensional transfer matrices}\label{sec:FiniteDim}
Here we present a crucial element of the whole construction -- the finite-dimensional transfer matrices. They will eventual lead us to the Baxter equation.

\paragraph{Lax and transfer matrices.}
To build finite-dimensional transfer matrices, we start by introducing the $\algsl(2)$ Lax matrices
\begin{equation}
    \hat L_{an}(u) = \left(\begin{array}{cc}
        u-i \hat S^{z}_n & -i  \hat S^+_n \\
        -i  \hat S^-_n & u+i  \hat S^{z}_n
    \end{array} \right)\,,
    \quad
    \hat{\dot{L}}_{an}({u}) =  \left(\begin{array}{cc}
        {u}-i  \hat {\dot{S}}^z_n & -i  \hat {\dot{S}}^+_n \\
        -i  \hat {\dot{S}}^-_n & {u}+i  \hat {\dot{S}}^z_n
    \end{array} \right)
\end{equation}
where $ \hat S^z_{n}, \hat S^{\pm}_{n}$ are defined in \eqref{eqn:sl2gensop}. Here the spin operators with subscript $n$ act on the $n$-th site of the spin chain, i.e. on the variables $z_n$ and $\bar z_n$.
It is easy to see that each Lax matrix satisfies the RLL relation
\begin{equation}\label{eqn:RLLrelation}
    R_{ab}(u-v)\hat L_{an}(u)\hat L_{bn}(v) = \hat L_{bn}(v)\hat L_{an}(u)R_{ab}(u-v)
\end{equation}
where the subscripts $a,b$ refer to two different copies of $\mathbb{C}^2$, and $R_{ab}$ denotes the R-matrix $R_{ab}(u) = u 1_{ab}+i P_{ab}$, with $1_{ab}$ and $P_{ab}$ denoting the identity and permutation operators. The same identity, of course, holds if we replace $\hat L$ with $\hat{\dot L}$ in \eq{eqn:RLLrelation}.

We now build the transfer matrices $\hat t(u)$ and $\hat{\dot{t}}({u})$ by tracing over the auxiliary $\mathbb{C}^2$ space as
\begin{equation}\label{eqn:transfermatrix}
\begin{split}
    & \hat t(u) = {\rm tr}_a(\hat T(u))\;\;,\;\;\hat T(u)=\hat L_{a1}(u-\theta_1)\dots \hat L_{aJ}(u-\theta_J)\,, \\
    & \hat {\dot{t}}({u}) = {\rm tr}_a(\hat {\dot T}(u))\;\;,\;\;\hat {\dot T}(u)=\hat {\dot{L}}_{a1}({u}-\theta_1)\dots \hat {\dot{L}}_{aJ}({u}-\theta_J)\,,
\end{split}
\end{equation}
where $\theta_n$ are complex inhomogeneities introduced already above, the same for both the dotted and undotted transfer matrices. Thanks to the RLL relation \eqref{eqn:RLLrelation} the transfer matrices commute at different values of the spectral parameter,
\begin{equation}
    [\hat t(u),\hat t(v)]=0\,,
    \quad
    [\hat {\dot{t}}({u}),\hat {\dot{t}}({v})]=0\,,
    \quad [\hat t(u),\hat {\dot{t}}({v})]=0\,,
\end{equation}
and hence generate a family of commuting integrals of motion. By examining the explicit form of the Lax matrices, we see that the transfer matrix $\hat t(u)$ is a polynomial in $u$ of degree $J$, and the coefficient of $u^J$ is proportional to the identity operator, while the coefficient of $u^{J-1}$ is fixed to be $-2\sum_{n=1}^J\theta_n$, and the others are non-trivial. So, $\hat t(u)$ generates $J-1$ non-trivial integrals of motion.
The same analysis holds for the dotted sector. In addition, we have two more integrals of motion coming from the global $\algsl(2)$ spin operators $\hat{\mathbb{S}}^z,\mathbb{\dot{S}}^z$ defined in \eq{bbs1}, in total yielding $2J$ integrals of motion (matching the total number of the real variables in the CFT wave-function).

Let us also note that the coefficient of $u^{J-2}$ in the transfer matrix $\hat t(u)$ (and similarly for the dotted case) can be written explicitly as
\beq
\label{c2ex}
    \hat c_2=-\left((\hat{\mathbb{S}}^z)^2+\frac{\hat{\mathbb{S}}^+\hat{\mathbb{S}}^-+\hat{\mathbb{S}}^-\hat{\mathbb{S}}^+}{2}\right)+\sum_{n=1}^Js_n(s_n-1)+2\sum_{m<n}^J\theta_m\theta_n\;,
\eeq
where the total raising and lowering spin operators $\hat{\mathbb{S}}^\pm$ are defined by summing over sites as in \eq{bbs1}. For conformal primary operators the CFT wave function is annihilated by $\hat{\mathbb{S}}^-$ so the first term (which is the global Casimir operator) simplifies and reduces to a constant
\beq
\label{c2hwt}
    c_2=-h(h-1)+\sum_{n=1}^Js_n(s_n-1)+2\sum_{m<n}^J\theta_m\theta_n \ .
\eeq

\paragraph{Discrete reparametrization symmetry.}
Our goal is to interpret the CFT wave function as an eigenvector of the transfer-matrix, that is
\begin{equation}
    \hat t(u) \varphi_\lO(z) = t(u) \varphi_\lO(z)\,.
\end{equation}
To do so, we associate a spin chain with the transfer matrix \eqref{eqn:transfermatrix} to a CFT wave function defined in \eqref{eq:CFTWaveFunction} using the dictionary in Table~\ref{tab:configs}. However, for such an identification to make sense we must verify that the definition of the CFT wave function is compatible with the integrability description. The CFT wave function is defined as a correlator of a non-protected conformal primary and a chain of
local probes. Clearly, there are several non-vanishing CFT wave functions one can define for each local operator. For example, one can change the order of the probe operators under the trace, or rearrange magnons and anti-magnons between the sites.
From the integrability/spin-chain picture, each choice corresponds to a different spin chain with a different order of sites or different values of inhomogeneities.

The equivalence of the different CFT wave functions was established in 4D through so-called {\it discrete reparametrization symmetries}
\cite{Gromov:2019jfh}, the name being due to their similarity to the reparametrization symmetry of a continuous string (for which the fishnet can be considered as a discretization). There are two types of such symmetries, the first involves annihilation of a magnon with an anti-magnon and the second corresponds to moving magnons along the chain. We claim that these symmetries are also present in our 2D biscalar fishnet theory, and that our identification is hence consistent. While both of these symmetries were treated operatorially in \cite{Gromov:2019jfh} we will be less ambitious and only consider moving magnons operatorially here. We postpone the magnon-anti-magnon annihilation discussion to Appendix~\ref{app:DiscreteReparam} where we will verify this discrete reparametrization symmetry functionally.

The magnon move operator $\hat\lM_k$ defined by
\begin{equation}
    [\hat \lM_k f](z) =  \frac{-1}{4\pi\xi^2}\displaystyle\int {\rm d}^{2}w_k\frac{[z_{k-1}-z_{k}]^{1/2}}{[z_{k-1}-w_{k-1}]^{3/2}} f(\dots,z_{k-2},w_{k-1},z_{k},\dots)\,,
\end{equation}
it has the following property
\begin{equation}
    \hat\lM_k \hat L_{a,k-1}(u-\theta_{k-1};s_{k-1})
    \hat L_{ak}(u-\theta_{k};s_{k})=\hat L_{a,k-1}(u-\theta_{k};s_{k})\hat L_{ak}(u-\theta_{k-1};s_{k-1})\hat\lM_k
\end{equation}
when site $k$ contains a magnon and site $k-1$ does not, i.e when $(\theta_{k-1},s_{k-1}) = (0,\tfrac{1}{4})$ and $(\theta_{k},s_{k}) = (\tfrac{i}{4},\tfrac{1}{2})$. For clarity of notation we have made the spin dependence $s_k$ in the Lax operators explicit. We see that the operator $\hat\lM_k$ moves the magnon from site $k$ to site $k-1$, which diagrammatically can be illustrated as follows:
\begin{center}
\begin{tikzpicture}[scale=1]

\tikzset{
    dot/.style = {circle, fill=black, inner sep=1.8pt},
    vline/.style = {line width=1pt},
    redline/.style = {line width=1.3pt, red}
}

\draw[redline] (0.8,0) -- (2,0);
\draw[redline] (3,0) -- (4.2,0);

\draw[redline] (2,0) -- (3,1.4);

\foreach \x in {1,2,3,4} {
    \draw[vline] (\x,0) -- (\x,1.4);
    \node[dot] at (\x,0) {};
    \node[dot] at (\x,1.4) {};
}

\node at (2,-0.35) {$k-1$};
\node at (3,-0.35) {$k$};

\draw[thick,->] (4.8,0.8) -- (5.8,0.8);

\draw[redline] (6.8,0) -- (7,0);
\draw[redline] (8,0) -- (10.2,0);

\draw[redline] (7,0) -- (8,1.4);

\foreach \x in {7,8,9,10} {
    \draw[vline] (\x,0) -- (\x,1.4);
    \node[dot] at (\x,0) {};
    \node[dot] at (\x,1.4) {};
}

\node at (8,-0.35) {$k-1$};
\node at (9,-0.35) {$k$};

\end{tikzpicture}
\end{center}

By using the discrete reparametrization symmetry, any CFT wave function can be brought to a standard form, featuring only magnons or only anti-magnons. It turns out, however, that the set-ups with just magnons or just anti-magnons are completely equivalent and are related by the spin chain inversion symmetry or ``antipode" map, see Appendix~\ref{app:Antipode}.

In conclusion, by discrete reparametrization symmetry, we can always map any CFT wave function to an equivalent one containing only magnons or anti-magnons. By the antipode map, these two configurations are completely equivalent and correspond to the same conformal dimension $\Delta$. Hence, it is enough to construct the operator formulation of the QSC in the case where we only have magnons.

\subsection{Infinite-dimensional transfer matrix}\label{sec:InfiniteDim}

While finite-dimensional transfer matrices, together with the global charges, are expected to contain the complete set of integrals of motion,
they do not contain the coupling $\xi$, or equivalently, the graph-building operator, in an obvious way in the 2D case\footnote{In 4D fishnet, the inverse of the graph-building operator sits inside the transfer matrix for the 6-dimensional representation in the auxiliary space~\cite{Gromov:2019jfh}.}. As we show here, the graph building operator can be extracted from a different transfer matrix corresponding to a certain infinite-dimensional auxiliary space, which nevertheless commutes with the finite-dimensional transfer matrices.

To build the required transfer matrix, we need another R-matrix, this time acting on two copies of the functional space as an integral operator
\begin{equation}
    [\hat{\mathbb{R}}_{12}(u,\dot u)\Phi](z_1,z_2) = \displaystyle \int {\rm d}^2 w_1 {\rm d}^2 w_2 \mathcal{R}(u,\dot u)(z_1,z_2,w_1,w_2) \Phi(w_1,w_2)\;.
\end{equation}
 The kernel $\mathcal{R}$ is fixed from the RLL relation
\beqa\la{RLL}
    \hat{L}_{a1}(u-v)\hat{L}_{a2}(u)\hat{\mathbb{R}}_{12}(v,\dot v)& =&\hat{\mathbb{R}}_{12}(v,\dot v)\hat{L}_{a2}(u)\hat{L}_{a1}(u-v)\\
        \hat{\dot L}_{a1}(\dot u-\dot v)\hat{\dot L}_{a2}(\dot u)\hat{\mathbb{R}}_{12}(v,\dot v)&=&\hat{\mathbb{R}}_{12}(v,\dot v)\hat{\dot L}_{a2}(\dot u)\hat{\dot L}_{a1}(\dot u-\dot v)\,.
\eeqa
The precise form of the kernel was found in \cite{Derkachov:2001yn} and is given by
\beqa\la{Rdef}
\nonumber\mathcal{R}(u,\dot u)(z_1,z_2,w_1,w_2)  &=&  [w_2 - z_1]^{i\{u,\dot u\} - s_1 + s_2 - 1}
[z_1 - z_2]^{-i\{u,\dot u\} - s_1 - s_2 + 1}\\
&\times& [w_1 - w_2]^{-i\{u,\dot u\} + s_1 + s_2 - 1}
[z_2 - w_1]^{i\{u,\dot u\} + s_1 - s_2 - 1}
\eeqa
where we have used the propagator \eqref{eqn:complexpropagator}.
It is convenient to represent \eq{Rdef} diagrammatically as:
\beq
{\cal R}(u,\dot u)(z_1,z_2,w_1,w_2)=
\begin{tikzpicture}[
  baseline={(current bounding box.center)},
  v/.style={circle, fill=red!80!black, inner sep=1.6pt},
  e/.style={draw=red!80!black, line width=1.1pt, -{Latex[length=3.2mm,width=2.2mm]}},
  lbl/.style={font=\scriptsize, inner sep=1pt}
]
  \node[v,label=left:{\color{red!80!black}$z_1$}]  (z1) at (0,0) {};
  \node[v,label=above:{\color{red!80!black}$z_2$}] (z2) at (1.9,1.5) {};
  \node[v,label=right:{\color{red!80!black}$w_1$}] (w1) at (3.8,0) {};
  \node[v,label=below:{\color{red!80!black}$w_2$}] (w2) at (1.9,-1.5) {};

  \draw[e] (z1) -- node[lbl, above left]  {$-iu-s_1-s_2+1$} (z2);
  \draw[e] (z2) -- node[lbl, above right] {$iu+s_1-s_2-1$} (w1);
  \draw[e] (w1) -- node[lbl, below right] {$-iu+s_1+s_2-1$} (w2);
  \draw[e] (w2) -- node[lbl, below left]  {$iu-s_1+s_2-1$} (z1);
\end{tikzpicture}\;
\eeq
where for simplicity we only represent the undotted part of the spectral parameter on the diagrams. The RLL relations \eq{RLL} can be verified by acting on a probe function and integrating by parts.  In order for the kernel ${\cal R}$ to be a single-valued function on the plane, individually for each of its $4$ arguments, we have to constrain $u-\dot{u}=i n$, $n\in \mathbb{Z}$.

With this R-operator we can build the following transfer matrices $\hat{\mathbb{T}}_s(u)$
\begin{equation}
    \hat{\mathbb{T}}_s(u,\dot u) = {\rm tr}_a(\hat{\mathbb{R}}_{a1}(u-\theta_1,\dot u-\theta_1)\dots \hat{\mathbb{R}}_{aJ}(u-\theta_1,\dot u- \theta_1))
\end{equation}
where $s$ denotes the spin parameter in the auxiliary space which is traced over. By abuse of notation we used the same symbol $\hat{\mathbb{R}}$ for each factor, but one should keep in mind that each tensor factor carries a different spin label, linked to the site on which it is acting.
It is useful to represent $\hat{\mathbb{T}}_s(u,\dot u)$ diagrammatically too,
\beq\la{Tdiagram}
\hat{\mathbb T}_s(u,\dot u)=
\begin{tikzpicture}[
  baseline={(current bounding box.center)},
  v/.style={circle, fill=red!80!black, inner sep=1.6pt},
  e/.style={draw=red!80!black, line width=1.1pt, -{Latex[length=3.2mm,width=2.2mm]}},
  ext/.style={draw=red!80!black, line width=1.1pt},
  extd/.style={draw=red!80!black, line width=1.1pt, dashed},
  lbl/.style={font=\scriptsize, inner sep=1pt}
]
  \node[v,label=left:{\color{red!80!black}$y_{k-1}$}] (yl) at (0,0) {};
  \node[v,label=above:{\color{red!80!black}$z_k$}]   (zk) at (2.3,1.5) {};
  \node[v,label=right:{\color{red!80!black}$y_k$}]   (yr) at (4.6,0) {};
  \node[v,label=below:{\color{red!80!black}$w_k$}]   (wk) at (2.3,-1.5) {};

  \draw[e] (yl) -- node[lbl, above left]  {$-iu+i\theta_k-s-s_k+1$} (zk);
  \draw[e] (zk) -- node[lbl, above right] {$iu-i\theta_k+s-s_k-1$}  (yr);
  \draw[e] (yr) -- node[lbl, below right] {$-iu+i\theta_k+s+s_k-1$} (wk);
  \draw[e] (wk) -- node[lbl, below left]  {$iu-i\theta_k-s-s_k-1$}  (yl);

  \draw[extd]  (yl) -- ++(-0.9, 0.5);
  \draw[extd]  (yl) -- ++(-0.9,-0.5);

  \draw[extd]  (yr) -- ++(0.9, 0.5);
  \draw[extd]  (yr) -- ++(0.9,-0.5);

\end{tikzpicture}\;.
\eeq
Note that this operator acts on  functions of $J$ 2D coordinates (unlike the finite dimensional $\hat t(u)$ and $\hat{\dot t}(\dot u)$, which only act on the combinations $z$ or $\bar z$).

These transfer matrices, for any $s$, generate a tower of mutually commuting quantities
\begin{equation}
    [\hat{\mathbb{T}}_s(u,\dot u),\hat{\mathbb{T}}_{s'}(u',\dot u')]=0
\end{equation}
which also commute with the charges contained in the finite-dimensional transfer matrices $\hat t$ and $\hat{\dot{t}}$ as follows from the RLL relation \eqref{RLL}. The graph-building operator belongs to the commutative family generated by $\hat{\mathbb{T}}_s$ and can be extracted from $\hat{\mathbb{T}}_{1/4}$ as was shown in \cite{Kazakov:2018qbr}. At the same time the eigenvalues of $\hat t$ and $\hat{\dot{t}}$ do contain sufficient information, but their relation to $\hat{\mathbb{T}}_s$ is rather non-trivial and is one of the main results of this paper. In order to extract it we have to explain how the Q-operator, introduced at the start of this section, was originally constructed.

\subsection{Q-operators}\label{sec:QOperator}

We now return to the Q-operator we introduced at the beginning of this section. There are several constructions for Q-operators known in the literature (see e.g. \cite{Bazhanov:1996dr,Bazhanov:2010ts,Derkachov:1999pz}). For our purposes we will define them as certain integral operators which factorize the transfer matrix $\hat{\mathbb{T}}_s(u)$. We will have two Q-operators $\hat{\mathbb{Q}}_\pm(u)$ with
\begin{equation}\label{eqn:Qopdefn}
    \hat{\mathbb{T}}_s(u,\dot u) = \rho(u+i s)\rho(\dot u+i s)
    \hat{\mathbb{Q}}_+(u-is+i,\dot u-is+i)\hat{\mathbb{Q}}_-(u+i s, \dot u+i s)
\end{equation}
where $\rho(u)$ is a complex function given by
\begin{equation}
    \rho(u)= \prod_{k=1}^J \frac{\Gamma(i(u-\theta_k)+s_k)}{\Gamma(i(u-\theta_k)+1-s_k)}\,.
\end{equation}
The Q-operators are defined by their integration kernels
\begin{equation}
    [\hat{\mathbb{Q}}_\pm(u,\dot u)\Phi](z_1,\dots,z_J) = \int {\rm d}^{2J} w {\mathcal{Q}}_\pm(u, \dot u)(z_1,\dots,z_J|w_1,\dots,w_J)\Phi(w_1,\dots,w_J)
\end{equation}
where we used the shorthand notation ${\rm d}^{2J} w:={\rm d}^2 w_1 \dots {\rm d}^2 w_J$ and the kernels ${\mathcal{Q}}_\pm$ are given by
\begin{equation}\label{eqn:Qopkernels}
\begin{split}
   &  {\mathcal{Q}}_+(u,\dot u)(z|w)=\displaystyle \prod_{k=1}^J  [w_{k}-z_k]^{\{\alpha^+_k,\dot\alpha^+_k\}}[w_{k-1}-z_k]^{\{\beta_k^+,\dot\beta_k^+\}} [w_{k}-w_{k-1}]^{\{\gamma^+_k,\dot\gamma^+_k\}}\,,\\
  &  {\mathcal{Q}}_-(u,\dot u)(z|w)=\frac{1}{\rho(u)\rho(\dot u)}\displaystyle \prod_{k=1}^J  [z_k-z_{k-1}]^{\{\gamma^-_k,\dot \gamma^-_k\}}[w_{k}-z_k]^{\{\alpha^-_k,\dot\alpha^-_k\}}[w_{k}-z_{k-1}]^{\{\beta_k^-,\dot \beta_k^-\}} \,,
  \end{split}
\end{equation}
where once again use the notation \eq{eqn:complexpropagator} for the spectral-parameter dependent propagators.
The exponents satisfy $\alpha^-_k = -\alpha_k^+ -1$, $\beta^-_k = - \beta^+_k-1$ and $\gamma_k^- = -\gamma_k^+$, and $\alpha_k^+$ and $\beta^+_k$ are given by
\begin{equation}\label{eqn:Qopexponents}
    \alpha_k^+ = i(u-\theta_k)-s_k,\quad \beta^+_k = -i(u-\theta_k)-s_k\,,
\end{equation}
and the same for the dotted exponentials with $u$ replaced by $\dot u$.
We  obtain $\gamma_k^+$ below from the condition that this operator commutes with the conformal symmetry. At the same time, the factorisation \eqref{eqn:Qopdefn} is true for arbitrary $\gamma^+_k$ as is clear from the diagram
\beq
\begin{tikzpicture}[
  baseline={(current bounding box.center)},
  v/.style={circle, fill=red!80!black, inner sep=1.6pt},
  e/.style={draw=red!80!black, line width=1.1pt, -{Latex[length=3.2mm,width=2.2mm]}},
  extd/.style={draw=red!80!black, line width=1.1pt, dashed},
  lbl/.style={font=\tiny, inner sep=1pt}
]
  \node[v] (ylU) at (0, 0.18) {};
  \node[v] (ylL) at (0,-0.18) {};
  \node[v] (yrU) at (4.6, 0.18) {};
  \node[v] (yrL) at (4.6,-0.18) {};

  \node[v,label=above:{\color{red!80!black}$z_k$}] (zk) at (2.3, 1.65) {};
  \node[v,label=below:{\color{red!80!black}$w_k$}] (wk) at (2.3,-1.65) {};

  \node at (-0.70,0) {\color{red!80!black}$y_{k-1}$};
  \node at (5.30,0)  {\color{red!80!black}$y_k$};

  \draw[e] (ylU) -- node[lbl, above left]  {$-iu+i\theta_k-s-s_k+1$} (zk);
  \draw[e] (yrU) -- node[lbl, above right] {$iu-i\theta_k+s-s_k-1$}  (zk);
  \draw[e] (wk)  -- node[lbl, below left]  {$iu-i\theta_k-s+s_k-1$}  (ylL);
  \draw[e] (wk)  -- node[lbl, below right] {$-iu+i\theta_k+s+s_k-1$} (yrL);

  \draw[e] (yrU) -- node[lbl, above] {$+\gamma_k^{+}$} (ylU);
  \draw[e] (yrL) -- node[lbl, below] {$-\gamma_k^{+}$} (ylL);

  \node at (-3.3, 0.65) {${\cal Q}_{+}(u-is+i)=$};
  \node at (-3.3,-0.85) {${\cal Q}_{-}(u+is)=$};

  \draw[extd] (ylU) -- ++(-1.05, 0);
  \draw[extd] (ylU) -- ++(-0.85, 0.45);
  \draw[extd] (ylL) -- ++(-1.05, 0);
  \draw[extd] (ylL) -- ++(-0.85,-0.45);

  \draw[extd] (yrU) -- ++( 1.05, 0);
  \draw[extd] (yrU) -- ++( 0.85, 0.45);
  \draw[extd] (yrL) -- ++( 1.05, 0);
  \draw[extd] (yrL) -- ++( 0.85,-0.45);

\end{tikzpicture}
\eeq
from which after cancelling the common propagators in the middle and adjusting the arrows' directions (by adding two times $(-1)^{u-\dot u}$ factors, which cancel each other) we reproduce \eq{Tdiagram}.

In order to constitute Q-operators, the operators $\hat{\mathbb{Q}}_\pm$ should satisfy a number of properties. In particular, they should satisfy an operatorial Baxter equation and commute with the finite-dimensional transfer matrices $\hat t(u)$ and $\hat{\dot{t}}(u)$. By direct computation we found that they satisfy the following Baxter equations
(notice the different ordering in the term with $\hat t$ for $\hat {\mathbb Q}_+$ and $\hat {\mathbb Q}_-$) for arbitrary spins $s_k$ and inhomogeneities $\theta_k$:
\begin{equation}
    \begin{split}\la{Qpbax}
   & \hat{\mathbb{Q}}_+(u+i,\dot{u})\prod_{k=1}^J (u-\theta_k+ i s_k)-\hat t(u)\hat{\mathbb{Q}}_+(u,\dot{u}) +\hat{\mathbb{Q}}_+(u-i,\dot{u})\prod_{k=1}^J (u-\theta_k-i s_k)=0 \\
     & \hat{\mathbb{Q}}_-(u+i,\dot{u})\prod_{k=1}^J (u-\theta_k+ i s_k)-\hat{\mathbb{Q}}_-(u,\dot{u})\hat t(u) +\hat{\mathbb{Q}}_-(u-i,\dot{u})\prod_{k=1}^J (u-\theta_k-i s_k)=0
    \end{split}
\end{equation}
and similarly for the dotted variables but with the shifts on $\dot{u}$ instead of $u$ and $\hat t(u)\to\hat{\dot t}(\dot u)$. However, $\hat{\mathbb{Q}}_{\pm}$ do not in general commute with the finite-dimensional transfer matrix $\hat t(u)$, that is why the ordering matters in \eqref{Qpbax}.

Luckily, the Q-operators can be made to commute with $\hat t(u)$ by tuning spins, inhomogeneities and $\gamma_k^+$ properly. In particular, for a chain without magnon–anti--magnon pairs, commutativity $[\hat t(v),\hat{\mathbb{Q}}_\pm(u,\dot{u})]=0$ can be proven using the methods of \cite{Derkachov:2001yn}, but it seems to be more complicated in general for more general configurations, as the kernels of the Q-operators are no longer just simple products of propagators but can involve additional integrations. For our purposes, however, the
kernels \eqref{eqn:Qopkernels} are completely sufficient.

From now on, let us restrict ourselves to the configurations with only magnons. The exponents $\alpha^+_k$, $\beta_k^+$ in \eqref{eqn:Qopexponents} can now be written as
\begin{equation}\label{eqn:simpleQexponents}
    \alpha_k^+ = iu-\frac{1}{4},\quad \beta^+_k = -i u-\frac{1}{4}-\frac{I_k}{2}
\end{equation}
where $I_k=0$ if site $k$ does not contain a magnon and $I_k=1$ if it does. Finally, the exponents $\gamma_k^+$ are fixed by requiring that the Q-operators commute with the global $\algsl(2)$ generators and are given by
$\gamma_k^+ = -1/2+I_k/2$.

\subsection{Properties of Q-operators}\label{sec:PropertiesOfQoperators}

We will now identify some key properties of the Q-operators, which will be necessary in what follows. The derivations of many of these results follow exactly the same steps as in \cite{Derkachov:2001yn} so we will keep it brief.

\paragraph{Commutativity.}
The Q-operators mutually commute among themselves
\begin{equation}
 [\hat{\mathbb{Q}}_\pm(u),\hat{\mathbb{Q}}_\pm(v)]=0,\quad [\hat{\mathbb{Q}}_+(u),\hat{\mathbb{Q}}_-(v)]=0
\end{equation}
and hence with the infinite-dimensional transfer matrices $\hat{\mathbb{T}}_s(u)$. Additionally, they commute with the finite-dimensional transfer matrices $[\hat{\mathbb{Q}}_\pm(u),\hat t(v)]=[\hat{\mathbb{Q}}_\pm(u),\hat{\dot{t}}(\dot{v})]=0$, as well as with the global $\algsl(2)$ generators
\begin{equation}\label{eqn:Qsl2commute}
    [\hat{\mathbb{Q}}_\pm(u,\dot{u}),\hat{\mathbb{S}}^\pm]=0,\quad [\hat{\mathbb{Q}}_\pm(u,\dot{u}),\hat{\mathbb{S}}^z]=0\,.
\end{equation}
These relations can be derived using the methods of \cite{Derkachov:2001yn}.

% belong to the tower of conserved quantities generated by the transfer matrices and as such 

As a consequence of these relations, the Q-operators can be diagonalised simultaneously with the transfer matrices $\hat t, \hat{\dot t}$ on the CFT wave functions $\varphi_{\lO}(z)$,
\begin{equation}
    \hat{\mathbb{Q}}_\pm(u,\dot{u})\varphi_{\lO}(z) = \mathbb{Q}_\pm(u,\dot{u})\varphi_{\lO}(z)\
\end{equation}
where we have denoted the eigenvalues of the Q-operators by $\mathbb{Q}_\pm(u,\dot{u})$.

\paragraph{Behaviour at special points.} Our primary aim in this section is to close the relations between the conserved charges generated by the Q-operators and finite-dimensional transfer matrices. For this we will need to know the behaviour of the Q-operators at some special points, in particular how the coupling $\xi$ is encoded in the Q-operators.

For our purposes we will be mainly interested in ${\mathbb{Q}}_+$. To obtain the relation between the graph-building operator $\hat{B}$ and $\hat{\mathbb{Q}}_+$, all we need to do is plug $u=\dot{u}=i/4$ into the expression for the Q-operator kernel $\hat{\mathcal{Q}}_+$ in \eqref{eqn:Qopkernels} and compare with the graph-building operator \eqref{eqn:generalgraphB}, immediately giving\footnote{when both dotted and undotted arguments are equal we only write one of them explicitly.}
\begin{equation}\la{QPtoB}
    \hat{\mathbb{Q}}_+\left(\frac{i}{4}\right)=\frac{\pi^J}{\xi^{2J}}\hat{B}\,.
\end{equation}

We also have the following relation to the identity operator. Setting $u=\dot{u} = 3i/4-i\epsilon$ and taking $\epsilon\rightarrow 0$ we obtain
\begin{equation}\label{eqn:Qopidentity}
    \lim_{\epsilon\rightarrow 0}\epsilon^J \hat{\mathbb{Q}}_+\left(\frac{3i}{4}-i\epsilon \right) = \pi^J \times 1
\end{equation}
after using the relation
\begin{equation}
      \delta^{(2)}(w) = \lim_{\epsilon\rightarrow 0} \frac{\epsilon}{\pi} \frac{1}{[w]^{1-\epsilon}}\;
\end{equation}
between the $\delta$-function and complex propagators.

Together, these relations imply the following simple normalization-independent relation between the eigenvalues of the Q-operator and the coupling constant $\xi$,
\begin{equation}\label{eq:QuantizationXi}
    \lim_{\epsilon\rightarrow 0}\epsilon^J \frac{\mathbb{Q}_+\left(\frac{3i}{4}-i\epsilon \right)}{\mathbb{Q}_+\left(\frac{i}{4} \right)}=\xi^{2J}\,.
\end{equation}
This relation generalises an analogous result found in 4D \cite{GrabnerGromovKazakovKorchemsky} (see \cite{Cavaglia:2021mft}) and is the key relation which will allow us to inject the coupling constant into the QSC equations.

\paragraph{Shift operator.}

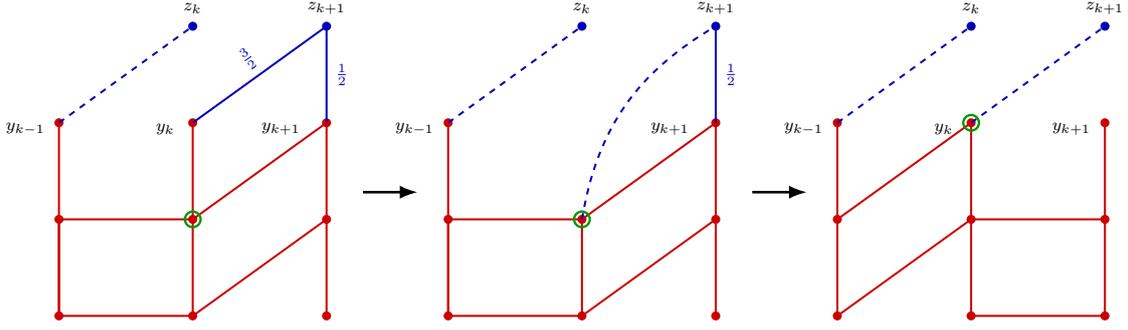
\begin{figure}[t]
\centering
\resizebox{\linewidth}{!}{
\begin{tikzpicture}[
  >=Latex,
  redpt/.style={circle, fill=red!80!black, inner sep=1.5pt},
  redptopen/.style={circle, draw=red!80!black, fill=white, line width=0.9pt, inner sep=1.5pt},
  redln/.style={draw=red!80!black, line width=0.95pt},
  bluept/.style={circle, fill=blue!70!black, inner sep=1.5pt},
  blueln/.style={draw=blue!70!black, line width=0.95pt},
  blueld/.style={draw=blue!70!black, line width=0.95pt, dashed},
  gmark/.style={draw=green!60!black, line width=1.1pt, circle, inner sep=2.6pt},
  lab/.style={font=\scriptsize},
  blab/.style={font=\scriptsize, text=blue!70!black}
]

\begin{scope}[xshift=0cm]
  \coordinate (xL) at (0,0);
  \coordinate (xM) at (2.2,0);
  \coordinate (xR) at (4.4,0);

  \coordinate (A) at (0,0);
  \coordinate (B) at (0,1.6);
  \coordinate (C) at (0,3.2);

  \coordinate (D) at (2.2,0);
  \coordinate (E) at (2.2,1.6);
  \coordinate (F) at (2.2,3.2);

  \coordinate (G) at (4.4,0);
  \coordinate (H) at (4.4,1.6);
  \coordinate (I) at (4.4,3.2);

  \coordinate (Zk)  at (2.2,4.8);
  \coordinate (Zkp) at (4.4,4.8);

  \draw[redln] (A)--(B)--(C);
  \draw[redln] (A)--(D)--(E)--(B)--cycle;
  \draw[redln] (D)--(H);
  \draw[redln] (E)--(I);
  \draw[redln] (F)--(E);
  \draw[redln] (I)--(H)--(G);

  \foreach \P in {A,B,C,D,E,F,G,H,I} \node[redpt] at (\P) {};
  \node[gmark] at (E) {};

  \node[bluept] at (Zk)  {};
  \node[bluept] at (Zkp) {};
  \draw[blueld] (C) -- (Zk);
  \draw[blueln] (F) -- node[blab, sloped, above] {$\tfrac{3}{2}$} (Zkp);
  \draw[blueln] (Zkp) -- node[blab, right] {$\tfrac{1}{2}$} (I);

  \node[lab] at (-0.55,3.1) {$y_{k-1}$};
  \node[lab] at ( 3.65,3.1) {$y_{k+1}$};
  \node[lab] at ( 1.75,3.1) {$y_k$};
  \node[lab, above=2pt] at (Zk)  {$z_k$};
  \node[lab, above=2pt] at (Zkp) {$z_{k+1}$};
\end{scope}

\draw[->, very thick] (5.0,2.05) -- (5.9,2.05);

\begin{scope}[xshift=6.4cm]
  \coordinate (A) at (0,0);
  \coordinate (B) at (0,1.6);
  \coordinate (C) at (0,3.2);

  \coordinate (D) at (2.2,0);
  \coordinate (E) at (2.2,1.6);

  \coordinate (G) at (4.4,0);
  \coordinate (H) at (4.4,1.6);
  \coordinate (I) at (4.4,3.2);

  \coordinate (Zk)  at (2.2,4.8);
  \coordinate (Zkp) at (4.4,4.8);

  \draw[redln] (A)--(B)--(C);
  \draw[redln] (A)--(D)--(E)--(B)--cycle;
  \draw[redln] (D)--(H);
  \draw[redln] (E)--(I);
  \draw[redln] (I)--(H)--(G);

  \foreach \P in {A,B,C,D,E,G,H,I} \node[redpt] at (\P) {};
  \node[gmark] at (E) {};

  \node[bluept] at (Zk)  {};
  \node[bluept] at (Zkp) {};
  \draw[blueld] (C) -- (Zk);
  \draw[blueld, bend left=25] (E) to (Zkp);
  \draw[blueln] (Zkp) -- node[blab, right] {$\tfrac{1}{2}$} (I);

  \node[lab] at (-0.55,3.1) {$y_{k-1}$};
  \node[lab] at ( 3.65,3.1) {$y_{k+1}$};
  \node[lab, above=2pt] at (Zk)  {$z_k$};
  \node[lab, above=2pt] at (Zkp) {$z_{k+1}$};
\end{scope}

\draw[->, very thick] (11.4,2.05) -- (12.3,2.05);

\begin{scope}[xshift=12.8cm]
  \coordinate (A) at (0,0);
  \coordinate (B) at (0,1.6);
  \coordinate (C) at (0,3.2);

  \coordinate (E0) at (2.2,0);
  \coordinate (E) at (2.2,1.6);
  \coordinate (F) at (2.2,3.2);

  \coordinate (G) at (4.4,0);
  \coordinate (H) at (4.4,1.6);
  \coordinate (I) at (4.4,3.2);

  \coordinate (Zk)  at (2.2,4.8);
  \coordinate (Zkp) at (4.4,4.8);

  \draw[redln] (A)--(B)--(C);
  \draw[redln] (B)--(F);
  \draw[redln] (A)--(E)--(E0);
  \draw[redln] (F)--(E);
  \draw[redln] (E)--(H);
  \draw[redln] (I)--(H)--(G);
  \draw[redln] (G)--(E0);

  \node[redpt] at (A) {};
  \node[redpt] at (B) {};
  \node[redpt] at (C) {};
  \node[redpt] at (E) {};
  \node[redpt] at (E0) {};
  \node[redpt] at (F) {};
  \node[redpt] at (G) {};
  \node[redpt] at (H) {};
  \node[redpt] at (I) {};

  \node[gmark] at (F) {};

  \node[bluept] at (Zk)  {};
  \node[bluept] at (Zkp) {};
  \draw[blueld] (C) -- (Zk);
  \draw[blueld] (F) -- (Zkp);

  \node[lab] at (-0.55,3.1) {$y_{k-1}$};
  \node[lab] at ( 3.85,3.1) {$y_{k+1}$};
  \node[lab] at ( 1.75,3.1) {$y_k$};
  \node[lab, above=2pt] at (Zk)  {$z_k$};
  \node[lab, above=2pt] at (Zkp) {$z_{k+1}$};
\end{scope}

\end{tikzpicture}
}
\caption{Action of $\hat{\mathbb Q}_+(-3i/4-i\epsilon)$ on a CFT wave function with a magnon at site $k\!+\!1$ adjacent to an empty site $k$.
Dashed lines denote $\delta$-functions.
In the first step we use that the $\(-\tfrac{3}{2}\)$-propagator, viewed as an integration kernel, annihilates the standard propagator. In the second step the remaining standard propagator cancels against the $\tfrac{1}{2}$-propagator.
After the action, the magnons remain at the same sites, while the integration variables on the vertical lines are shifted by one index.
This operation can therefore be viewed as a generalised shift operator. The highlighted node carries the extra factor $4\pi\xi^2$ from the Feynman rules.\la{figureU}}
\end{figure}

Another important special value is $u=-\tfrac{3i}{4}$. At this point the operator
$\hat{\mathbb Q}_+(-\tfrac{3i}{4}-i\epsilon)$ simplifies considerably.
On an empty site, $I_k=0$, it reduces to
$
 -\frac{\pi}{\epsilon}\,
\int {\rm d}^{2} y_{k-1}\; \delta^{(2)}(y_{k-1}-z_k)\,,
$
so that its action amounts to a relabelling (a one-site shift of the integration variables).

On a magnon site, $I_k=1$, one instead obtains a non-trivial integral operator with kernel
$
[y_{k-1}-z_k]^{-3/2}\,[y_k-z_k]^{1/2}\,.
$
As illustrated in Fig.~\ref{figureU}, when this kernel acts on a magnon site adjacent to an empty site (say $I_{k-1}=0$), the net effect is again a shift operator. The same manipulation can then be applied sequentially to all magnon sites.
This argument requires the presence of at least one empty site; in the ``full-house'' configuration (all $I_k=1$) one must instead use a different special value of~$u$.
Carefully tracing all factors we get
\beq
\hat{\mathbb Q}_{+}\left(-\frac{3\ii}{4}-i\epsilon\right)=\frac{(-\pi)^J (4\xi^2)^M}{\epsilon^{J-M}}\hat U
\eeq
where the generalised shift operator should satisfy $(\hat U)^J=1$ as is clear from the discussion above.
Finally the normalization independent version of $\hat U$ can be written as
\beq\la{UshiftDer}
\hat U=
\frac{(-1)^J}{ (4\xi^2)^M}
\lim_{\epsilon\to 0}
\frac{1}{\epsilon^M}
\frac{\hat {\mathbb Q}_{+}\left(-\frac{3\ii}{4}-i\epsilon\right)}
{\hat {\mathbb Q}_{+}\left(+\frac{3\ii}{4}-i\epsilon\right)}\;.
\eeq
For the remaining case $J=M$ we can consider a different point $u=-1/4-i\epsilon$, which just reduces $\hat{\mathbb Q}_+$ at all sites to an index-shifting $\delta$-function, so we have
\beq
\hat{\mathbb Q}_+\(-\frac{i}{4}-i\epsilon\)\simeq \(-\frac{\pi}{\epsilon}\)^J \hat U^{M=J}
\eeq
and the normalization independent version is
\beq\la{defUJM}
\hat U^{M=J}=
\(-1\)^J
\lim_{\epsilon\to0}
\frac{\hat{\mathbb Q}_+\(-\tfrac{i}{4}-i\epsilon\)}{
\hat{\mathbb Q}_+\(+\tfrac{3i}{4}-i\epsilon\)
}\;.
\eeq
\paragraph{Baxter equation.}

As already outlined at the beginning of this subsection, the operators $\hat{\mathbb{Q}}_\pm(u,\dot{u})$ satisfy the Baxter equation \eqref{eqn:bax1} \textit{as an operator equation} in both $u$ and $\dot{u}$ separately. For the set-up with only magnons, it takes the form
\begin{equation}
   \hat t(u)\hat{\mathbb{Q}}_\pm(u,\dot{u}) =\left(u+\tfrac{i}{4} \right)^J  \hat{\mathbb{Q}}_\pm(u+i,\dot{u})+   \left(u-\tfrac{3i}{4} \right)^{M}\left(u-\tfrac{i}{4} \right)^{J-M}\hat{\mathbb{Q}}_\pm(u-i,\dot{u})
\end{equation}
and similarly with $\dot{t}(\dot{u})$ but now with the shifts on the r.h.s. appearing in the $\dot{u}$ variable. These Baxter equations are easily confirmed by checking that the kernels ${\mathcal{Q}}_\pm(u,\dot{u})$ themselves satisfy the Baxter equation, which can be done by merely acting on them with the transfer matrix $\hat t(u)$ as a differential operator (see \cite{Derkachov:2001yn} for more details).

\paragraph{Asymptotics.} Now let us work out the asymptotics of the Q-operator eigenvalues as $u,\dot{u}\rightarrow \infty$, following \cite{Derkachov:2001yn}. For simplicity we focus on $\hat{\mathbb{Q}}_+$ -- it is enough for our purposes and a similar calculation can be carried out for $\hat{\mathbb{Q}}_-$.

Computing the asymptotics of these eigenvalues on a CFT wave function $\varphi_{\lO}$ is quite tricky. It is useful to take a linear combination of solutions $\Psi_p(z)$ with ``momentum" $p$ given by\footnote{We assume there is a range of values of the conformal dimension $\Delta$ and local spins where this integral is convergent.}
\begin{equation}\label{eqn:nicefns}
    \Psi_p(z_1,\dots,z_J) = \displaystyle \int {\rm d}^2 z_0\, e^{i p z_0+i \dot{p} \bar{z}_0} \varphi_{\lO}(z_1-z_0,\dots,z_J-z_0)\,.
\end{equation}

The function $\Psi_p(z)$ diagonalises the global $\algsl(2)$ generators $\hat{\mathbb{S}}^+ =\hat S_1^+ +\dots + \hat S_J^+$ with eigenvalue $i p$ and similarly for the dotted sector. Since the Q-operators commute with all global $\algsl(2)$ generators \eqref{eqn:Qsl2commute} we conclude that their eigenvalues on the space of functions \eqref{eqn:nicefns} cannot depend on $p$. Hence we can compute the asymptotics of their eigenvalues for $p=0$ to deduce the behaviour for all $p$, and hence conclude the same asymptotic behaviour of the eigenvalues should hold for the original CFT wave functions.

The zero-momentum function $\Psi_0(z)$ has some nice properties. First, by  construction it is translation invariant under simultaneous shift of all arguments
\begin{equation}
    \Psi_0(z_1+\epsilon,\dots,z_J+\epsilon) = \Psi_0(z_1,\dots,z_J)\,.
\end{equation}
Next, since the original CFT wave function $\varphi_{\lO}(z)$ has $\hat{\mathbb{S}}^z$ eigenvalue $h$, we conclude it transforms under rescalings as $\varphi_{\lO}(\lambda z_1,\dots,\lambda z_J)=\lambda^{-h-s_{\rm tot}}\dot{\lambda}^{-\dot{h}-s_{\rm tot}}\varphi_{\lO}(z_1,\dots,z_J)$, where we remind the reader that $s_{\rm tot}=s_1+\dots+s_J=\frac{J+M}{4}$
denotes the sum of all spins of the individual sites. It then follows that the zero-momentum state transforms as
\begin{equation}\label{eqn:scaletransform}
    \Psi_0(\lambda z)=\lambda^{1-h-s_{\rm tot}}\dot{\lambda}^{1-\dot{h}-s_{\rm tot}}\Psi_0(z)\,,
\end{equation}
with the additional powers of $\lambda,\dot{\lambda}$ coming from the rescaling of the integration variables $z_0$ and $\bar{z}_0$.

We will now act with the Q-operator $\hat{\mathbb{Q}}_+$ on $\Psi_0(z)$ using the exponents \eqref{eqn:simpleQexponents}, obtaining
\begin{equation}
   \int \prod_{k=1}^J \displaystyle{\rm d}^2 w_k[w_{k-1}-w_k]^{(I_k-1)/2} [w_{k}-z_k]^{i \{u,\dot u\}-1/4}[w_{k-1}-z_k]^{-i \{u,\dot u\}-1/4-I_k/2} \Psi_0(w)\,.
\end{equation}
Recall that one must require $u-\dot{u}=i n$ with $n$ an integer. To make this constraint more explicit, we  define $u=v+i n/2$ and $\dot{u} = v-i n/2$. The goal is to find the asymptotics of the above integral when $v\to\infty$.

As in \cite{Derkachov:2001yn}, there are two regions $(1)$ and $(2)$ which contribute to the integral when $v\rightarrow\infty$. The first is when all $w_k = \lO(v)$ and the second is when $w_k -w_{k+1}=\lO(1/v)$. Let's examine the contribution from the first region.

First, we rescale $w_k= v y_k$ and consider
\begin{equation}
    [v y_k-z_k]^{i u-1/4}[v y_k-z_{k+1}]^{-i u-1/4-I_{k+1}/2}\,.
\end{equation}
Taking $v\rightarrow \infty$, this yields an expression of the form $v^{-1-I_k}\times \lO(1)$, keeping in mind that the complex propagators are products of a holomorphic and anti-holomorphic propagator. The remaining contributions in this region come from the change of variables under the integral ${\rm d}^2 w_k = v^{2} {\rm d}^2 y_k$, the scaling behaviour \eqref{eqn:scaletransform} of $\Psi_0(v y)$ and the remaining factor $[v y_{k-1}-v y_k]^{(I_k-1)/2}$. At the leading order in $v$ this integration region then gives the contribution which scales as follows
\begin{equation}
\sim \(\prod_{k=1}^J v^{-1-I_k}\)\times v^{2J}\times v^{2-h-\dot h-2 s_{\rm tot}}\times
\(\prod_{k=1}^J v^{-1+I_k}\)
=    v^{2-h-\dot{h}-2s_{\rm tot}}\,.
\end{equation}

We now examine the contribution from region $(2)$. In this case all $w_k$ approach the same point $w_0$, which can be arbitrary and hence should be integrated over. We can account for this by swapping the order of integrations and writing the action of the Q-operator on $\Psi_0(z)$ (using the definition \eq{eqn:nicefns} with $p=0$) as
\beqa
  \int  {\rm d}^2 z_0  \int &&\prod_{k=1}^J
  \displaystyle{\rm d}^2 w_k
  [w_{k-1}-w_k]^{(I_k-1)/2} [w_{k}-z_k]^{i u-1/4}\times\\
\nonumber  &&[w_{k-1}-z_k]^{-i u-1/4-I_k/2} \varphi_{\cal O}(w_1-z_0,\dots,w_J-z_0)\,.
\eeqa
Now we introduce the variables $y_k = v(w_k-z_0)$ where $y_k$ is $\lO(1)$ and take the limit $v\rightarrow \infty$. The contribution from the propagators involving $z$'s are $\lO(1)$, while $[w_{k-1}-w_k]^{(I_k-1)/2}$ contributes $v^{1-I_k}$. A factor of $v^{-2J}$ comes from the measure and finally a factor of $v^{h+\dot{h}+2s_{\rm tot}}$ comes from the scaling of CFT wave function $\varphi(w)$. Combining, we find a total contribution of the form
\begin{equation}
    v^{h+\dot{h}-2s_{\rm tot}} \times \lO(1)\,.
\end{equation}

Combining the contributions from both regions, we see that the eigenvalues of the Q-operator must have the following asymptotics as $v\rightarrow \infty$
\begin{equation}\label{eqn:Qopasymptotics}
    \hat{\mathbb{Q}}_+(v+\tfrac{i n}{2},v-\tfrac{i n}{2}) \simeq v^{h+\dot{h}-2s _{\rm tot}} \times \lO(1) + v ^{2-h-\dot{h}-2 s_{\rm tot}}\times \lO(1)\,.
\end{equation}
This will be very helpful in establishing the form of $ \mathbb{Q}_+$ as a bilinear form $\Gamma^{i j} q_i \dot q_j$, as is discussed in detail in section~\ref{sec:OperatorialQSC}.

\paragraph{Singularities.} Finally, we establish the singularities of $\hat{\mathbb{Q}}_+$. We remind the reader that requiring the operator to be well-defined on the complex plane leads to the constraint $u-\dot{u}=i n$ with integer $n$, which we assume to be satisfied. Then the only possible singularities are poles which come from the integrations.
Following \cite{Derkachov:2001yn} we find that these poles occur when $|w_k-z_k|\rightarrow 0$ or when $|w_k-z_{k+1}|\rightarrow 0$. In the first region, we can write $w_k-z_k=\delta_k$ and expand the integrand in small $\delta_k$ and then perform the integral, and similarly in the second region\footnote{In more detail, in the first region, we expand the function on which the Q-operator acts as $\sum_{m_k,\dot{m_k}\geq 1}c_{m_k,\dot{m}_k}\delta_k^{m_k-1}\bar{\delta}_k^{\dot{m}_k-1}$, and the pole comes from integrals potentially divergent at the origin, of the kind $\int_{|\delta_k|\leq \epsilon} d^2 \delta_k \;\delta_k^{\alpha_k} \bar{\delta}_k^{\dot{\alpha}_k}\delta_k^{m_k-1}\bar{\delta}_k^{\dot{m}_k-1}\sim \frac{1}{\alpha_k+m_k}\delta_{\alpha_k+m_k,\dot{\alpha}_k+\dot{m}_k}$. This fixes $u$ and $\dot{u}$ in terms of $m_k, \dot{m}_k$. After some simplification, combining with similar results for the 2nd region $w_k\to z_{k+1}$, we get \eq{qqpole1}.}.

Let us use the parametrization $u=v+in/2,\ \dot{u}=v-in/2$ with integer $n$. Then we find that ${\mathbb Q}_+(u,\dot{u})$ has poles for
\beq
\label{qqpole1}
    n=m-\dot{m}, \ \ \ v=\theta_k
    \pm i\left(\frac{m+\dot{m}}{2}-s_k\right)
\eeq
where $m,\dot{m}=1,2,\dots$. This can be rewritten as
\beq\la{poles_locattions}
    v= \theta_k
    \pm i\left(\frac{|n|}{2}+l-s_k\right) \ \ , \ \ l=1,2,\dots
\eeq
The order of these poles is determined by how many values given by this expression can coincide (noting that typically some of $\theta_k$ and $s_k$ are equal for different sites $k$). E.g. for the case with $M$ magnons, there are poles of order at most $J-M$ at  $v=\pm i\left(\frac{|n|}{2}+l-\frac{1}{4}\right)$ and poles of order at most $M$ at $v=\frac{i}{4}\pm i\left(\frac{|n|}{2}+l-\frac{1}{2}\right)$, with integer $l\geq 1$.

\subsection{Properties of the Baxter equation}\label{sec:BaxterEquation}

We have established that $\hat{\mathbb{Q}}_\pm(u,\dot{u})$ satisfies a Baxter equation in both $u$ and $\dot{u}$ as an operatorial identity. Since the Q-operators commute with the finite-dimensional transfer matrices, the same equation must also be satisfied by their eigenvalues. This implies that we can expand $\mathbb{Q}_\pm(u,\dot{u})$ as a bilinear combination of solutions to the ``holomorphic'' and ``anti-holomorphic'' Baxter equations. In this section, we explore the properties of these solutions.

The ``holomorphic'' Baxter equation takes the form
\begin{equation}\label{eqn:bax1}
    \left(u+\tfrac{i}{4} \right)^J q(u+i) - t(u) q(u) + \left(u-\tfrac{3i}{4} \right)^{M}\left(u-\tfrac{i}{4} \right)^{J-M}q(u-i)=0\,,
\end{equation}
and the same equation holds in the dotted sector, with $t \rightarrow \dot{t}$ and $q\rightarrow \dot{q}$. Since the Baxter equation is a second-order finite difference equation, it has two linearly independent solutions. A natural basis of solutions are the solutions with power-like behaviour at $u\rightarrow \infty$ which we denote as $q_1$ and $q_2$ with
\begin{equation}\label{eqn:qasympts}
    q_1(u) \simeq u^{h-(J+M)/4},\quad q_2(u) \simeq u^{1-h-(J+M)/4}
\end{equation}
and similarly in the dotted sector with $h\rightarrow \dot{h}$. These asymptotics follow from the first few leading coefficients of the transfer matrix discussed in section \ref{sec:FiniteDim} (see \eq{c2hwt}).

We can further refine this basis of solutions by specifying their analyticity domains. We can demand that $q_{i}$ are analytic in the upper half-plane, then as a consequence of \eqref{eqn:bax1} the solution will in general have poles in the lower half-plane. We denote these solutions as $q_i^\downarrow$. On the other hand, we could have started with analytic solutions below the real axis, which will then feature poles in the upper half plane. We denote these solutions as $q_i^\uparrow$.

More precisely, the upper-half plane analytic (UHPA) solutions $q_i^\downarrow$ possess an infinite sequence of poles in the lower-half-plane, which are located at $u=-i/4-i n$ and $n=0,1,\dots$, of order not more than $M$. There is another sequence of poles at $u=-3i/4-i n$ and $n=0,1,\dots$, of order not more than $J-M$. On the other hand, the lower-half plane analytic (LHPA) solutions $q_i^\uparrow$ possess an infinite sequence of poles in the upper half plane at $u=3i/4+i n$, $n=0,1,\dots$, all of which are of order at most $J$. This is dictated by the factors in the first term of \eq{eqn:bax1}.

Since there can only be two linearly independent solutions of the Baxter equation, $q_i^\uparrow$ and $q_i^\downarrow$ must be related by a linear transformation with $i$-periodic coefficients. We write
\begin{equation}
\label{qom11}
    q_i^\uparrow(u) = \Omega_i{}^{j}(u) q_j^\downarrow(u)
\end{equation}
where $\Omega_i{}^{j}$ is an $i$-periodic matrix, $\Omega_i{}^{j}(u+i) = \Omega_i{}^{j}(u)$.  The matrix $\Omega$  plays an important role in the formulation of the quantisation condition, as we argue below.

\paragraph{Wronskian relations.}
Let $p_i$ be two independent solutions of the Baxter equation \eqref{eqn:bax1}, they must then satisfy $W(u)= \epsilon^{ij}p_i(u+\tfrac i2)p_j(u-\tfrac i2)$ where
\begin{equation}
    \frac{W^-}{W^+} = \left(\frac{u+\frac{\ii}{4}}{u-\frac{\ii}{4}}\right)^{J-M}\left(\frac{u+\frac{\ii}{4}}{u-\frac{3\ii}{4}}\right)^{M}\,.
\end{equation}
In the case $p_i=q_i^\downarrow$ and $p_i=q_i^\uparrow$ one can fix $W$ uniquely from the asymptotics \eqref{eqn:qasympts} and analyticity of $q$. This gives
\beqa\label{eq:WronskianDown}
W^\downarrow &=& q_1^{\downarrow, +} q_2^{\downarrow,-}-
q_1^{\downarrow,-} q_2^{\downarrow,+}=
 C^{\downarrow}_W
\left(\frac{\Gamma \left(-i u+\frac{1}{4}\right)}{\Gamma \left(-i u+\frac{3}{4}\right)}\right)^{J-M}
\left(\frac{\Gamma \left(-i u-\frac{1}{4}\right)}{\Gamma \left(-i u+\frac{3}{4}\right)}\right)^M\,,\\
\label{eq:WronskianUp}
W^\uparrow &=& q_1^{\uparrow, +} q_2^{\uparrow,-}-
q_1^{\uparrow,-} q_2^{\uparrow,+} =
C^{\uparrow}_W
\left(\frac{\Gamma \left(+i u+\frac{1}{4}\right)}{\Gamma \left(+i u+\frac{3}{4}\right)}\right)^{J-M}
\left(\frac{\Gamma \left(+i u+\frac{1}{4}\right)}{\Gamma \left(+i u+\frac{5}{4}\right)}\right)^M\,,
\eeqa
where explicitly
\beq\la{eqCW}
C^{\downarrow}_W= i e^{-\frac{1}{4} i \pi  (J+M)}\left(2h-1\right)\quad{\rm and}\quad C^{\uparrow}_{W} = e^{\frac{\ii \pi}{2}\left(J+M\right)} C^{\downarrow}_{W}\;.
\eeq
It will be also convenient to introduce the ratio
\beq\la{Ddef}
{\mathbb D}^-\equiv \frac{W^\uparrow}{W^\downarrow}=
\frac{\dot W^\uparrow}{\dot W^\downarrow}
= \[\tanh \pi\left(u+\frac{\ii}{4}\right)\]^{J-M}\;.
\eeq
This notation also allows us to write the periodic matrix $\Omega$ in a useful and explicit form. Evaluating \eq{qom11} at two values of $u$ differing by $i$, and using $i$-periodicity of $\Omega$, we find a linear system for components of this matrix, whose solution can be written as
\begin{align}\la{OmegaW}
    \Omega_{i}{}^{j}=\frac{q^\uparrow_{i}q^{\downarrow[2]}_{k}-q^{\uparrow[2]}_{i}q^{\downarrow}_{k}}{W^{\downarrow +}}\epsilon^{kj}\,.
\end{align}
Using the $\ii$-periodicity of $\Omega$ we also have
\beq\la{detOm}
\det \Omega =\det \dot\Omega = \left(\frac{W^\uparrow}{W^\downarrow }\right)^{+}={\mathbb D}\;.
\eeq

\subsection{Quantum Spectral Curve}\label{sec:OperatorialQSC}

We now have all the ingredients necessary to close the system of equations by writing the quantisation conditions and relating the conserved charges coming from the finite-dimensional transfer matrix $\hat t(u)$ and the Q-operator $\hat{\mathbb{Q}}_+$. This will allow us to relate the conformal dimension $\Delta$ to the coupling $\xi$.

\paragraph{Q-operator eigenvalues in terms of Q-functions.}

Using the solutions $q_{i},\dot{q}_i$ introduced in the previous section, we can now expand our original Q-operator as follows
\begin{equation}\label{eqn:Qintoq}
    \mathbb{Q}_+(u,\dot{u}) ={\mathbb c}\; \Gamma^{ij}(u)\, q_i(u)\, \dot{q}_j(\dot{u}) \,,
\end{equation}
where $\Gamma^{ij}(u)$ is an $i$-periodic matrix
and ${\mathbb c}$ is a constant, which we will choose later to fix a particularly convenient normalization of $\Gamma^{ij}(u)$.
Note that the diagonal terms in $\Gamma^{ij}(u)$ are constrained by the asymptotics~\eq{eqn:Qopasymptotics}. Indeed, by comparing~\eq{eqn:qasympts} with~\eq{eqn:Qopasymptotics}, we see that all off-diagonal terms in $\Gamma^{ij}(u)$ must vanish at infinity, whereas diagonal elements should tend to a constant.

Furthermore, we know that for $u=\dot u+in$, with $n$ a positive integer, there exists an analyticity strip for $\mathbb{Q}_\pm(u,\dot{u})$ without poles, whose width grows with $n$. To ensure this property, we choose a particular ansatz for $q_i$ and $\dot q_j$, namely $q_i^\downarrow$ and $\dot q_j^\uparrow$, which are analytic above and below the real axis, respectively. In this case, the only possible poles can come from the periodic coefficients $\Gamma^{ij}(u)$, and thus the required analyticity holds automatically for all $n>0$ provided we assume that $\Gamma^{ij}(u)$ is analytic\footnote{Since $\Gamma^{ij}(u)$ is $i$-periodic, analyticity in a single $i$-strip implies analyticity on the whole complex plane.}. Finally, we have already established that the diagonal elements of $\Gamma^{ij}(u)$ have constant asymptotics, while the off-diagonal elements must vanish at infinity. Since $\Gamma^{ij}(u)$ is $i$-periodic, it follows that it must be a constant function for $i=j$, and identically zero otherwise.

\paragraph{The quantisation condition.}
Note that the above argument can be repeated when $n$ is a negative integer as well. In this case we take the basis $q_i^\uparrow$ and $\dot q_i^\downarrow$ instead.
Since asymptotically $q_i^\uparrow \simeq q_i^\downarrow$, we conclude that there are two ways of writing~\eqref{eqn:Qintoq} in terms of solutions of the holomorphic/anti-holomorphic Baxter equations:
\begin{equation}\label{eqn:twochoices}
    {\mathbb Q}_+(u,\dot{u})
    ={\mathbb c}\; \Gamma^{ij}\, q_i^\downarrow(u)\, \dot{q}_j^\uparrow(\dot{u})
    ={\mathbb c}\;  \Gamma^{ij}\, q_i^\uparrow(u)\, \dot{q}_j^\downarrow(\dot{u})\,.
\end{equation}
However, the compatibility of these two representations is non-trivial and requires a fine-tuning of the parameters in the Baxter equations. In other words, requiring these two representations to coincide provides a quantisation condition for the integrals of motion!\footnote{Note that writing $T_n(v) = {\mathbb Q}_+(u,\dot u)=\Gamma^{ij}q_i^\downarrow(v+\tfrac{i n}{2}) \dot{q}_j^\uparrow(v-\tfrac{i n}{2})$ suggests that the functions $T_n(v)$ themselves satisfy a Hirota equation, and that there should exist a representation in which these $T_n(v)$ become eigenvalues of the corresponding transfer matrices. Since $T_n(v)$ has an analyticity strip, it is also natural to expect that the $T_n$'s are related to the $Y$-functions appearing in the TBA equations of~\cite{Basso:2019xay}.}

In particular, these constraints relate $h$ and $\dot h$, which have been completely independent until now. Furthermore, using the relation~\eqref{eq:QuantizationXi} between the coupling $\xi$ and the Q-operators (and hence their eigenvalues), this infinite family of constraints allows us to close the system and obtain a discrete spectrum for all integrals of motion (for fixed $\xi$), as we explore numerically in section~\ref{sec:FiniteCoupling}. As a result, we express the eigenvalues of all conserved charges, including the conformal dimension $\Delta$, in terms of the coupling constant $\xi$ and derive the simplified ABA equations at weak coupling in section~\ref{sec:ABA}.

\paragraph{Quantisation condition as a gluing condition.}

To estimate the number of independent constraints hidden in the requirement~\eq{eqn:twochoices}, we reformulate it in a much simpler form, involving only finitely many relations. These are very similar in spirit to the quantisation conditions of other known Quantum Spectral Curves, such as those appearing in the AdS/CFT context. The idea is to use the $i$-periodic $\Omega$-matrix relating the UHPA and LHPA Q-functions \eqref{qom11}. We then ensure \eqref{eqn:twochoices} by demanding that
\begin{equation}\label{eqn:quant2}
    \Gamma^{ik}\,\Omega_{k}{}^{j} = \Gamma^{jk}\,\dot{\Omega}_{k}{}^{i}\,.
\end{equation}

We do not need to track the exact overall normalisation of ${\mathbb Q}_+$ (which in principle can be done from its integral representation). We absorb our ignorance about the normalization of ${\mathbb Q}_+$ into the overall constant factor ${\mathbb c}$, which we choose so that $\Gamma^{11}=1$, leaving us with only one non-trivial constant,
\beq\la{gluingMatrix}
\Gamma^{ij}=\left(
\bea{cc}
1&0\\
0& c
\eea
\right)\;.
\eeq
Then in components \eq{eqn:quant2} becomes
\beq\la{quantcomponents}
\Omega_{11}=\dot\Omega_{11}\,,
\quad
\Omega_{22}=\dot\Omega_{22}\,,
\quad
c=\frac{\dot\Omega_{1}{}^2}{\Omega_{2}{}^1}=\frac{\Omega_{1}{}^2}{\dot\Omega_{2}{}^1}\,.
\eeq

We can now count the number of constraints. Since $\Omega_i{}^{j}$ has poles of order at most $J$, it can be parametrized in terms of $(J+1)\times 4$ residues, which are implicit functions of the $J$ coefficients in $t(u)$. From this we see that~\eq{eqn:quant2} imposes $(J+1)\times 4$ conditions on $\sim 2J$ coefficients in $t(u)$ and $\dot t(u)$, so it is far from obvious that it is possible to satisfy all constraints~\eq{eqn:quant2}. Nevertheless, by checking explicit examples as well as performing numerical studies, we find in all cases a discrete spectrum for a given value of $\xi$, fixed by the additional condition~\eq{eq:QuantizationXi}. Part of these constraints are guaranteed to be linearly dependent due to the Wronskian relations on the determinants of $\Omega$'s, derived above in~\eq{detOm}.

In this way, we also see that the quantisation condition of the bi-scalar fishnet theory in 2D is remarkably similar to the quantisation condition in 4D, and also to the quantisation conditions of QSCs in ${\rm AdS}_3$ integrable systems.
This completes the derivation of the QSC equations presented in section~\ref{sec:DefinitionAndQSC}. We now turn to tests of the proposal.

\section{Example of an analytic solution for $J=2$} \label{sec:ExactL2Solution}

In order to demonstrate explicitly how the equations we derived in the previous section work, here we give an exact analytic solution of the QSC in the simple case $J=2,M=0$, with all ingredients worked out explicitly in Appendix~\ref{app:ExactJ2}. As a result, we reproduce the result of~\cite{Grabner:2017pgm,Kazakov:2018qbr} for the spectrum $\Delta$ as a function of the spin $S$ and the coupling. Then in the rest of the paper, we explore the general $J$ case, for which the methods of~\cite{Grabner:2017pgm,Kazakov:2018qbr} are no longer applicable.

\paragraph{The exact Q-functions }
Let us start by writing out the Baxter equation for the undotted system for $J=2,M=0$. It takes the form
\begin{equation}\label{eq:L2Baxter}
    \left(u+\tfrac{\ii}{4}\right)^{2}\,q^{[2]} - \left(2 u^2-(h-1) h-\tfrac{3}{8}\right) q -\left(u-\tfrac{\ii}{4}\right)^2 q^{[-2]}=0\,
\end{equation}
where we note that the transfer matrix eigenvalues $t(u)$ are completely fixed by requiring that the Q-functions have asymptotics given by~\eq{eqn:qasympts}, which in this case read
\begin{equation}\label{eq:AsymptoticsL2M2}
    q_{1} \sim u^{h-\frac{1}{2}}\,,
    \quad
    q_{2} \sim u^{-h+\frac{1}{2}}\,.
\end{equation}
The equation for $\dot{q}$ is obtained by sending $h\to \dot{h}$ in \eqref{eq:L2Baxter}. Since the treatment of the dotted system is fully analogous, we will focus on the undotted one.

Using the methods of \cite{Derkachov:2001yn,Derkachov:2002pb} it is possible to find a solution of \eqref{eq:L2Baxter} by taking its Mellin transform, solving the resulting differential equation and then transforming back, which gives the following result
\begin{equation}\label{eq:FUHP}
    q_{\rm seed}(u)\equiv i \frac{\Gamma(-i u+\frac{3}{4})}{\Gamma(-i u+\frac{1}{4})}F(u)
    \,,
    \quad
    F(u) = {}_3 F_2(i u+3/4,1-h,h;\tfrac{3}{2},1;1)\,.
\end{equation}
After that one can use $u\to -u$ symmetry to generate the second linearly independent solution and find a particular combination of the two which has the specified analytic properties and asymptotics \eq{eq:AsymptoticsL2M2}. We present the details in Appendix~\ref{app:ExactJ2} and here we just quote the result for the
UHPA and the LHPA bases
\begin{equation}
\begin{split}
    &q_1^\downarrow/A^{\downarrow}_{1} = \,\left[- \ii \cos(h\pi)- \sin(h \pi)\coth{\pi(u- \ii \tfrac{3}{4})}\right] q_{\rm seed}(u)+q_{\rm seed}(-u)\tanh{\pi(u- \ii \tfrac{3}{4})}   \\
    &q_2^\downarrow/A^{\downarrow}_{2} = \,\left[+ \ii \cos(h\pi)- \sin(h \pi)\coth{\pi(u- \ii \tfrac{3}{4})}\right] q_{\rm seed}(u)+q_{\rm seed}(-u)\tanh{\pi(u- \ii \tfrac{3}{4})}   \\
    &q_1^\uparrow/A^{\uparrow}_{1}=q_{\rm seed}(u)\tanh{\pi(u+ \ii \tfrac{3}{4})}  + \left[+\ii \cos(h\pi)-\sin(h\pi)\coth{\pi(u+ \ii \tfrac{3}{4})}\right] q_{\rm seed}(-u)\\
    &q_2^\uparrow/A^{\uparrow}_{2}=q_{\rm seed}(u) \tanh{\pi(u+ \ii \tfrac{3}{4})} + \left[-\ii \cos(h\pi)-\sin(h\pi)\coth{\pi(u+ \ii \tfrac{3}{4})}\right] q_{\rm seed}(-u)
\end{split}
\end{equation}
Here the explicit form of the normalization factors is
\beqa
    &&A_{1}^{\downarrow} = +e^{\pi\ii h}\,A^{\uparrow}_{1}= \frac{ 4^{1-h} (2 h-1) \pi }{\Gamma (1-h)\sin{\left(2\pi h\right)}}e^{\frac{ i \pi}{2}  (h-\frac{1}{2})} \,, \\
    &&A_{2}^{\downarrow} = - e^{-\pi\ii h}\,A^{\uparrow}_{2}= \frac{ 4^{h} (2 h-1) \pi }{\Gamma (h)\sin{\left(2\pi h\right)}}e^{\frac{i \pi }{2}  (-h+\frac{1}{2})}\,.
\eeqa
Now it is straightforward to check that the two bases are related by the following periodic matrix:
\begin{equation}
    \Omega_i{}^j = \frac{1}{\left(e^{2 \pi  u}+i\right)^2}\left(
\begin{array}{cc}
 e^{4 \pi  u}+e^{-2 i \pi  h} & \ii \, e^{2 \pi  u}\, 2^{3-4 h}\sin (\pi  h)\, \frac{ \Gamma (h)}{\Gamma (1-h)} \\
 i\, e^{2 \pi  u} \,2^{4 h-1}\,\sin (\pi  h)\,\frac{  \Gamma (1-h)}{\Gamma (h)} & e^{4 \pi  u}+e^{2 i \pi  h} \\
\end{array}
\right)\,.
\end{equation}
With $\Omega$ at our disposal, we impose \eqref{quantcomponents} to deduce that
\beq
e^{2\pi i h}=e^{2\pi i \dot h}\;\;,\;\;
c = 2^{-4 h-4 \dot{h}+4} \frac{\sin (\pi  h)}{\sin(\pi \dot{h})} \frac{\Gamma (h) \Gamma (\dot{h})}{\Gamma (1-h) \Gamma (1-\dot{h})}\,.
\eeq
Here, the first identity simply implies that the spin $S$  (given by $S=\dot h-h$, see \eqref{dshh}) is an integer. Thus, to explore non-integer $S$, one has to modify the structure of $\Gamma$. It would be fascinating to derive the structure of $\Gamma$ for non-integer spin from our operators $\hat{\mathbb{Q}}_{\pm}$, we leave this for future work.

\paragraph{Relation to the coupling constant and the spectrum.} Note that the sum $\dot h+h=\Delta$ so far remains unconstrained. To fix $\Delta$ we have to inject the coupling constant $\xi$ via \eq{eq:QuantizationXi}. For that we have to construct ${\mathbb Q}_+(u)$ with \eq{eqn:twochoices}, where we know now all the ingredients. Then we have to simply evaluate it at $u=i/4$ and expand around $u=3i/4$.
After massive cancellations we find

\begin{equation}
    \xi^{4} = \frac{1}{4}\left(1+S-\Delta\right)\left(-1+S+\Delta\right)\,,
\end{equation}
perfectly reproducing the result of \cite{Kazakov:2018qbr}! This gives a quadratic equation for the dimension of the local operator $\tr \phi_1 D_+^S\phi_1$ and its shadow operator. Of course, in this particular case one can diagonalize the graph-building operator explicitly to verify the result, we give the details in Appendix~\ref{app:DirectDiag}.

\section{Numerical solutions of the fishnet QSC\label{sec:FiniteCoupling}}
In this section, we study the isotropic fishnet QSC using numerical methods. The numerical algorithm is an adaptation of the 4D algorithm described in \cite{Gromov:2019jfh}, itself inspired by the more sophisticated algorithm developed for the spectrum of ${\cal N}=4$ SYM in \cite{Gromov:2015wca}.

Let us briefly review the main steps in the numerical evaluation of the spectrum. The starting point is the Baxter equations \eqref{eq:Baxter1ABA}, \eqref{eq:Baxter2ABA}, which we readily solve at $u\rightarrow \pm i \infty$ to find both $q^\downarrow_i(u)$ and $q^\uparrow_i(u)$ as functions of the parameters in $t(u)$ and $t(\dot u)$. We then use the Baxter equations to evaluate these functions at any finite value of the spectral parameter. In particular, we compute $\Omega_i{}^j(u)$ and $\dot\Omega_i{}^j(u)$ using \eqref{OmegaW} and impose the quantisation conditions \eq{quantcomponents} and the relation to $\xi$ in \eq{eq:QuantizationXi}. These are solved using a Newton-type method, in which we adjust the parameters in $t(u)$ and $\dot t(u)$ to minimize the mismatch in these conditions.

Using this numerical algorithm, we now study the non-perturbative spectrum of 2D bi-scalar fishnets. We present three applications: first, in section~\ref{subsec:FiniteCouplingJ3} we study the non-perturbative spectrum of the simplest non-trivial family of operators that cannot be solved exactly; in section~\ref{subsec:Luscher}, we compute  Lüscher corrections (or equivalently, one-wheel diagrams) numerically and compare with analytic predictions as a test of our results; and finally, in section~\ref{subsec:MagnonNumerics}, we study operators with magnons.

\subsection{Finite coupling solution for $J=3$, $M=0$ case}\label{subsec:FiniteCouplingJ3}
As conformal symmetry does not fix the form of the CFT wave function for $J>2$,  already at $J=3$ the problem of finding the spectrum becomes considerably more complicated and, except for some special regimes, there is very little hope this problem can be solved analytically at finite values of $\xi$. Motivated by this, we consider in this section
\begin{equation}
    J=3\,,
    \quad
    M=0\,,
    \quad
    S = 0\,,
\end{equation}
as the simplest demonstrative example where other methods do not apply. Furthermore, we impose cyclicity, that is, $U=1$, defined in \eqref{introShiftMnJ}.

The Baxter equation in this particular case becomes
\begin{equation}\label{eq:BaxterJ3}
    \left(u+\frac{i}{4}\right)^{3} q^{[2]}+ q \,t(u) + q^{[-2]}\left(u-\frac{i}{4}\right)^{3}  = 0\,,
    \quad
     t(u) = 2\,u^3-\frac{u}{16}\left(4 (\Delta -2) \Delta +9\right) + c_{-1}\,.
\end{equation}
The only unfixed parameter in \eqref{eq:BaxterJ3} is $c_{-1}$. For $\dot{t}$, we have an equivalent eigenvalue equation with $c_{-1}\leftrightarrow
\dot c_{-1}$. In all cases we considered, the cyclicity requirement fixed $\dot c_{-1}=-c_{-1}$. Finally, the asymptotics of the $q$-functions are $q_{1},\dot{q}_{1} \simeq u^{\frac{1}{4}\left(2\Delta-3\right)}$.

To find a starting point for our numerical algorithm, we begin at weak coupling. Let us write $\Delta_0 =\Delta\big|_{\xi=0}$. For states with $\Delta_0 = \frac{3}{2},\frac{7}{2},\dots$, the asymptotics of $q_{1}$ and $\dot{q}_{1}$ are integer. This inspires us to consider a polynomial ansatz for $q_{1},\dot{q}_1$ in this case. When $\Delta_0=\frac{5}{2},\frac{9}{2},\dots$, the asymptotics of $q_{1},\dot{q}_1$ become half-integer and clearly a polynomial ansatz is no longer appropriate. We recall that, in general, $q^{\downarrow}$ are functions with poles at $-\frac{3i}{4}-i n$, $n\in \mathbb{Z}_{>0}$. To find an appropriate parametrization of $q$, we use intuition from magnon-anti-magnon annihilation. Indeed, as we show in Appendix~\ref{app:DiscreteReparam}, magnon-anti-magnon annihilation redefines $q$ with a gauge factor given as a ratio of $\Gamma$-functions with precisely half-integer asymptotics.

Summarizing the above discussion, we introduce polynomials
\begin{equation}\label{eq:NonPolynomialQAnsatz}
    \mathcal{Q} = \prod_{i=1}^{K}\left(u-v_{i}\right)\,,
    \quad
    \dot{\mathcal{Q}} = \prod_{i=1}^{\dot{K}}\left(u-\dot{v}_{i}\right)\,,
\end{equation}
and consider the following ansatz:
\begin{equation}\label{eq:TreeLevelAnsatz}
    q_{1}(u) = \mathcal{Q}(u)\,,\,\Delta_0=\frac{3}{2},\frac{7}{2},\dots
    \quad
    q_{1}(u) = \frac{\Gamma(-i u+\tfrac{3}{4})}{\Gamma(-i u+\tfrac{1}{4})}\mathcal{Q}\,,\,\Delta_0=\frac{5}{2},\frac{9}{2},\dots
\end{equation}
We could also in principle consider an ansatz involving more $\Gamma$-functions, but we will here limit ourselves to the above configuration\footnote{We looked for states with $q_1 \sim \left(\Gamma\left(-\ii u+\frac{3}{4}\right)/\Gamma\left(-\ii u+\frac{1}{4}\right)\right)^{2}\mathcal{Q}$ but didn't find any examples for which our numerics converged. It is unclear to us if this is a feature for $J=3, M=0$ or if we simply need better initial data.}.

We then plug \eqref{eq:TreeLevelAnsatz} into the Baxter equation to deduce $t(u)$, giving us a starting point around which to initialize numerics. We enumerate the states we studied at finite coupling in Table~\ref{tab:WeakStarting}. We cannot rule out that there might exist additional states for $\Delta_0\leq 11/2$; it would be interesting to find a complete classification of states. The consideration in the section~\ref{sec:ABA}, where we derive the ABA equations, in principle could provide a reliable way of counting the states and finding their weak coupling starting points for numerics.

\definecolor{navyblue}{HTML}{1E3A8A}
\definecolor{darkgreen}{HTML}{2D5016}
\definecolor{yellowgreen}{HTML}{84CC16}
\definecolor{salmon}{HTML}{F87171}
\definecolor{darkred}{HTML}{991B1B}
\definecolor{teal}{HTML}{14B8A6}

\begin{table}[h]
    \centering
    \begin{tabular}{c|c|c|c|c}
         $\Delta_0$ & $\mathcal{Q}$ & $\dot{\mathcal{Q}}$ & $c_{-1}$ & Plot colour in figure~\ref{fig:LowLyingSpectrum} \\
         \hline
         $3/2$ & $1$ & $1$ & $0$  & \color{darkred}dark red\\
         $5/2$ & $1$ &  $1$ & $0$&
         \color{salmon}salmon
         \\
         $7/2$ & $u\mp\tfrac{1}{4\sqrt{3}}$ & $u\pm\tfrac{1}{4\sqrt{3}}$ & $\mp\tfrac{\sqrt{3}}{8}$&\color{yellowgreen}yellow-green\\
         $9/2$ & $u\mp\sqrt{\frac{3}{5}}/4$ & $u\pm\sqrt{\frac{3}{5}}/4$ & $\mp\frac{\sqrt{15}}{8} $ & \color{darkgreen}dark green \\
         $11/2$ & $u^2-\frac{7}{80}$ &$u^2-\frac{7}{80}$ & $0$& \color{gray}gray
         \\
         $11/2$ & $u^2\mp\frac{3 \sqrt{3}}{14}u+\frac{1}{112}$ & $u^2\pm\frac{3 \sqrt{3}}{14}u+\frac{1}{112}$ & $\mp\tfrac{3 \sqrt{3}}{4}$ &
         \color{navyblue}navy blue
    \end{tabular}
    \caption{All starting $q$-functions at $\xi=0$ for which we studied $\Delta$ at finite coupling, see figure~\ref{fig:LowLyingSpectrum}. }
    \label{tab:WeakStarting}
\end{table}

\begin{figure}[]
    \centering
    \includegraphics[width=0.6\linewidth]{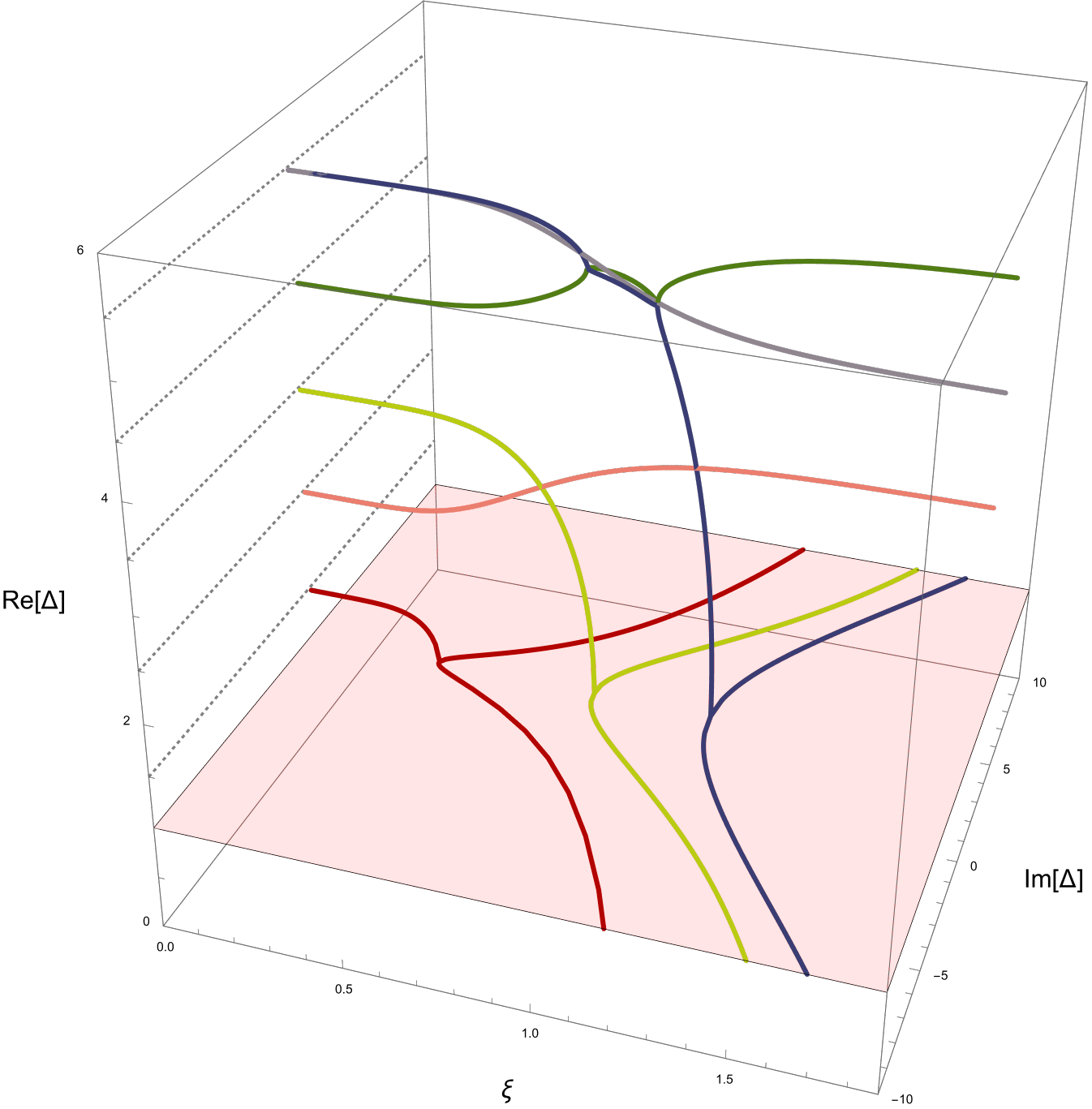}
    \caption{We plot the low-lying spectrum for 2D bi-scalar fishnet theory from weak to strong coupling for operators with $J=3$, $M=0$, and $S=0$. As expected, we find collisions among states and complex energy levels.}
    \label{fig:LowLyingSpectrum}
\end{figure}

We plot the resulting energy levels in figure~\ref{fig:LowLyingSpectrum}. As can be clearly seen in that figure, the spectrum at finite coupling is highly non-trivial, sharing many features with its 4D counterpart \cite{Gromov:2017cja} as well as the BFKL regime of $\mathcal{N}=4$ \cite{Ekhammar:2024neh,Ekhammar:2025vig}. In particular, at finite coupling, various trajectories collide, either amongst themselves or with their shadows at $\Delta=1$, resulting in complex energies.

\begin{figure}[]
    \centering
    \includegraphics[width=0.7\linewidth]{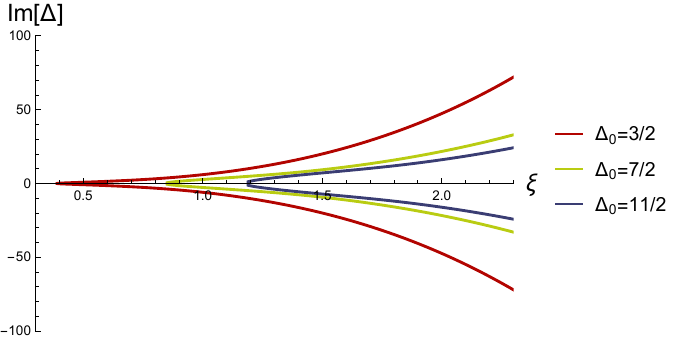}
    \caption{The strong coupling behaviour of the imaginary part of $\Delta$ for the states that become complex at $\Delta=1$; see figure~\ref{fig:LowLyingSpectrum}. We find numerically that $\Im[\Delta] \sim \xi^{3}$ for all these states.}
    \label{fig:StrongCouplingFigDelta}
\end{figure}

\begin{figure}
    \centering
    \includegraphics[width=0.7\linewidth]{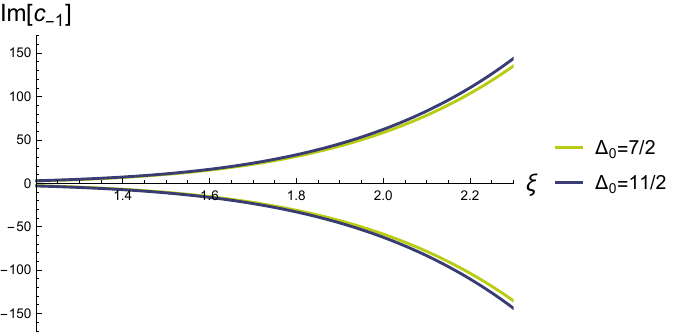}
   %\begin{tikzpicture}
        %\node[inner sep=0] (img) \includegraphics[width=0.7\linewidth]{StrongPlotQ_2.pdf}}
        %\node[
        %    anchor=north west,
        %    xshift=-2mm, yshift=0mm,
        %    fill=white, fill opacity=1,
        %    text opacity=1,
        %    inner sep=1.5pt,
        %    rounded corners=1pt
        %] at (img.north west) {$\Im[c_{-1}]$};
    %\end{tikzpicture}
    \caption{The strong coupling behaviour of the imaginary part of $c_{-1}$; see \eqref{eq:BaxterJ3}. We show the states that become complex at $\Delta=1$; see figure~\ref{fig:LowLyingSpectrum}. We find numerically that $\Im[c_{-1}] \sim \xi^{6}$ for all these states.}
    \label{fig:StrongCouplingFigQ}
\end{figure}

\paragraph{Strong coupling data.}
Let us comment on the strong coupling behaviour of the spectrum obtained for $J=3$. We find two distinct behaviours. The first is that an operator approaches a finite real value of $\Delta$ as $\xi\rightarrow\infty$; the second is that an operator collides with its shadow at $\Delta=1$ and develops an imaginary part that subsequently grows with the coupling. We present a plot of $\Im[\Delta]$ in figure~\ref{fig:StrongCouplingFigDelta} and of $c_{-1}$ in figure~\ref{fig:StrongCouplingFigQ}. At strong coupling, we found the following fits for $\Delta$:
\begin{align}
    \Delta\bigg|_{\xi=0}&=\frac{3}{2}\,, \quad
    \Im[\Delta] = 5.917350 \xi ^3+\frac{0.06337296}{\xi ^3}+\frac{0.003280404}{\xi ^9}+\dots\\
    \Delta\bigg|_{\xi=0}&=\frac{7}{2} \,, \quad
    \Im[\Delta]=2.703912 \xi ^3+\frac{0.5085222}{\xi ^3}+\frac{0.5212628}{\xi ^9}+\dots
    \\
    \Delta\bigg|_{\xi=0}&=\frac{11}{2}\,, \quad
    \Im[\Delta]=1.972450 \xi ^3+\frac{2.220859}{\xi ^3}+\frac{4.692624}{\xi ^9}+\dots
\end{align}
We have not been able to guess an exact expression for any of these coefficients. We note that the strong coupling spectrum appears significantly more difficult compared to that of 4D bi-scalar fishnet \cite{Gromov:2017cja}. This is indeed expected since we can only define the coupling through the quantisation condition \eqref{eq:QuantizationXi}, while in 4D it appears explicitly in the Baxter equation unless $J=M$.
It is fully possible that there exist further distinct behaviours at strong coupling; it would be interesting to fully classify the spectrum in this regime and attempt to understand it analytically with some suitable version of WKB formalism, for example.

\subsection{The first Lüscher correction}\label{subsec:Luscher}
In this section, we study the first corrections to the conformal dimension of operators of the form $\tr\bar{\partial}^{S} \phi^{J}_{1}+\text{permutations}$, usually called Lüscher corrections. In a Feynman diagram approach to 2D bi-scalar fishnet, this would amount to computing a one-wheel diagram and hence these corrections are expected to appear at order $\xi^{2J}$, which is precisely what we find from our numerical solutions of the QSC. To be precise, we expect the dimension of this operator to be of the form:
\begin{equation}
    \Delta = \frac{J}{2}+\abs{S}+\xi^{2J}\,\gamma_{2J}+\dots\,,
\end{equation}
and our objective is to calculate $\gamma_{2J}$. In the special case of $S=0$, the computation of $\gamma_{2J}$ from explicit evaluation of the one-wheel graph was performed in \cite{Derkachov:2018rot}, where a final formula was given as\footnote{We have slightly modified the expression $(6.7)$ in \cite{Derkachov:2018rot} to account for what we believe are notation differences, namely we have set $\xi^{2}_{\cite{Derkachov:2018rot}} = \frac{(-1)^{J+1}}{\pi}\xi^{2}$.}
\begin{equation}\label{eq:WrappingFormula}
    \gamma_{2J}\bigg|_{S=0} = 2\frac{(-1)^{J-1}}{(J-1)!}\frac{d^{J-1}}{d\epsilon^{J-1}}\bigg|_{\epsilon=0}\left(\frac{\Gamma\left(1+\epsilon\right)\Gamma\left(1-\epsilon\right)}{\Gamma\left(3/2+\epsilon\right)\Gamma\left(-1/2-\epsilon\right)}\right)^{J}\left(\sum_{k=0}^{\infty}\frac{\Gamma^{J}\left(1/2+k-\epsilon\right)}{\Gamma^{J}\left(1+k-\epsilon\right)}\right)^2\,.
\end{equation}
Solving our QSC numerically, we perfectly reproduced \eqref{eq:WrappingFormula} to high precision for $J=3,4,5$. Furthermore, for $J=3$ we managed to simplify \eqref{eq:WrappingFormula} to
\begin{equation}
    \Delta \bigg|_{J=3,M=0,S=0}= \frac{3}{2}-\frac{4\pi^4}{\Gamma(\frac{3}{4})^{8}}\xi^6+\dots\,.
\end{equation}
Using the $S=0$ result as a clue, we computed $S=2,4,\dots 14$ numerically which we fitted to obtain the following expression:
\begin{equation}
    \gamma_{2J}\bigg|_{J=3,M=0,\Delta_0 = \frac{3}{2}+\abs{S}} =-\frac{2 \,\sqrt{2}\, \pi ^{5/2} }{\Gamma
   \left(\frac{3}{4}\right)^6}
    \frac{\Gamma \left(\frac{\abs{S}}{2}+\frac{1}{4}\right)
   \Gamma \left(\frac{\abs{S}}{2}+\frac{1}{2}\right)}{ \Gamma \left(\frac{\abs{S}}{2}+\frac{3}{4}\right)
   \Gamma \left(\frac{\abs{S}}{2}+1\right)}\,.
\end{equation}
While this result was obtained using numerics, we are highly confident in the analytic form due to our very high precision. It would be fascinating to develop an efficient perturbative algorithm to solve the Baxter equations systematically. This would require a novel basis of functions, as the so-called $\eta$-function basis, applicable in 4D \cite{Leurent:2013mr,Marboe:2014gma}, does not seem to be suitable in this case. It would also be interesting to elucidate the algebraic/motivic interpretation (see e.g. \cite{Gurdogan:2020ppd}) of the $\Gamma$-function values which as we see appear here in the anomalous dimension, in contrast to N=4 SYM where only multiple zeta values can appear.

\subsection{The $J=3$ case with magnons}\label{subsec:MagnonNumerics}
In this section, we consider operators with magnons, that is the operators of schematic form $\tr \phi^{J-M}_1\left(\phi_1\phi_2\right)^{M}$. These states are amenable to the Asymptotic Bethe Ansatz approach; we already presented the main equations in section~\ref{subsec:IntroducingABA} and will derive them in section~\ref{sec:ABA}. Hence, in this section we will combine our ABA and numerics, finding a perfect match every time. As is usual when considering ABA equations we will discard any solution that features equal Bethe roots.

With ABA or numerics, we can straightforwardly consider any value of $J$ and $M$, but for presentation purposes, we will here constrain ourselves to $J=3,M=1,2,3$ since this is the first non-trivial case. Thus, the quantum numbers that we consider in this subsection are
\begin{equation}
    \Delta\big|_{\xi \rightarrow 0} = \frac{J+M}{2}\,,
    \quad
    J=3\,,
    \quad
    M=1,2,3\,,
    \quad
    S=0\,.
\end{equation}
Note that the coupling enters into the system only in the combination $\xi^{2J}$ via \eq{eq:QuantizationXi}. At the same time, for the states with magnons we expect, in the generic situation, to get expansion in $\xi^2$.
This implies that there is a $Z_J$ symmetry, generated by $\xi^2\to \xi^2 e^{2\pi i/J}$.
Due to this it is convenient to introduce the notation
\begin{equation}
   \hat{\xi}^2_n = e^{\frac{2\pi \ii}{3}n}\xi^2\,,\, n=0,1,2\,.
\end{equation}

\begin{figure}[h]
    \centering
    \includegraphics[width=0.7\linewidth]{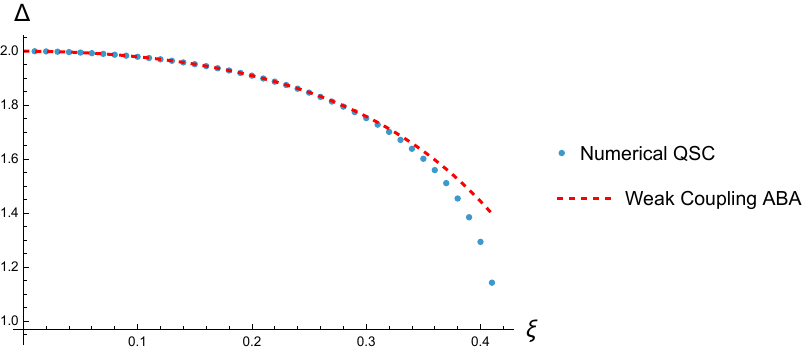}
    \caption{We plot the conformal dimension for $J=3,M=1$ and $U=1$ obtained by numerically solving the QSC and compare it to the weak coupling prediction coming from ABA up to $\xi^{6}$. One can see excellent agreement for small values of $\xi$ and a deviation as $\Delta$ starts to rapidly approach $1$.}
    \label{fig:J3M1}
\end{figure}

\paragraph{The single magnon case.} For $J=3,M=1$ we found $3$ states (where only one of them satisfies the cyclicity condition). These states are distinguished by the eigenvalue of $\hat{U}$. Let us enumerate the anomalous dimensions of these states as $\gamma_{(n)},\,n=0,1,2$. From the ABA we find the following expansion
\begin{equation}\label{eq:J3M1ABA}
   \gamma_{(n)} = -2\,
   \hat{\xi}_n^2\,-4\, \log 4\,\hat{\xi}_n^{4}-12 \log ^2 4 \,\hat{\xi}^{6}_n +\mathcal{O}\left(\xi^{8}\right)\,,
   \quad
   U = e^{\frac{2\pi \ii}{3}n}\,,
\end{equation}
which matches perfectly with our numerics to very high accuracy. At $\xi^8$ the ABA fails, showing that wrapping effects kick in at $\xi^{2(L+M)}$. For $U=1$ we plot the energy using the numerical QSC and compare it to the weak coupling prediction obtained from ABA, i.e truncating \eqref{eq:J3M1ABA} at $\xi^{6}$, in figure~\ref{fig:J3M1}. As can be clearly seen in that figure, the state will eventually collide with is shadow at $\Delta=1$.

\begin{figure}[h]
    \centering
    \includegraphics[width=0.7\linewidth]{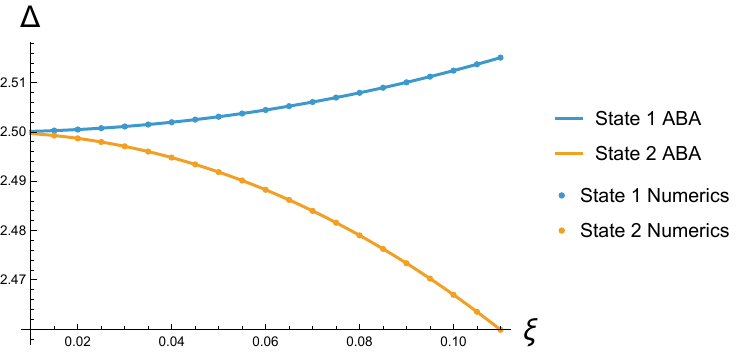}
    \caption{The dimensions of the two states satisfying cyclicity condition for $J=3,M=2$, see \eqref{eq:J3M2ABA1} and \eqref{eq:J3M2ABA2}.}
    \label{fig:J3M2}
\end{figure}

\paragraph{The case of two magnons.}
For $M=2$ there are $6$ states in total, we will label the anomalous dimension as $\gamma_{(n),i}, i=1,2,n=0,1,2$. The leading order solution is rather messy, we will therefore present the results using numerical digits, we find from the ABA
\begin{align}\label{eq:J3M2ABA1}
\gamma_{(n),1} &= 1.23606 \hat{\xi}^2_n+0.947229 \hat{\xi}^4_n -4.31140 \hat{\xi}^6_n+
1.34130 \hat{\xi}^8_n+\dots,  \quad
U=e^{\frac{2\pi \ii}{3}n}\,,
\quad
\\ \label{eq:J3M2ABA2}
\gamma_{(n),2} &=-3.23606 \hat{\xi}^2_n-6.49240 \hat{\xi}^4_n-18.75034 \hat{\xi}^6_n-58.1775 \hat{\xi}^8_n+\dots  \quad
U=e^{\frac{2\pi \ii}{3}n}\,.
\end{align}
These results are once again in perfect agreement with numerics and wrapping effects are now first present at $\xi^{10}=\xi^{2(J+M)}$. We have plotted $\Delta$ for small values of $\xi$ in figure~\ref{fig:J3M2} for $U=1$.

\paragraph{A fully filled $J=3$ spin chain.}
Finally, we consider $J=M=3$. In this case, we find $10$ states from solving the ABA. For $U=1$ there are $4$ states, let us label them as $\gamma_{(0),n}\,, n=0,1,2$ and $\gamma_{(0),3}$. We observed numerically that the ABA correctly predicts $\Delta$ to $\mathcal{O}\left(\xi^{10}\right)$, thus consistently we are getting wrapping at $\xi^{2(J+M)}$ for $M=1,2,3$. We present the first few coefficients below for $U=1$
\begin{align}\label{eq:GammaJ3M31}
    \gamma_{(0),n} &= -4\,\hat{\xi}_n^2-\frac{16}{3}
   \log 4\,\hat{\xi}_{n}^4 -\frac{32}{3} \log ^24 \,\hat{\xi}_n^6+\left(\frac{8}{3}\zeta_3-\frac{2048}{81} \log ^34\right)\hat{\xi}_n^8+\dots \,, \\
   \gamma_{(0),3} &= \mathcal{O}\left(\xi^{12}\right)\,.
\end{align}
Note that in this case the index $n$ is not related to the eigenvalue of the shift operator.
Our numerics perfectly reproduces \eqref{eq:GammaJ3M31}, however, we were not able to find initial points, which would converge to the last state, i.e $\gamma_{(0),3}$.

The conformal dimensions for $U=e^{\pm 2\pi \ii/3}$ are quiet similar up to a phase that can be reabsorbed into the coupling constant. Explicitly, for $U=e^{2\pi \ii/3}$  we obtain
\begin{equation}
\begin{split}
    \gamma_{(1),n} =-2\,e^{\tfrac{\ii \pi}{9}}\, \hat{\xi}^{2}_{n} -&\frac{4}{3}  \log(4) e^{\tfrac{2\ii \pi}{9}}\, \hat{\xi}^{4}_{n}\\
   &-\frac{4}{3}\log ^2(4)e^{\tfrac{3\ii \pi}{9}}\, \hat{\xi}^{6}_{n}+ \left(\frac{8}{3}\zeta_3-\frac{128}{81} \log
   ^3(4)\right)e^{\tfrac{4\ii \pi}{9}}\, \hat{\xi}^{8}_{n}+\dots\,,
\end{split}
\end{equation}
while for $U=e^{4\pi \ii /3}$ the anomalous dimension takes the form
\begin{equation}
\begin{split}
    \gamma_{(2),n} =-2\,e^{\tfrac{11\ii \pi}{9}}\, \hat{\xi}^{2}_{n} -&\frac{4}{3}  \log
   (4) e^{2\tfrac{11\ii \pi}{9}}\, \hat{\xi}^{4}_{n}\\
   &-\frac{4}{3}\log ^2(4)e^{3\tfrac{11\ii \pi}{9}}\, \hat{\xi}^{6}_{n}+ \left(\frac{8}{3}\zeta_3-\frac{128}{81} \log
   ^3(4)\right)e^{4\tfrac{11\ii \pi}{9}}\, \hat{\xi}^{8}_{n}+\dots\,.
\end{split}
\end{equation}

\section{Derivation of the Asymptotic Bethe Ansatz}\label{sec:ABA}
Based on experience with other holographic integrable models such as ${\cal N}=4$ SYM or, more closely related, bi-scalar fishnet theory in higher dimensions, one expects massive simplifications in the large $J$ limit, or at weak coupling up to the so-called wrapping order, in our case $\sim \xi^{2J}$. In this regime, the exact Baxter equations alongside the quantisation conditions (to which we refer as QSC) are expected to reduce to a set of algebraic so-called Asymptotic Bethe Ansatz (ABA) equations.

In this section, we derive these equations from the QSC proposal. We find a perfect match for a subsector of operators considered in \cite{Basso:2019xay} (which have bare dimension $\frac{J+M}{2}$). The ABA we derived also includes ``excited states,'' i.e., states with large bare dimension $\frac{J+M}{2}+K+\dot K$ for the same operators with $J$ and $M$ R-charges and spin $\dot K-K$, corresponding to the operators of the type $\tr \d^K\bar\d^{\dot K}\phi_1^J\phi_2^M$.

\subsection{Key assumptions in the ABA regime}
Let us now spell out the key assumptions that will allow us to analytically investigate the ABA regime. These conditions are inspired by the fishnet limit of the ABA of ${\cal N}=4$ SYM, in the style of \cite{Caetano:2016ydc}, and verified numerically for a number of states.

We start by assuming that when $\xi\to 0$, the functions $q_1^\downarrow$ and $\dot q_1^\downarrow$ have $M$ zeroes at $u_{k}-\tfrac{i}{2}$ and $\dot u_{k}-\tfrac{i}{2}$, which behave as $u_{k},\dot{u}_k \simeq \tfrac{i}{4}+{\cal O}(\xi^2)$. As we will later see, it turns out that $u_k = \dot{u}_k$. In addition, $q_1^{\downarrow},\dot{q}_1^{\downarrow}$ could have additional zeroes, $v_i,\dot{v}_i$, which do not approach $-i/4$. We arrange them into the Baxter-style monic polynomials
\begin{equation}
    {\cal Q}=\prod_{i=1}^{K}(u-v_i)\,,
    \quad
    \dot{\cal Q}=\prod_{i=1}^{\dot{K}}(u-\dot v_i)\,.
\end{equation}

\paragraph{Key simplification.}
Consider the Baxter equation \eqref{eq:Baxter1ABA} for $q_{1}^{\downarrow}$. If we evaluate this equation at $u = u_k-\tfrac i2$, the middle term drops out since $q_1^{\downarrow}$ vanishes by assumption. At the same time, the last term is suppressed as $\xi^{2J}$ since $u_k-\tfrac i4 = {\cal O}(\xi^2)$. From this, we conclude that
\beq
q^{\downarrow}_1(u_k-\tfrac{3i}{2})={\cal O}(\xi^{2J})\,.
\eeq
Now iterating this argument, we conclude that for any $n\geq 0$ we have
\beq
q^{\downarrow}_1(u_k-\tfrac i2-i n)={\cal O}(\xi^{2J})\;\;,\;\;n=0,1,\dots\;.
\eeq
Thus, we have infinitely many zeroes at the leading order in $\xi$. To obtain power-like asymptotics, it must be that we also have infinitely many poles. We recall that $q^\downarrow_1$ can have poles of maximal degree $M$ at $-\tfrac{i}{4}- in$ or poles at $-\tfrac{3i}{4}- in$ of maximal degree $J-M$. In this section, we will restrict ourselves to solutions which start from polynomial $q_{1},\dot{q}_1$ at $\xi=0$; this implies that the pole at $-\frac{3i}{4}-i n$ is necessarily suppressed.

As we discovered in section~\ref{sec:FiniteCoupling}, there is a more general class of states which violates this condition in a controllable way. We consider these states later in section~\ref{subsec:StrangeStates}.

To summarize, we will take the following ansatz:
\begin{equation}\label{eq:q1ABA}
q_1^\downarrow = {\bf q}^\downarrow_1+{\cal O}(\xi^{2J})\,,
\quad
{\bf q}_1^\downarrow(u)  \equiv
e^{+\tfrac{1}{4} i \pi  \gamma}
{\cal Q}(u)
\prod_{k=1}^{M}
\frac{\Gamma(-i u+\tfrac 14)}{
\Gamma(-i u+ i u_{k}+\tfrac 12)}\,,
\end{equation}
and a similar expression for $\dot q_1^\downarrow$. In this section in general we use the bold font to distinguish ABA approximations from the exact quantities (e.g. like in \eq{eq:q1ABA}), we will define $\gamma$ shortly below, see \eqref{eq:Dispersion}. The above approximation should be valid everywhere except in the vicinity of the order $J-M$ poles at $-3i/4-i n$ of the exact $q_1^\downarrow$ (where ${\bf q}_1^\downarrow$ is regular). We will now proceed to outline how to find equations that fix all $u_{k},v_{k},\dot{v}_k$.

\paragraph{The ABA dispersion relation.}
As \eq{eq:q1ABA} is valid, in particular, at large real $u$,
we can read off the asymptotics of $q_1^\downarrow$ and
$\dot q_1^\downarrow$
from \eqref{eq:q1ABA}, leading to
\begin{equation}
    q_{1} \simeq u^{K+\gamma/2}\,,
    \;
    \dot{q}_{1} \simeq u^{\dot{K}+\dot\gamma/2}\,,
\end{equation}
where $\gamma$ in \eqref{eq:q1ABA} is given by the following sum over roots:
\beq\label{eq:Dispersion}
\gamma = \sum_{k=1}^{M}\(-2 i u_k-\frac{1}{2}\)\;\;,\;\;
\dot\gamma = \sum_{k=1}^{M}\(-2 i \dot u_k-\frac{1}{2}\)\;.
\eeq
Comparing with the general asymptotics in terms of quantum numbers \eqref{eqn:qasymptotics}, we obtain $\gamma=\dot\gamma$ (which already hints towards $u_k=\dot u_k$) and
\begin{equation}
    S = \dot{K}-K\,,
    \quad
    \Delta = \frac{J+M}{2}+K+\dot{K}+  \gamma\,.
\end{equation}
This shows that we reproduce perfectly the dispersion relation expected from the TBA analysis in \cite{Basso:2019xay}.

\paragraph{Finding ${q}^{\uparrow}_i$.}
Below we will see that the constant $c$ in the gluing matrix \eq{gluingMatrix} is very small at weak coupling, $c= {\cal O}(\xi^{2J})$. Then due to \eq{quantcomponents} we see that if $\Omega_{2}{}^1\sim 1$, then $\Omega_{1}{}^{2}$ is suppressed up to the wrapping order. This behaviour of $\Omega$ can be anticipated from the fact that as $\xi\rightarrow 0$, $\mathbf{q}_{1}$ becomes polynomial, and hence is both upper and lower half-plane analytic. The vanishing of $\Omega_{1}{}^{2}$ implies that $\mathbf{q}_{1}^\uparrow = \mathbf{\Omega}_{1}{}^{1}\mathbf{q}_{1}^{\downarrow}$, a very constraining equation since $\mathbf{\Omega}_{1}{}^{1}$ must be a periodic function with constant asymptotics. This leads to
\beq\label{eq:q1ABAUp}
q_1^\uparrow = {\bf q}^\uparrow_1+{\cal O}(\xi^{2J})\,,
\quad
{\bf q}_1^\uparrow(u)  \equiv
e^{-\tfrac{1}{4} i \pi  \gamma}
{\cal Q}(u)
\prod_{k=1}^{M}
\frac{\Gamma \left(+i u-i u_{k}+\frac{1}{2}\right)}{\Gamma \left(+i u+\tfrac{3}{4}\right)}\;.
\eeq
In order to see that this is the case, we notice that
${\bf q}_1^\uparrow(u)$ is analytic in the lower half-plane, has the correct asymptotics and is related to $\mathbf{q}^{\downarrow}_1$ by a periodic function.

\paragraph{The periodic matrix $\Omega$.} Let us now deduce as much information as possible about the periodic matrix $\Omega$ away from its poles. We already know that this matrix takes the form
\begin{equation}\label{eq:ABAOmega}
    \Omega \simeq \begin{pmatrix}
    \mathbf{\Omega}_{1}{}^{1} & 0 \\
    \mathbf{\Omega}_{1}{}^{2} & \mathbf{\Omega}_{2}{}^{2} \\
    \end{pmatrix}+\mathcal{O}(\xi^{2J})\,.
\end{equation}
Furthermore, from \eqref{eq:q1ABA} and \eqref{eq:q1ABAUp} it follows immediately that
\begin{equation}\label{om11}
{\bf \Omega}_1{}^1=e^{-\frac{1}{2} i \pi \gamma} \prod_{k=1}^{M} \frac{\cosh (\pi(u-i / 4))}{\cosh \left(\pi (u- u_{ k})\right)}\;.
\end{equation}
Now we can finally establish the relation between $u_k$
and $\dot u_k$. Note that $\dot{\bf \Omega}_1{}^1$ would have the same form as \eq{om11} with $u_k\to\dot u_k$. But at the same time, the first quantisation condition in \eqref{eq:OmegaQuantisation} sets them equal and thus indeed $u_k=\dot{u}_{k}$.

To find $\mathbf{\Omega}_{2}{}^{2}$ we now notice that
the $(1,1)$ component of the inverse matrix $[\Omega^{-1}]_{1}{}^{1}$, which maps in the opposite direction $q^\uparrow\to q^\downarrow$, can also be deduced. Indeed, it must be equal to $1/{{\bf \Omega}_1{}^1}$ in our approximation, which then implies
\beq
\Omega_2{}^2 = \det \Omega\;[\Omega^{-1}]_1{}^1\simeq
\det \Omega\frac{1}{{\bf \Omega}_{1}{}^1}\equiv {\bf \Omega}_2{}^2\;.
\eeq
Note that we know $\det\Omega$ exactly from \eq{detOm}, and hence
\begin{equation}\label{om22}
    \mathbf{\Omega}_{2}{}^{2} = e^{\frac{1}{2}i \pi \gamma}\left[\tanh \pi\left(u-\frac{i}{4}\right)\right]^{J-M}\prod_{k=1}^{M} \frac{\cosh(\pi(u-u_{k}))}{\cosh(\pi(u-i/4))}\,.
\end{equation}
We note that whereas ${\bf\Omega}_1{}^1$ from \eq{om11} does approximate ${\Omega}_1{}^1$ with wrapping order precision for generic $u$, it fails to do so near the poles as it has poles at $u=u_k$ rather than at $u=i/4$.
At the same time, the expression ${\bf \Omega}_2{}^2$ has the correct pole structure and is a good approximation of $\Omega_2{}^2$ even in the vicinity of the poles. We verified this statement numerically for various states.

In particular, evaluating \eq{om22} at $u=u_k-\tfrac{i}{2}$ we find ${\bf \Omega}_2{}^2(u_k-\tfrac{i}{2})=0$ and thus
\beq
\Omega_2{}^2(u_k-\tfrac{i}{2})={\cal O}(\xi^{2J})\;.
\eeq
Now we show how to convert the knowledge of the above quantities in the ABA limit to find a closed set of equations for the Bethe roots.
\subsection{The asymptotic Bethe equations}
Having established explicit forms of $q_1,\dot{q}_1$ as well as $\Omega,\dot{\Omega}$ in the ABA limit we are now in a position to find Bethe equations that fix the exact location of the zeros $u_k,v_k,\dot{v}_k$.

\paragraph{The auxiliary Bethe equations.}
Let us start by verifying that \eqref{eq:q1ABA} indeed leads to a polynomial transfer-matrix and hence is compatible with our Baxter equation. For that we use \eqref{eq:Baxter1ABA} to write
\begin{equation}
\begin{split}
t(u) = &(u-\tfrac i4)^{J-M}
\frac{\cal Q^{--}}{\cal Q}
\prod_{k=1}^{M}( u- u_{k}-\tfrac{i}{2})+
(u+\tfrac i4)^{J+M}
\frac{\cal Q^{++}}{\cal Q}
\prod_{k=1}^{M} \frac{1}{(u- u_{k}+\tfrac{i}{2})}\,.
\end{split}
\end{equation}
We see that there are potential poles at $u=v_k$ and $u=u_k-\frac{\ii}{2}$ that could prevent $t$ from being polynomial. As we noticed before when deducing the structure of zeros of $q_1^{\downarrow}$ when $u\rightarrow u_k-\frac{\ii}{2}$ we also have $(u+\frac{\ii}{4})^{J+M} \sim \xi^{2(J+M)}$ hence the first pole is suppressed by $\mathcal{O}(\xi^{2J+2})$.  This is compatible with the the naive wrapping order, that is the order of validity of our Asymptotic Bethe Ansatz. To cancel the potential pole at $u=v_i$ we need to require the roots $v_i$ to satisfy
\begin{equation}\label{eq:AuxiliaryABA}
    \frac{{\cal Q}^{++}}{{\cal Q}^{--}}\bigg|_{u=v_i}\prod_{k=1}^{M} \frac{(v_i+\frac{\ii}{4})(v_i-\frac{\ii}{4})}{(v_i-u_k-\frac{\ii}{2})(v_i-u_k+\frac{\ii}{2})} = -\left(\frac{v_i-\frac{\ii}{4}}{v_i+\frac{\ii}{4}}\right)^{J}\,.
\end{equation}
This, and the equation obtained by sending $v_i \rightarrow \dot{v}_i$ and ${\cal Q}\rightarrow\dot{\cal Q}$ , are the auxiliary Bethe equations determining $v_i,\dot{v}_i$.

\paragraph{The momentum carrying equation.}
Next we fix the Bethe roots $u_k$. To do this, we need an equation of schematic form $f(u_k) = \text{constant}$. We will now show that such an equation is provided by the quantisation condition
\begin{equation}\la{cspecial}
    c = \frac{\Omega_{1}{}^{2}(u_{k}-\tfrac{i}{2})}{\dot{\Omega}_{2}{}^{1}(u_{k}-\tfrac{i}{2})}\,,
\end{equation}
The reason for evaluating at $u_{k}-\tfrac{i}{2}$ becomes clear by looking at the Wronskian-type expression for $\Omega$ \eq{OmegaW} evaluated at this special point, where one of the terms in the numerator is exactly zero leading to
\beq\label{eq:Omega12Zero}
\Omega_1{}^2(u_k-\tfrac{i}{2})=\frac{q_1^\uparrow
(u_k-\tfrac{i}{2})
q_1^\downarrow
(u_k+\tfrac{i}{2})
}{W^\downarrow(u_k)}\,,
\eeq
Note that in the r.h.s. of \eqref{eq:Omega12Zero} we have good control over all the terms: the denominator we know exactly from \eq{eq:WronskianDown}, and in the numerator, being far from all poles, we can readily replace $q_1$'s by ${\bf q}_1$'s, given in  \eq{eq:q1ABAUp} and \eq{eq:q1ABA}. This results in the following expression
\begin{equation}
\begin{split}
    \Omega_{1}{}^{2}\left(u_{i}-\tfrac i2\right) = &(-1)^{M}\left(\frac{\Gamma\left(-\ii u_i+\tfrac{3}{4}\right)}{\Gamma\left(-\ii u_i+\tfrac{1}{4}\right)}\right)^{J}\left(\frac{\Gamma\left(-\ii u_{i}+\tfrac{1}{4}\right)}{\Gamma\left(\ii u_{i}+\frac{1}{4}\right)}\right)^{M}\\
    &\times \frac{{\cal Q}\left(u_{i}-\tfrac i2\right) {\cal Q}\left(u_{i}+\tfrac i2\right)}{C^{\downarrow}_W}\prod_{j=1}^{M}\frac{\Gamma\left(1+\ii\left(u_{i}-u_{j}\right)\right)}{\Gamma\left(1-\ii\left(u_{i}-u_{j}\right)\right)}
\end{split}
\end{equation}
Again one has an identical equation with dots.
In order to find an expression for $c$ (the non-trivial component in the gluing matrix $\Gamma^{ij}$ \eqref{eq:quantizationOmega}) in terms of the coupling $\xi$, we also need to determine $\Omega_2{}^1(u_i-\tfrac{i}{2})$, as then we can use the gluing condition. For this we recall that $\Omega_2{}^2(u_i-\tfrac{i}{2})=0$, meaning that the determinant of $\Omega$ reduces to
\beq
\det\Omega(u_i-\tfrac{i}{2}) = -
\Omega_{1}{}^2(u_i-\tfrac{i}{2})\Omega_{2}{}^1(u_i-\tfrac{i}{2})={\mathbb D}^-\(u_i\)\;,
\eeq
where ${\mathbb D}$ is a fixed exact function, given in \eq{Ddef}. Together with \eq{eq:Omega12Zero}
this gives us $\Omega_{2}{}^1(u_k-\tfrac{i}{2})$, and then from \eq{cspecial} we find
\beq
c = \frac{\Omega_{1}{}^2(u_{i}-\tfrac i2)}{\dot\Omega_{2}{}^1(u_{i}-\tfrac i2)}
 = \frac{\dot\Omega_{1}{}^2(u_{i}-\tfrac i2)}{\Omega_{2}{}^1(u_{i}-\tfrac i2)}=-
 \frac{\Omega_{1}{}^2(u_{i}-\tfrac i2)\dot\Omega_{1}{}^2(u_{i}-\tfrac i2)}{
\[\tanh{\pi\left(u_{i}+\frac{\ii}{4}\right)}\]^{J-M}}\;.
\eeq
Simplifying this expression by rewriting the $\tanh$ in terms of $\Gamma$-functions we finally obtain
\begin{equation}\label{eq:ABAwithC}
\begin{split}
    {\cal C} &= {\cal Q}\left(u^+_{i}\right){\cal Q}\left(u^-_{i}\right)\dot{{\cal Q}}\left(u^+_{i}\right)\dot{{\cal Q}}\left(u^-_{i}\right)\\
    &\times\left(\frac{\Gamma\left(\ii u_{i}+\tfrac{3}{4}\right)}{\Gamma\left(-\ii u_{i}+\tfrac{1}{4}\right)}\right)^{J-M}\left(\frac{\Gamma\left(-\ii u_{i}+\tfrac{3}{4}\right)}{\Gamma\left(\ii u_{i}+\tfrac{1}{4}\right)}\right)^{J+M}\\
    &\times\left(\prod_{\substack{k=1\\k\neq j}}^{M}\frac{\Gamma[1+\ii (u_{i}-u_{k})]}{\Gamma[1-\ii (u_{i}- u_{k})]}\frac{\Gamma[\ii (u_{i}-u_{k})]}{\Gamma[-\ii (u_{i}- u_{k})]}\right)\;\;,\;\;i=1,\dots,M\,.
\end{split}
\end{equation}
where the constant ${\cal C}$ does not depend on the index $i$,
\beq\la{calC}
{\cal C}\equiv  c\; C^{\downarrow}_W \dot{C}^{\downarrow}_W e^{\frac{\ii \pi}{2}(J+M)}\;.
\eeq
Since the l.h.s. of \eqref{eq:ABAwithC} does not depend on $i$, at this stage we obtained a complete set of Asymptotic Bethe Ansatz equations. Indeed, \eqref{eq:ABAwithC} supplemented with the auxiliary equations \eqref{eq:AuxiliaryABA} constitutes a set of algebraic equations on $\{u_{k}\}_{k=1}^{M}$, $\{v_{k}\}_{k=1}^{K}$ and $\{\dot v_{k}\}_{k=1}^{\dot K}$, fixing them in terms of a single yet to be established number, $c$.

To make contact with the Feynman integral picture of 2D bi-scalar fishnet theory, we still need to re-express the only unknown constant $c$ in terms of the coupling constant $\xi$. This is done in the next section by constructing ${\mathbb Q}_+$ in the ABA limit, which requires some additional set of tricks.

\subsection{Coupling constant in the ABA equations and the cyclicity condition}\la{ABACyc}
In this section, we relate the unknown coefficient $c$ in \eqref{calC} to the coupling constant $\xi$.
The only connection to the coupling we have is via ${\mathbb Q}_+$ i.e. \eqref{IntroquantisationXi} and \eq{introShiftMnJ} (we assume $M<J$). Furthermore, a very similar constraint gives our proposed momentum operator. For the reader's convenience, let us reproduce these definitions here. We have
\begin{equation}\la{Qplocal}
    \lim_{\epsilon\rightarrow 0} \epsilon^J \frac{{\mathbb Q}_+(\tfrac{3i}{4}-i\epsilon)}{{\mathbb Q}_+(\tfrac{i}{4})} =  \xi^{2J}\,,
    \quad
\lim_{\epsilon\to 0}
\frac{1}{\epsilon^M}
\frac{{\mathbb Q}_{+}\left(-\frac{3\ii}{4}-i\epsilon\right)}
{{\mathbb Q}_{+}\left(+\frac{3\ii}{4}-i\epsilon\right)}=
(-1)^J{ (4\xi^2)^M}
\;,
\end{equation}
where we used that for the fishnet QFT realization of the current spin chain one has the cyclicity condition $U=1$. Finally, for convenience let us recall that, up to an overall constant factor (irrelevant in the above equations)  we have
\begin{equation}
      {\mathbb Q}_+(u)\propto q_1^\downarrow(u) \dot{q}^\uparrow_{1}(u)+
    c\; q_2^\downarrow(u) \dot{q}^\uparrow_{2}(u)\;.
\end{equation}
To evaluate these expressions, we will need to compute the leading residues of $q^{\downarrow/\uparrow}_{i}$ at $u=\pm \frac{3\ii }{4}$.

Note that, in order to complete the ABA and fix the remaining constant ${\cal C}$ \eq{calC} it is sufficient to use only one of these relations, so the second one will provide a non-trivial cross-check of the consistency of our approximation.

\paragraph{Finding the residue of ${\mathbb Q}_+$ at $\frac{3\ii}{4}$.}
We start by considering ${\mathbb Q}_{+}(\frac{3\ii}{4}+\epsilon)$. Since $q^{\downarrow}_i$ are analytic in the upper half-plane they will be regular at $u=\frac{3\ii}{4}$. However, $q^{\uparrow}$ will have poles as dictated by the Baxter equation. We introduce the following notation for the residue at these poles:
\begin{equation}
    C_i=\lim_{\epsilon\to 0}\epsilon^J q^{\uparrow}_{i}\left(\tfrac{3\ii}{4}+\epsilon\right)\,,
    \quad
    \dot C_i=\lim_{\epsilon\to 0}\epsilon^J \dot q^{\uparrow}_{i}\left(\tfrac{3\ii}{4}+\epsilon\right)\,.
\end{equation}
Note that while $q_1$'s are under good analytic control, at least far from the poles, so far we have not established much about $q_2$'s, which is one of the challenges of the current calculation.
In order to constrain $C_i$ and $\dot C_i$ we use the relation
\beq\label{qOq}
q_i^{\uparrow} = \Omega_{i}{}^{j}q^{\downarrow}_j\;.
\eeq
The $i$-periodic functions $\Omega_i{}^j$ themselves have poles of degree $J$ at $3i/u+i n$ for $n\in{\mathbb Z}$. We denote their residues as
\begin{equation}
    R_i{}^j = \lim_{\epsilon\to 0}\epsilon^J \Omega_i{}^j\left(\tfrac{3i}{4}+\epsilon\right)\,.
\end{equation}
Since $q^\downarrow$ are regular in the upper-half-plane, near $u=3i/4$ equation \eq{qOq} becomes, in the ABA approximation
\begin{equation}\label{eq:CandR}
    C_{i} = R_{i}{}^{j} \,\mathbf{q}^{\downarrow}_{j}\left(\tfrac{3\ii}{4}\right)\,,
    \quad
    \dot{C}_{i} = \dot{R}_{i}{}^{j} \,\dot{\mathbf{q}}^{\downarrow}_{j}\left(\tfrac{3\ii}{4}\right)\,.
\end{equation}

Note that as $\Omega$'s are constrained by the quantisation conditions \eq{quantcomponents}, we also have similar relations on their leading singularities
\begin{equation}\la{Rccond}
    R_{1}{}^{1} = \dot{R}_{1}{}^{1}\,,
    \quad
    R_{2}{}^{2} = \dot{R}_{2}{}^{2}\,,
    \quad
    R_{2}{}^{1} = \frac{1}{c}\,\dot{R}_{1}{}^{2}\,,
    \quad
    R_{1}{}^{2} = {c}\,\dot{R}_{2}{}^{1}\,.
\end{equation}
Furthermore, the determinant of $\Omega$ vanishes slower than $\frac{1}{\epsilon^{J}}$, see \eqref{detOm}, and hence we must have
\beq\la{Rdet}
\det R_{i}{}^{j} = \det \dot{R}_{i}{}^{j} = 0\;.
\eeq
Combining all these constraints \eq{Rccond}, we are left with $3$ undetermined coefficients which we will take to be $R_{2}{}^{2},R_{2}{}^{1},\dot{R}_{2}{}^{1}$, as well as the $4$ residues $C_{i},\dot{C}_i$, constrained by $4$ relations \eq{eq:CandR}. We can, thus, solve all these relations in terms of $C_2,\dot C_2$ and $R_{2}{}^2$.
For example, for $\dot C_1$ we obtain
\begin{equation}\label{eq:C1toC2}
\begin{split}
    \dot{C}_{1}
    =c\frac{\dot C_2 (C_2-{\bf q}_2^\downarrow(\tfrac{3i}{4}) R_2{}^2)}{{\bf q}_1^\downarrow(\tfrac{3i}{4}) R_2{}^2}\;.
\end{split}
\end{equation}
Plugging this relation into the definition of ${\mathbb Q}_{+}$ we get
\begin{equation}
    \lim_{\epsilon\rightarrow 0}\epsilon^{J}\,{\mathbb Q}_{+}(\tfrac{3\ii}{4}+\epsilon) \propto \left( \mathbf{q}_{1}^{\downarrow}\left(\tfrac{3\ii}{4}\right)\dot{C}_{1} + c \,\mathbf{q}^{\downarrow}_2\left(\tfrac{3\ii}{4}\right)\dot{C}_{2} \right) = c\frac{ C_{2}\dot{C}_2}{R_{2}{}^{2}}\,.
\end{equation}
We are hence left with the task of computing $C_{2},\dot{C}_2$ and $R_{2}{}^{2}$ explicitly.

\paragraph{Finding $C_{2},\dot{C}_2$ and $R_{2}{}^{2}$.}
To find $C_{2},\dot{C}_2$ we start by noticing that from \eqref{eq:C1toC2} it follows from the explicit factor $c\sim \xi^{2J}$ that $C_{1}$ is suppressed compared to $C_{2}$. This allows us to use the Wronskian equation \eqref{eq:WronskianUp}, evaluate at the pole and drop the residue coming from $q_1$, that is
\beq
C_2=\lim_{\epsilon\to 0}\epsilon^J q_2^\uparrow(\tfrac{3\ii}{4}+\epsilon)\simeq-
\lim_{\epsilon\to 0}\epsilon^J
\frac{W^{\uparrow}(\frac{\ii}{4}+\epsilon)}{\mathbf{q}_1^{\uparrow}(-\frac{\ii}{4})}\,,
\eeq
and with an analogous expression for $\dot{q}_{2}\left(\frac{3\ii}{4}+\epsilon\right)$. From the exact explicit form of $W^{\uparrow}$ \eq{eq:WronskianUp} we obtain
\begin{equation}
    C_{2} = -\frac{(-\ii)^{J}C_{W}^{\uparrow}}{\mathbf{q}_1^{\uparrow}\left(-\frac{\ii}{4}\right)}\left(\frac{1}{\sqrt{\pi}}\right)^{J-M}\,,
    \quad
    \dot{C}_2=-\frac{(-\ii)^{J}\dot{C}_{W}^{\uparrow}}{\dot{\mathbf{q}}_1^{\uparrow}\left(-\frac{\ii}{4}\right)}\left(\frac{1}{\sqrt{\pi}}\right)^{J-M}\,.
\end{equation}

To find $R_{2}{}^{2}$ we can use the ABA expression for $\Omega_{2}{}^{2}$ \eq{om22}
and simply compute the corresponding residue, which results in
\begin{equation}
\begin{split}
    R_{2}{}^{2} &= e^{\frac{\ii}{2}\pi\gamma}\frac{\ii^{M}}{\pi^{J}}\prod_{i=1}^{M} \cosh{\pi\left(u_{i}+\frac{\ii}{4}\right)} = e^{\frac{\ii}{2}\pi \gamma}\frac{\ii^{M}}{\pi^{J-M}}\prod_{i=1}^{M}\frac{1}{\Gamma\left(\ii u_{i}+\frac{1}{4}\right)\Gamma\left(-\ii u_{i}+\frac{3}{4}\right)}\,.
\end{split}
\end{equation}
Combining all these quantities together we find
\begin{equation}\label{eq:QpRes1}
    \lim_{\epsilon\rightarrow 0}\epsilon^{J}\,{\mathbb Q}_{+}(\tfrac{3\ii}{4}+\epsilon) \propto \frac{\ii^{2J-M}\,c\, C_{W}^{\uparrow}\dot{C}_{W}^{\uparrow}}{\mathcal{Q}\left(-\frac{\ii}{4}\right)\dot{\mathcal{Q}}\left(-\frac{\ii}{4}\right)} \prod_{k=1}^{M}\frac{\Gamma\left(+\ii u_{k}+\tfrac{1}{4}\right)}{\Gamma\left(-\ii u_{k}+\tfrac{3}{4}\right)}\,
\end{equation}
where the constants $C^\uparrow_{W}$ and $\dot C^\uparrow_{W}$ are given in \eq{eqCW}.

\paragraph{Fixing the constant in the momentum carrying Bethe equation.}
Finally, in order to fix the unknown proportionality coefficient in \eq{eq:QpRes1} we use the ratio \eq{Qplocal}. For that we need to evaluate ${\mathbb Q}_+$ at $i/4$. Luckily, that is a lot easier to do as we can neglect the term with $c$, as it is now suppressed by wrapping, and evaluate the $q$ functions using known ABA expressions for
$q_{1}^{\uparrow/\downarrow}$, e.g. \eq{eq:q1ABAUp} and \eq{eq:q1ABA} to obtain immediately
\beq
{\mathbb Q}_+(\tfrac i4) \propto {\bf q}_1^\downarrow(\tfrac i4)
\dot {\bf q}_1^\uparrow(\tfrac i4) =
\mathcal{Q}\left(\tfrac{i}{4}\right) \dot{\mathcal{Q}}\left(\tfrac{i}{4}\right)
\prod_{k=1}^{M}\frac{\Gamma \left(-i u_k+\frac{1}{4}\right) }{\Gamma \left(i u_k+\frac{3}{4}\right)}\,.
\eeq
Then combining with \eqref{eq:QpRes1}, we get the only missing relation, connecting the constant ${\cal C}$ from \eq{calC} with the coupling $\xi$
\begin{equation}
\begin{split}
    \xi^{2J}
    &= \frac{
    {\cal C}
    }{\mathcal{Q}\left(\frac{\ii}{4}\right)\dot{\mathcal{Q}}\left(\frac{\ii}{4}\right)\mathcal{Q}\left(-\frac{\ii}{4}\right)\dot{\mathcal{Q}}\left(-\frac{\ii}{4}\right)} \prod_{i=1}^{M} \frac{\Gamma\left(\ii u_i+\frac{1}{4}\right)\Gamma\left(\ii u_i+\frac{3}{4}\right)}{\Gamma\left(-\ii u_i+\frac{3}{4}\right)\Gamma\left(-\ii u_i+\frac{1}{4}\right)}\,.
\end{split}
\end{equation}
We now put together all the equations obtained so far. We find 
\begin{equation}
    \begin{split}
    \xi^{2J}\lambda^{J}_{0}(u_{i}) \nu^{M}_{0}(u_{i}) &= \frac{{\cal Q}(u_{i}+\frac{\ii}{2}){\cal Q}(u_{i}-\frac{\ii}{2})\dot{{\cal Q}}(u_{i}+\frac{\ii}{2})\dot{{\cal Q}}(u_{i}-\frac{\ii}{2})}{{\cal Q}(+\frac{\ii}{4}){\cal Q}(-\frac{\ii}{4})\dot{{\cal Q}}(+\frac{\ii}{4})\dot{{\cal Q}}(-\frac{\ii}{4})}\\
     &\times \nu_{f}\prod_{\substack{j=1\\j\neq i}}^{M}\frac{\Gamma\(1+\ii(u_{i}-u_{j})\)}{\Gamma\(1-\ii(u_{i}-u_{j})\)}\frac{\Gamma\(+\ii(u_{i}-u_{j})\)}{\Gamma\(-\ii(u_{i}-u_{j})\)}
    \end{split}
\end{equation}
where we have used notation
\begin{align}\label{eq:LambdaNuDefEqLocal}
    \lambda(u) &= \frac{\Gamma(-\ii u+\frac{1}{4})\Gamma(\ii u+\frac{1}{4})}{\Gamma(-\ii u+\frac{3}{4})\Gamma(\ii u+\frac{3}{4})}\,,
    \quad
    \nu(u) = \frac{\Gamma(\frac{1}{4}+\ii u)\Gamma(\frac{3}{4}+\ii u)}{\Gamma(\frac{1}{4}-\ii u)\Gamma(\frac{3}{4}-\ii u)}\,,
    \quad
    \nu_{f} = \prod_{i=1}^{M}\nu(u_{i})\,.
\end{align}
This completes our derivation of the Asymptotic Bethe Ansatz for 2D bi-scalar fishnet theory. We collected the final results in \eqref{eq:BetheAux}, \eqref{eq:BetheMiddle} and \eqref{eq:BetheAuxDot}. In the case ${\cal Q}=1$ it reduces to a result of \cite{Basso:2019xay} derived with a very different method.

\paragraph{Cyclicity condition.}
In the fishnet CFT local operators obey the cyclicity condition, originating from the fact that these are build out of fundamental fields under a trace. This implies that the shift operator $U=1$, whereas for the general states in the spin chain one has a weaker requirement $U^J=1$. In order to find this condition we use the second equation from \eq{Qplocal}.

We now proceed to the computation of the pole of ${\mathbb Q}_{+}(u)$ at $u=-\frac{3\ii}{4}$. The discussion is essentially the same as in the previous paragraph and we will hence be brief.

In Appendix~\ref{app:cyc} we derive a relation for ${\mathbb Q}_{+}\left(-\tfrac{3\ii}{4}+\epsilon\right)$ in analogy with our calculation above for  ${\mathbb Q}_{+}\left(+\tfrac{3\ii}{4}+\epsilon\right)$. Using this relation we can assemble the ratio in the expression for $U$ (for the $M<J$ case) to obtain
\beqa\la{UABA}
U&=&
\frac{(-1)^J}{{ (4i\xi^2)^M}}\lim_{\epsilon\to 0}
\frac{1}{\epsilon^M}
\frac{{\mathbb Q}_{+}\left(-\frac{3\ii}{4}+\epsilon\right)}
{{\mathbb Q}_{+}\left(+\frac{3\ii}{4}+\epsilon\right)}
=
\frac{{\cal Q}\left(-\frac{\ii}{4}\right)\dot{{\cal Q}}\left(-\frac{\ii}{4}\right)}
{{\cal Q}\left(\frac{\ii}{4}\right)\dot{{\cal Q}}\left(\frac{\ii}{4}\right)}
\prod_{i=1}^{M} \frac{1}{\xi^2\lambda\left(u_{i}\right)}\,.
\eeqa
We notice that one can also obtain $U^J=1$ by computing the product of all ABA equations in the Bethe ansatz which is yet another test of our results. In addition, we have made several numerical tests of our results as presented in section~\ref{sec:FiniteCoupling}.

\subsection{Including $\phi_2^\dagger\phi_2$ into the ABA}
\label{subsec:StrangeStates}

Let us consider operators with fixed  $J$, number of magnons $M$ and spin $S$. There are in general infinite number of operators for each such configuration, a natural way to obtain a finite number is to also specify the bare dimension, $\Delta\big|_{\xi\rightarrow0}$.

Operationally, to increase the bare dimension we can add a pair of derivatives $\partial\bar\partial$, this would shift the bare dimension by $2$. We have already considered this case, it corresponds to shifting $K$ and $\dot K$ by $1$. Another option is to add a neutral pair of fields, for example $\phi_2\phi_2^\dagger$, again this would not change $J,M$ nor $S$ but would shift the bare dimension by $1$\footnote{adding pair $\phi_1\phi_1^\dagger$ would change the structure of the Feynman diagrams as that would prevent the field $\phi_2$ from spiralling around in the correlator resulting in a trivial anomalous dimension.}. Note that there are certain similarities between these two options, more precisely the equation of motion of the fishnet theory implies $(\partial\bar\partial)^{1/2}\phi_1\propto \phi_2^\dagger\phi_1\phi_2$. So one can argue that insertion of $(\phi_2^\dagger\phi_2)^2$ should be similar to $(K,\dot K)\to (K+1,\dot K+1)$.

In Appendix~\ref{app:DiscreteReparam} we show that magnon-anti-magnon annihilation requires one to redefine the Q-function with a ratio of $\Gamma$-functions whose purpose essentially is to trade poles at $-\frac{\ii}{4}+\ii n$ with poles at $-\frac{3\ii}{4}+\ii n$ for UHPA Q-functions. We furthermore in section~\ref{sec:FiniteCoupling} numerically studied Q-functions that reduce to ratios of $\Gamma$-functions at tree level and showed that they mix non-perturbatively with the more standard solutions that have Q-functions that approach polynomials in the spectral parameter as $\xi\rightarrow0$.

To account for states like this in the ABA-set up we are naturally lead to consider the following generalisation of the ansatz \eqref{eq:q1ABA}
\begin{equation}\label{eq:q1ABAGen}
{\bf q}_1^\downarrow(u)  \equiv
e^{+\tfrac{1}{4} i \pi  \left(\gamma+N\right)}
{\cal Q}
\prod_{k=1}^{M}
\frac{\Gamma(-i u+\tfrac 14)}{
\Gamma(-i u+ i u_{k}+\tfrac 12)}\prod_{k=M+1}^{M+N}
\frac{\Gamma(-i u+\tfrac 34)}{
\Gamma(-i u+ i u_{k}+\tfrac 12)}\;
\end{equation}
and with an analogous expression for the dotted counterpart. It is a straightforward task to now repeat the same argument all the way to \eqref{eq:ABAwithC}, which generalises to
\begin{equation}\label{eq:ABAwithCGen}
\begin{split}
    c &C^{\downarrow}_W \dot{C}^{\downarrow}_W e^{\frac{\ii \pi}{2}(L+M)}(-1)^{N} = \mathcal{Q}\left(u^+_{i}\right)\mathcal{Q}\left(u^-_{i}\right)\dot{\mathcal{Q}}\left(u^+_{i}\right)\dot{\mathcal{Q}}\left(u^-_{i}\right)\\
    &\times\left(\frac{\Gamma\left(\ii u_{i}+\tfrac{3}{4}\right)}{\Gamma\left(-\ii u_{i}+\tfrac{1}{4}\right)}\right)^{J-M}\left(\frac{\Gamma\left(-\ii u_{i}+\tfrac{3}{4}\right)}{\Gamma\left(\ii u_{i}+\tfrac{1}{4}\right)}\right)^{J+M}\left(\frac{\Gamma\left(-\ii u_{i}+\frac{5}{4}\right)}{\Gamma\left(\ii u_{i}+\frac{3}{4}\right)}\right)^{2N}\\
    &\times\left(\prod_{\substack{k=1\\k\neq j}}^{M+N}\frac{\Gamma[1+\ii (u_{i}-u_{k})]}{\Gamma[1-\ii (u_{i}- u_{k})]}\frac{\Gamma[\ii (u_{i}-u_{k})]}{\Gamma[-\ii (u_{i}- u_{k})]}\right)\,.
\end{split}
\end{equation}
However, at this stage we reache an impasse. The constant $c$ turns out to be suppressed far beyond the ABA limit. One way of seeing this is that $\mathbb{Q}_{+}\left(\frac{\ii}{4}\right) = \mathcal{O}(\xi^{\text{wrapping}})$ since the pole of $\mathbf{q}^{\downarrow}_{1}$ at $-\frac{\ii}{4}$ leads to a zero in $\mathbf{q}^{\uparrow}$ at $\frac{\ii}{4}$. Nevertheless, \eqref{Qplocal}  must still be true forcing $c$ to cancel the effect of these zeros and hence pushing it beyond the ABA regime.

We presently do not know of a way around this, and it would be fascinating to try and develop an ABA framework to also understand these more exotic solutions, especially since there exists no similar phenomenon in $4D$ \footnote{We however expect states of this kind  to be present for bi-scalar fishnets in other dimensions where the equations of motions are also non-local. }. We leave a complete investigation of these general states for future work, but before closing this section we will display some encouraging numerical results.

\paragraph{Numerical example.} To investigate states with Q-functions of the form \eqref{eq:ABAwithCGen} we will consider $J=4,M=1$. In this particular case, we found a numerical solution from which we fitted the zeros of $q_{1}$ as
\begin{align}
    u_{1} = \left(\sqrt{2 \left(5+\sqrt{5}\right)}+i \left(1-\sqrt{5}\right)\right)\xi^{6}+\mathcal{O}\left(\xi^{12}\right)\,,
    \\
    u_{2} = \left(-\sqrt{2 \left(5+\sqrt{5}\right)}+i \left(1-\sqrt{5}\right)\right)\xi^{6}+\mathcal{O}\left(\xi^{12}\right)\,,
\end{align}
and found that $\Delta$ indeed was reproduced by the standard dispersion relation \eqref{eq:Dispersion}. Furthermore, $c$ is indeed suppressed far beyond the ABA regime, to be precise, we found
\begin{equation}
    \Delta = \frac{7}{2}+4(1 - \sqrt{5})\,\xi^{6}+\mathcal{O}\left(\xi^{10}\right)\,,
    \quad
    c \simeq -(1024/25) \xi^{30}+\dots\,.
\end{equation}

\section{Twisted Quantum Spectral Curve}\label{sec:Twisting}

A crucial ingredient for applying the Separation of Variables (SoV) framework to correlation functions is the introduction of a twist parameter, which encodes quasi-periodic boundary conditions on the underlying spin chain. Not only does it make the SoV expressions more concise, but it also lifts degeneracies in the spectrum, allowing for a one-to-one map between the Q-functions and the operators.
In the context of integrable spin chains, quasi-periodic (``twisted'') boundary conditions have been well studied since Baxter's seminal work on the eight-vertex/XYZ chain~\cite{Baxter1972EightVertex,Baxter1973EightVertexI,Sklyanin1992QISM}.
The Q-functions of the compact spin chains
are typically twisted polynomials parametrized as $Q_a(u) = \lambda_a^{iu} q_a(u)$, where $\lambda_a$ are twist matrix eigenvalues and $q_a$ are polynomials. In the case of integrable QFTs the polynomiality is replaced by a non-trivial quantisation condition but the twists still enter through large $u$ asymptotics in the same way (see e.g. \cite{Gromov:2010dy,Kazakov:2015efa}).

In this section, we outline the main changes to the QSC construction in the twisted case. At the operatorial level, the twist manifests itself as a phase rotation applied to propagators crossing a designated cut on the surface of the planar diagram of the correlator~(see figure \ref{fig:TwistCut}).  We discuss how the twisted QSC takes a form analogous to the untwisted case, with modified analyticity properties and quantisation conditions that incorporate the twist angle, and we also provide some non-trivial tests.

\subsection{Colour-twist fields and graph-building operator}

\paragraph{Colour-twist fields and the twist operator.}
The colour-twist construction, introduced in \cite{Cavaglia:2020hdb}, provides a  generalization of local operators in planar gauge theories by inserting a twist operator into the trace. This twist operator acts as a non-local modification that changes the planar diagrams, by replacing the propagators with twisted propagators when crossing the cuts which originate from the twist operator.
In principle one can use any symmetry of the theory as a twist (see \cite{Cavaglia:2020hdb} for details), but we restrict ourselves to the twist manifesting as a phase rotation applied to propagators that cross a designated "twist cut" in the correlation functions.

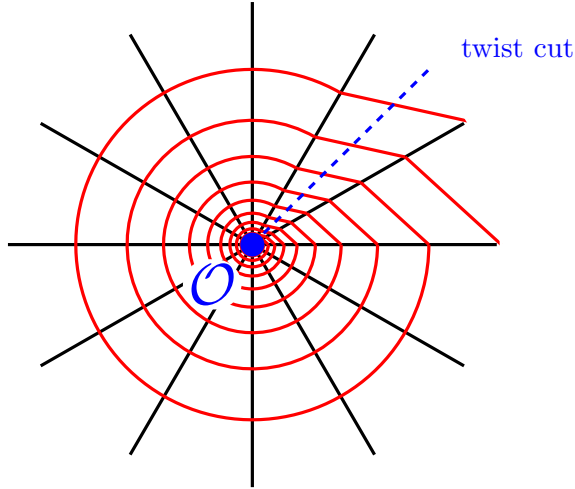
\begin{figure}
    \centering
\begin{tikzpicture}[scale=1.5]
  \def\rmin{2}
  \def\rstep{10}
  \def\q{1.41}
  \def\rmax{51}
  \def\a{0}
  \def\da{30}
  \def\magnon{1}

  \foreach \a in {0,30,...,150}{
    \draw[black,very thick] ({-(\rmax+\rstep pt)*cos(\a)},{-(\rmax+\rstep pt)*sin(\a)})--({(\rmax+\rstep pt)*cos(\a)},{(\rmax+\rstep pt)*sin(\a)});
  }

  \pgfmathsetmacro{\Nreal}{ln(\rmax/\rmin)/ln(\q)}
  \pgfmathtruncatemacro{\N}{floor(\Nreal)}

  \foreach \k in {0,...,\N}{
    \pgfmathsetmacro{\r}{\rmin*pow(\q,\k)}
    \pgfmathsetmacro{\rnext}{\rmin*pow(\q,\k+1)}

    \ifnum\magnon=1
      \ifnum\k<\N
        \draw[red,very thick]
        ({\a+2*\da}:\r pt) arc[start angle=\a+2*\da,end angle=\a+360,radius=\r pt];
        \draw[red,very thick] ({\a+\da}:\r pt) -- ({\a}:\rnext pt);
        \draw[red,very thick] ({\a+2*\da}:\r pt) -- ({\a+\da}:\rnext pt);
      \fi
      \ifnum\k=\N
        \draw[red,very thick]
        ({\a+2*\da}:\r pt) arc[start angle=\a+2*\da,end angle=\a+360,radius=\r pt];
        \draw[red,very thick] ({\a+\da}:\r pt) -- ({\a}:\rnext pt);
        \draw[red,very thick] ({\a+2*\da}:\r pt) -- ({\a+\da}:\rnext pt);
      \fi
    \else
      \draw[red,very thick] (0,0) circle[radius=\r pt];
    \fi
  }

  \pgfmathsetmacro{\rfinal}{\rmin*pow(\q,\N+1)}
  \draw[white,very thick] (0,0) circle[radius=\rfinal pt];

  \draw[blue,very thick,dashed] (0,0) -- (45:\rfinal pt);

  \pgfmathsetmacro{\rlabel}{\rfinal+8}
  \node[blue,anchor=west] at (45:\rlabel pt) {twist cut};

  \fill[blue] (0,0) circle (3pt);
  \node[blue,scale=2,fill=white,inner sep=1pt,rounded corners=10pt] at (-10pt,-10pt) {$\mathcal{O}$};

\end{tikzpicture}
\caption{Fishnet correlator with magnons and a twist cut defining the CFT wave-function for an operator with a twist. The dashed blue line represents the twist cut emanating from the center, crossing infinitely many propagators.}
\label{fig:TwistCut}
\end{figure}

Formally, we consider operators of the form $\tr(\mathcal{T}_\phi \, \phi_1^{J-M}(\phi_1\phi_2)^M)$, where $\mathcal{T}_\phi$ denotes the twist operator implementing the transformation. The twist angle $\phi$ parametrizes the strength of this rotation. When propagators $[z-w]^{\{\alpha,\dot\alpha\}}$ cross the twist cut (illustrated in figure~\ref{fig:TwistCut}), their $w$ coordinate acquires a phase  factor $e^{2i\phi}$ and gets replaced with $[z-e^{2i\phi} w]^{\{\alpha,\dot\alpha\}}$. Note that as we limit ourselves to the rotation, there is no additional Jacobian factor, which has to be added in general~\cite{Cavaglia:2020hdb}.

The integrability structure remains preserved under twisting, with the commuting family of transfer matrices and Q-operators modified in a controlled manner as we explain below.

\paragraph{Twisted graph-building operator.}
Quite naturally, the presence of the twist modifies the periodic boundary conditions. In our holomorphic and anti-holomorphic coordinates this is simply realized by imposing the twisted periodic boundary conditions $w_0 = e^{2i \phi}w_J$. Apart from this change in boundary conditions, the expression \eqref{eqn:generalgraphB} remains unchanged.

\paragraph{Twisting the finite-dimensional transfer matrices and Baxter equations.}

Twisting the finite-dimensional transfer matrices is performed by including a twist matrix $g$ in the definition \eqref{eqn:transfermatrix} leading to
\begin{equation}\la{ttwisted}
    \hat t(u) = {\rm tr}_a(\hat L_{a1}(u-\theta_1)\dots \hat L_{aJ}(u-\theta_J)g),\quad g={\rm diag}(e^{i \phi},e^{-i\phi})\,.
\end{equation}
The twisted transfer matrix continues to generate a commutative family of operators $[\hat t(u),\hat t(v)]=0$. In order to be consistent with the twisting of the Feynman diagrams, the dotted and undotted transfer matrices need to be twisted differently, namely in the dotted sector we should twist with $\dot{g} = {\rm diag}(e^{-i\phi}, e^{i\phi}) = g^{-1}$.

Twisting modifies the leading coefficients of the transfer matrix which are now given by
\begin{equation}\label{eqn:twistedtransferasympt}
    \hat t(u) = 2\cos(\phi)u^J -2u^{J-1}(\cos(\phi)(\theta_1+\dots +\theta_J)-\sin(\phi)\hat{\mathbb{S}}^z)+\dots\,.
\end{equation}
The twisted Baxter equation is exactly the same as in the untwisted case \eqref{eqn:bax1} which we repeat here for convenience assuming $M$ magnons:
\begin{equation}\label{eqn:twistedbax}
    \left(u+\tfrac{i}{4} \right)^J q(u+i) - t(u) q(u) + \left(u-\tfrac{3i}{4} \right)^{M}\left(u-\tfrac{i}{4} \right)^{J-M}q(u-i)=0\,.
\end{equation}
The locations of the poles of the Q-functions are also unchanged -- twisting only modifies the asymptotic behaviour of the Q-functions at $u\rightarrow \infty$ as can be see directly from \eq{eqn:twistedtransferasympt}.
Indeed, \eq{eqn:twistedbax} implies that at large $u$ we have
\begin{equation}
    q_1(u) \simeq e^{-\phi u}u^{h-(J+M)/4},\quad q_2(u) \simeq e^{\phi u} u^{-h-(J+M)/4}\,.
\end{equation}
Note that because the dotted and undotted sectors are twisted differently the asymptotics of the dotted and undotted Q-functions are not identical, and we have
\begin{equation}
    \dot{q}_1(u) \simeq e^{\phi u}u^{h-(J+M)/4},\quad \dot{q}_2(u) \simeq e^{-\phi u} u^{-h-(J+M)/4}\,.
\end{equation}

As before, we will be mainly interested in the UHPA and LHPA solutions $q_i^\downarrow$ and $q_i^\uparrow$, which are related by the $i$-periodic matrix $\Omega(u)$:
\begin{equation}
    q_i^\uparrow(u) = \Omega_{i}^{\ j}(u)q_j^\downarrow(u)\,,
\end{equation}
and similarly in the dotted sector.

\subsection{Twisted Q-operator}

Like the graph-building operator, the twisted Q-operator is obtained by simply modifying the propagators that cross the twist cut. Unlike the graph-building operator, we also need to include a simple exponential factor, in order for ${\mathbb Q}_+$ to satisfy the Baxter equations \eq{Qpbax} but with the twisted $\hat t$ and $\dot{\hat t}$ given by \eq{ttwisted}. The result is that the twisted Q-operator $\hat{\mathbb{Q}}_+(u,\dot{u})$ is specified by its action on functions $f(z)$
\begin{equation}
   [{\hat{\mathbb{Q}}}_+(u,\dot{u})f](z) = \displaystyle \int{\rm d}^2 w \mathcal{Q}_+(u,\dot u)(z|w) f(w)
\end{equation}
where now the kernel is given by
\begin{equation}\la{Qptwisted}
{\mathcal{Q}}_+(u,\dot u)(z|w)=  e^{-\phi(u-\dot{u})}  \displaystyle \prod_{k=1}^J  [w_{k}-z_k]^{\{\alpha^+_k,\dot{\alpha}^+_k\}}[w_{k-1}-z_k]^{\{\beta^+_k,\dot{\beta}^+_k\}} [w_{k}-w_{k-1}]^{\{\gamma^+_k,\dot{\gamma}^+_k\}}\,,
\end{equation}
but with $w_0 = e^{2i\phi} w_J$. The exponents $\alpha^+_k$, $\beta_k^+$ and $\gamma_k^+$ are exactly the same as \eqref{eqn:simpleQexponents}. The twisted Q-operator commutes with the twisted transfer matrix $t(u)$ and satisfies the Baxter equation \eqref{eqn:twistedbax} as an operator equation in both $u$ and $\dot{u}$.

\paragraph{Reduction to graph-building operator.} In order to relate the conformal dimension $\Delta$ to the coupling we need to identify the relation between the Q-operator and the coupling, or equivalently the graph-building operator $\hat{B}$. Comparing with the graph-building operator \eqref{eqn:generalgraphB} with twisted boundary conditions $w_0 = e^{2i\phi}w_J$, it is clear that this relation is exactly the same as in the untwisted case \eq{QPtoB}
\begin{equation}
    \hat{\mathbb{Q}}_+\left( \frac{i}{4}\right) = \frac{\pi^J}{\xi^{2J}}\hat{B}\,.
\end{equation}
The relation \eqref{eqn:Qopidentity} to the identity operator is also unchanged and hence the eigenvalues $\mathbb{Q}_+(u,\dot{u})$ of the Q-operator satisfy the same relation to the coupling as the untwisted case
\begin{equation}
    \lim_{\epsilon\rightarrow 0}\epsilon^J \frac{\mathbb{Q}_+\left(\frac{3i}{4}-i\epsilon \right)}{\mathbb{Q}_+\left(\frac{i}{4} \right)}=\xi^{2J}\,.
\end{equation}
The expressions for the shift operator are also unchanged and given by \eq{UshiftDer} for $M<J$ and  \eq{defUJM} for $J=M$. What changes is that instead of $U^J$ being the identity operator it is a rotation operator by the $2\phi$ angle. For the operators with spin $S$ we thus have 
\beq
U^J = e^{-2i \phi S} \;.
\eeq
Thus the analogue of the cyclicity condition should be
\beq\la{twistedcyclicity}
U = e^{-2i \phi S/J}\;.
\eeq
\paragraph{Asymptotics and singularities.} The asymptotics of the Q-operator eigenvalues can be deduced in the same way as the untwisted case and so we will not repeat the calculation. The end conclusion is that \eq{eqn:Qopasymptotics} still holds.
 The location and order of the poles is also exactly the same as in the untwisted case i.e. \eq{poles_locattions}.

\paragraph{Twisted QSC equations.}

We now have all the ingredients necessary to write down the twisted QSC equations. Like in the untwisted case, the fact that $\mathbb{Q}_+(u,\dot{u})$ solves the Baxter equation, together with its asymptotics and singularities, is consistent with the decomposition
\begin{equation}
    \mathbb{Q}_+(u,\dot{u}) = \Gamma^{ij}q_i^\downarrow(u) q_j^\uparrow(\dot{u}) = \Gamma^{ij}q_i^\uparrow(u) q_j^\downarrow(\dot{u})
\end{equation}
with $\Gamma^{ij}$ diagonal. We see that the the exponential factors in $q_i$ cancel leading to unchanged asymptotics of ${\mathbb Q}_+$. So the quantisation conditions \eq{quantcomponents} remain unchanged in the twisted case.

\paragraph{Twisted ABA.}
The changes in the derivation of the ABA are also minimal.
One has to simply include the factor $e^{-\phi u}$ into the ansatz for $q_1$ and $e^{+\phi u}$ for $\dot q_1$ in \eq{eq:q1ABA}, which only affects the auxiliary BAE \eq{eq:AuxiliaryABA} by the extra $e^{2i\phi}$ factor in the l.h.s. (and an opposite factor for the dotted equation). The momentum carrying equation remains unchanged. Finally, there is also a modification for
the cyclicity condition in the ABA limit \eq{UABA}, where we now set $U=e^{-2i\phi S/J}$, in accordance with \eq{twistedcyclicity}.

\subsection{Explicit example}

We will now demonstrate an explicit example of the twisted QSC equations in action. We begin by constructing the twisted CFT wave function.

For $J=1$ the CFT wave function $\varphi_\lO(z)$ is fixed by requiring that it is an eigenfunction of the global spin operator $\hat{\mathbb{S}}^z$ which fixes its form to be
\begin{equation}\label{eqn:twistedCFTwave}
    \varphi_{\lO}(z) = \frac{1}{[z]^{h+1/4}}
\end{equation}
assuming the site does not contain a magnon.
We now apply the graph-building operator. For $J=1$ the general expression simplifies to
\begin{equation}
    [\hat{B}f](z) =\frac{\xi^2}{2\pi\sin(\phi)} \displaystyle\int {\rm d}^2 w \frac{1}{[w -z]^{1/2}[w]^{1/2}} f(w)
\end{equation}
where we denoted $w_1=w$ and $z_1=z$ for simplicity.
We now act with $\hat{B}$ on the CFT wave function \eqref{eqn:twistedCFTwave}. Using the chain relation for propagators \eqref{eqn:chainreln} we find
\begin{equation}
    [\hat{B}\varphi_{\lO}](z) = \frac{1}{2}\xi^2(-1)^S \csc(\phi) \frac{\Gamma(\tfrac{1}{4}+\dot{h})\Gamma(\tfrac{1}{4}-\dot{h})}{\Gamma(\tfrac{3}{4}+h)\Gamma(\tfrac{3}{4}-h)}\varphi_{\lO}(z)\,.
\end{equation}
Imposing that the CFT wave function is stationary under the action of the graph-building operator we are then led to the quantisation condition
\begin{equation}\label{eq:SpectrumJ1Twisted}
  \frac{1}{2}\xi^2(-1)^S \csc(\phi) \frac{\Gamma(\tfrac{1}{4}+\dot{h})\Gamma(\tfrac{1}{4}-\dot{h})}{\Gamma(\tfrac{3}{4}+h)\Gamma(\tfrac{3}{4}-h)}=1\,.
\end{equation}
As an example, we can expand this result analytically at weak coupling assuming $\Delta = \frac{1}{2}+\lO(\xi^2)$ and $S=0$ for simplicity. Denoting the rescaled coupling constant $\hat{\xi}^2 = \xi^2 \csc(\phi)$ we find
\begin{equation}
    \Delta=\frac{1}{2} -\hat{\xi}^2 - 2\hat{\xi}^4\; {\rm log}(2)- 6\, \hat{\xi}^6\; {\rm log}(2)^2-\hat{\xi}^8\left(\frac{64}{3}{\rm log}(2)^3 + \frac{\zeta_3}{2}\right)+\lO(\hat{\xi}^{10})\,.
\end{equation}
We finish this section by noting that the numerical method works just as before, indeed we numerically verified \eqref{eq:SpectrumJ1Twisted} for a few states.

\section{Conclusions}\label{sec:Conclusions}

In this paper, we have formulated a set of equations for 2D bi-scalar fishnet conformal field theory, to which we refer as Quantum Spectral Curve (QSC), due to its reminiscence to the QSC's found for more complicated holographic models. Our QSC takes the form of two coupled $\algsl(2)$ Baxter equations, connected by a set of quantisation conditions that select the physical spectrum. Previous investigations using the Thermodynamic Bethe Ansatz into the spectrum of bi-scalar fishnet theory away from four dimensions only dealt with a subsector of operators and are not strictly applicable to the case of two dimensions \cite{Basso:2019xay}\footnote{We are grateful to G. Ferrando for informing us that the $2D$ bi-scalar fishnet TBA has been derived, but has not yet been published. It would be highly interesting to compare our results with the TBA.}. Hence, our construction provides the first complete non-perturbative description of the spectrum for a fishnet theory beyond four dimensions,  opening up a wealth of possibilities for exploring integrability in lower-dimensional field theories and related systems.

Our results are based on the fact that in 2D there exists an operatorial construction of the Q-functions, rooted in the representation theory of the $SL(2,\mathbb{C})$ spin chain. This allowed us to derive many of their properties rigorously from first principles, with the methods of~\cite{Derkachov:2001yn,Derkachov:2002wz,Derkachov:2002pb}. The operatorial approach also enabled us to understand crucial aspects such as the quantisation conditions. 

While we closely follow~\cite{Derkachov:2001yn} in parts of the operatorial construction,
an important feature of our set-up is that we have relaxed assumptions about the complex conjugation properties of the spins at each site and other parameters. The operatorial formalism and resulting QSC equations remain well-defined even for complex values of the coupling constant $\xi$ and complex conformal dimensions. One of the reasons why this is important is that the dimensions of local operators are not expected to be of the (unitary) principal series $1+i\nu$ form.
Furthermore, to study Regge trajectories, it is crucial to have a framework that can extend quantum numbers, in particular spin, away from integer values to arbitrary complex numbers.

\paragraph{Inclusion of magnon excitations.}
In this paper, we systematically introduced magnon excitations---corresponding to insertions of $\phi_2$ fields---into the spin chain picture. While the vacuum state $\tr(\phi_1^J)$ provides the natural starting point, physical operators of interest involve arbitrary combinations of $\phi_1$ and $\phi_2$ fields (and $\phi_2^\dagger$). Our formalism allows us to study the full non-trivial operator spectrum\footnote{In addition to the operators we study here, whose dimensions are governed by the graph-building operator, one expects that in analogy with 4D case (see e.g.~\cite{Ahn:2020zly}) there exist non-dynamical ``protected" operators with zero anomalous dimensions, which we are not considering here.} of the theory and we give examples both numerically and analytically. 

\paragraph{Shift operator and operator selection.}
An important ingredient in our construction is the derivation of the shift operator. While its derivation is quite straightforward in the non-magnon case, it becomes more complicated when magnons are present. Magnons are implemented as an inhomogeneity and representation choice at a given site, and when cyclically shifting the chain we have to readjust the inhomogeneities and representations with a suitable discrete reparametrization transformation~\cite{Gromov:2019jfh}. Furthermore, we show that the case $M=J$ (operators with equal numbers of $\phi_1$ and $\phi_2$ fields) and $M<J$ require separate treatment. The shift operator is essential for correctly selecting local operators from the full spectrum of the spin chain Hamiltonian. The cyclicity condition, which is setting the shift operator to identity, ensures that only states corresponding to genuine local CFT operators are included, filtering out unphysical solutions (present naturally in the spin chain). Beyond its immediate utility for local operators, the cyclicity condition may prove important for studying non-local operators.

\paragraph{Derivation of the Asymptotic Bethe Ansatz.}
We have developed a novel method for deriving the Asymptotic Bethe Ansatz (ABA) directly from our functional QSC equations. Similar methods were developed in $\mathcal{N}=4$ SYM and its lower-dimensional cousins, but the case of fishnets requires several new tricks and insights. This derivation is conceptually important because it demonstrates the consistency between the exact QSC framework and the more traditional Bethe ansatz methods, while also extending the validity of the ABA to a broader class of states. Similar techniques could potentially be adapted to study non-local operators, i.e fluctuations around horizontal Regge trajectories for example.

\paragraph{Implications for Separation of Variables.}
This work lays the groundwork for a comprehensive Separation of Variables programme in 2D biscalar fishnet theory, with the ultimate goal of computing correlation functions of non-protected operators. The operatorial formalism we have developed will prove essential for this next step: computing OPE coefficients and correlation functions through a full SoV framework. The 2D case provides an ideal playground that is still tractable, yet moves away from highest-weight spin chains toward a more complex integrability setting similar to those encountered in $\mathcal{N}=4$ super Yang-Mills, ABJM theory, and other higher-dimensional CFTs. The key advantage of working in 2D is the rank-one $\mathfrak{sl}(2)$ structure, where many aspects of the SoV construction are expected to simplify significantly, allowing for more detailed and rigorous analysis while maintaining full operatorial control. Additionally, the theory admits a non-trivial twist parameter associated with spacetime rotations---a feature that is crucial for the SoV approach to correlation functions \cite{Cavaglia:2018lxi,Cavaglia:2021mft}. We plan to pursue this direction in upcoming work, using the detailed treatment of the twisted case in section~\ref{sec:Twisting} as a natural starting point. Once the SoV measure and overlap formulas are worked out, one can express correlation functions in terms of the $Q$-functions obtained from the QSC. This path has been successfully followed for simpler integrable systems and for certain aspects of 4D fishnet theory \cite{Gromov:2020fwh,Gromov:2022waj,Ekhammar:2023iph,Cavaglia:2021mft}. Pursuing this programme in 2D will not only yield explicit results for fishnet correlators, but also provide valuable lessons that can inform SoV approaches in higher-rank cases.

\paragraph{Connections to AdS$_3$/CFT$_2$ integrability.}
An exciting aspect of our results is their potential relevance for AdS$_3$/CFT$_2$ integrability. Our QSC equations bear striking similarities to recently proposed AdS$_3$ QSCs \cite{Ekhammar:2021pys,Cavaglia:2021eqr,Cavaglia:2022xld,Chernikov:2025jko,Cavaglia:2025icd}. In particular, our operatorial derivation sheds light on structural elements such as the gluing matrix, which are difficult to constrain based solely on symmetry considerations. An interesting question is whether 2D fishnet theory might arise as a limiting case of an AdS$_3$ holographic system, analogous to how the 4D fishnet emerges from the strong $\gamma$-deformation limit of $\mathcal{N}=4$ SYM. If such a connection exists, it could also provide a concrete realization of a fishchain-like dual description \cite{Gromov:2019aku,Gromov:2019bsj,Gromov:2019jfh,Gromov:2021ahm} for AdS$_3$ backgrounds. Moreover, exploring generalisations of our setup to include analogues of NSNS flux could reveal new integrable structures relevant for string theory in AdS$_3 \times S^3$ backgrounds.

\paragraph{Other directions.} Other avenues for future research include exploring connections to Calabi-Yau geometry of Feynman integrals that arises in 2D \cite{Duhr:2022pch,Duhr:2023eld}, BFKL physics in QCD which is governed by a very similar spin chain \cite{Lipatov:1993qn,Faddeev:1994zg,Alfimov:2020obh,Ekhammar:2024neh,Alfimov:2023vev}, and the broader landscape of fishnet theories with additional fields or interactions, as well as deformations by non-abelian twists that recently attracted interest in $\mathcal{N}=4$ SYM \cite{Driezen:2025izd}. Ultimately, we believe the 2D fishnet theory stands as a valuable laboratory where powerful integrability methods can be developed and tested.

\paragraph{Acknowledgements.} We are grateful to T. Bargheer, C. Bercini, G. Ferrando, V. Kazakov, V. Mishnyakov and D. Volin for discussions.
We are also grateful to A. Cavagli\`a for discussions and participation in this project at its initial stage.
We thank all participants of the ``Fishnet QFTs: Integrability, periods and beyond'' workshop at the University of Southampton in July 2025, where a preliminary version of these results were presented, for stimulating discussions,
and in particular B. Basso, V. Kazakov, G. Korchemsky, J. Krivorol, M. Preti, D. Serban, K. Zarembo, and D. Zhong for fruitful exchanges.
We are especially grateful to Ömer Gürdoğan for continuous encouragement to write this paper.

The work of S.E. was conducted with funding awarded by the Swedish
Research Council, Vetenskapsrådet, grant VR 2024-00598. The work of F.L.-M. was supported by the STFC grant APP69281. The work of P. R. is funded by the Deutsche Forschungsgemeinschaft (DFG, German Research Foundation) Grant No. 460391856. The work of N.G.\ was supported by the European Research Council (ERC) under the European Union's Horizon 2020 research and innovation programme (grant agreement No.~865075) EXACTC.
N.G.'s research is supported in part by the Science Technology \& Facilities Council (STFC) under the grants ST/P000258/1 and ST/X000753/1.

\appendix

\section{Antipode}\label{app:Antipode}

\paragraph{Magnon/anti-magnon symmetry and antipode.}
From the definition of the Lagrangian, it is clear that, for each operator, we can obtain another one with the same anomalous dimension if we exchange $\phi_2\leftrightarrow\phi_2^\dagger$ and reverse the order of the fields under the trace.  Let us argue that, at the level of integrability, this symmetry translates to the antipode map.

In general the Yangian antipode operation \cite{molev2007yangians} maps the monodromy matrix  $\hat T(u)$, defined in \eq{eqn:transfermatrix}, to its inverse, up to an overall factor and shift of $u$. In our case we can define the antipode of $\hat T(u)$ as the operator $\hat{T}^*(u)$ given by
\begin{equation}
\label{th1}
    \hat{T}^*(u) = \sigma_y \hat T^{\mathsf{t}}(u) \sigma_y^{-1},\quad \sigma_y = \left(\begin{array}{cc}
       0  & -i  \\
       i   & 0
    \end{array} \right)
\end{equation}
where $\hat T^{\mathsf{t}}(u)$ denotes the transpose of the matrix $\hat T(u)$ viewed as a $2\times 2$ matrix. This definition in fact means that $\hat T^*(u)$ is the cofactor matrix for $\hat T(u)$, so roughly speaking it is indeed the inverse of $\hat T(u)$. More precisely, their product gives the quantum determinant and can be written as
\begin{equation}
\label{tth1}
    \hat T(u)\hat{T}^*(u-i) =\prod_{n=1}^J((s_n-1)s_n+(u-\theta_n)(u-\theta_n-i))\cdot {\mathbb 1}_{2\times 2} \
\end{equation}
(where ${\mathbb 1}_{2\times 2}$ is the identity matrix) which we prove explicitly below.

Let us see how this new operator $\hat T^*(u)$ helps us to realise a symmetry mapping magnons to anti-magnons. First, to derive \eq{tth1}, let us examine the effect of the antipode map on the local Lax operators. Since transposition reverses the order of products, we have
\begin{equation}
    \hat{T}^*(u) = \hat{L}^*_{aJ}(u-\theta_J)\dots \hat{L}^*_{a1}(u-\theta_1)
\end{equation}
where we denoted
\begin{equation}\label{eqn:hatlax}
    \hat{L}^*_{an}(u)=\sigma_y \hat L^{\mathsf{t}}_{an}(u) \sigma_y^{-1}=\left(\begin{array}{cc}
        u-i \hat S^{z}_n & -i \hat S^-_n \\
        -i \hat S^+_n & u+i \hat S^{z}_n
    \end{array} \right)\,.
\end{equation}
We find explicitly, using \eq{scas}, that
\beq
    \hat L_{an}(u)\hat L^*_{an}(u-i)=((s_n-1)s_n+u(u-i))\cdot {\mathbb 1}_{2\times 2} \
\eeq
which indeed implies the relation \eq{tth1} given above. Clearly, we also have $\hat{L}^*_{an}(u)=-\hat L_{an}(-u)$ and hence
\begin{equation}
    \hat{T}^*(-u) = (-1)^J \hat L_{aJ}(u+\theta_J)\dots \hat L_{a1}(u+\theta_1)\,.
\end{equation}
We see that, in total, the antipode map reverses the order of the spin chain sites and swaps $\theta\rightarrow -\theta$. Nicely, the values of $\theta$'s for magnons and anti-magnons are precisely related by a change of the overall sign as can be seen in conserved charges contained in Table \ref{tab:configs}. Hence, if we start with a configuration with $M$ magnons, the antipode map transforms us to a configuration with $M$ anti-magnons. Furthermore,
the conserved charges contained in $\hat t(u)$ for these configurations coincide up to a sign because $\hat{t}^*(-u) = (-1)^J \hat t(u)$ (since the operation \eq{th1} preserves the trace). Note that for conformal primary operators the value of $h$ can be found from the coefficient of $u^{J-2}$ in $\hat t(u)$, which as we see from \eq{c2ex} is moreover invariant under flipping the sign of $\theta$'s. This means that for our two configurations the value of $h$ is the same, and the same is true for $\dot{h}$ as well. Therefore the scaling dimension $\Delta=h+\dot{h}$ is also the same for the two configurations.

\section{Discrete reparametrization symmetry from the Baxter equation}\label{app:DiscreteReparam}
In this appendix, we discuss discrete reparametrization symmetry for the isotropic 2D bi-scalar fishnet. Let us first record the Baxter equation in a set-up with $M$ magnons and $\bar{M}$ anti-magnons. Using the dictionary from Table~\ref{tab:configs} and the Baxter equation \eqref{Qpbax} we find
\begin{equation}
    \left(u+\tfrac{\ii}{4}\right)^{J-\bar{M}}\left(u+\tfrac{3\ii}{4}\right)^{\bar{M}} \,q^{[2]}- t(u) q + \left(u-\tfrac{\ii}{4}\right)^{J-M}\left(u-\tfrac{3\ii}{4}\right)^{M} q^{[-2]} = 0\,
\end{equation}
where $M$ counts magnons and $\bar{M}$ anti-magnons.

\paragraph{Magnon-anti-magnon annihilation.} Let us show that magnons and anti-magnons annihilate each other. To that end, let us send $M\rightarrow M+1$ and $\bar{M}\rightarrow M+1$. This gives a new Baxter equation
\begin{equation}
    \left(u+\tfrac{\ii}{4}\right)^{J-\bar{M}}\left(u+\tfrac{3\ii}{4}\right)^{\bar{M}} \,\frac{u+\tfrac{3\ii}{4}}{u+\tfrac{\ii}{4}}q^{[2]}- t(u) q + \left(u-\tfrac{\ii}{4}\right)^{J-M}\left(u-\tfrac{3\ii}{4}\right)^{M}\frac{u-\tfrac{3\ii}{4}}{u-\tfrac{\ii}{4}} q^{[-2]} = 0\,.
\end{equation}
To compensate for the new factors we now redefine $q(u)$ according to $q \rightarrow \sigma \, q$ and demand that
\begin{equation}
    \frac{\sigma^{[2]}}{\sigma} = \frac{u+\tfrac{\ii}{4}}{u+\frac{3\ii}{4}}\,,
    \quad
    \frac{\sigma^{[-2]}}{\sigma} = \frac{u-\tfrac{\ii}{4}}{u-\frac{3\ii}{4}}\,.
\end{equation}
These equations admit many solutions, but upon demanding that $q$ is UHP-analytic  and only has poles at the appropriate places we find
\begin{equation}\label{eq:GammaFactor}
    \sigma = \frac{\Gamma[-\ii u+\frac{1}{4}]}{\Gamma[-\ii u+\frac{3}{4}]}\,.
\end{equation}
We hence conclude that the Baxter equation is left invariant under the simultaneous operations $M\rightarrow M+1,\bar{M}\rightarrow \bar{M}+1,q\rightarrow \sigma q$. Alternatively, simply redefining $q$ with $\sigma$ annihilates a magnon and an anti-magnon.

\section{Analytic solution for $J=2$}\label{app:ExactJ2}

In this appendix, we give all the details regarding the exact $J=2, M=0$ calculation presented in section~\ref{sec:ExactL2Solution}. We are starting from the following object
\begin{equation}\label{eq:qSeed1App}
    q_{\rm seed}(u)\equiv i \frac{\Gamma(-i u+\frac{3}{4})}{\Gamma(-i u+\frac{1}{4})}F(u)
    \,,
    \quad
    F(u) = {}_3 F_2(i u+3/4,1-h,h;\tfrac{3}{2},1;1)\,,
\end{equation}
which is a solution to the Baxter equation \eqref{eq:L2Baxter}. The solution $q_{\rm seed}(u)$ is not normalized to have the pure asymptotics~\eq{eq:AsymptoticsL2M2} as it should be for the standard basis. In order to find the correct basis we start from two linearly independent solutions $q_{\rm seed}(+u)$ and $q_{\rm seed}(-u)$ and analyse their analytic properties.

We note the standard series representation for $F(u)$ is convergent for ${\rm Im}\, u >-\frac{3}{4}$:
\begin{equation}\label{eq:FUHPapp}
    F(u) = \displaystyle \sum_{n=0}^\infty \frac{(h)_n(1-h)_n(iu+\tfrac{3}{4})_n}{n!(\frac{3}{2})_n (1)_n}\,.
\end{equation}
To investigate $\Im(u)<-\frac{3}{4}$ one can in principle start from the expression for $F(u)$ in the upper half-plane and then repeatedly use the Baxter equation \eqref{eq:L2Baxter}. However, we find it convenient to obtain an explicit analytic form of $F(u)$ also in the lower-half-plane.

To do so we start from the following integral representation of $F(u)$
\begin{equation}\label{intrep}
    F(u)=\frac{1}{\Gamma \left(\frac{1}{4}-i u\right) \Gamma \left(i
   u+\frac{3}{4}\right)}\displaystyle \int^{1}_0 {\rm d}t\  t^{-\frac{1}{4}+i u}(1-t)^{-\frac{3}{4}-i u}{}_2 F_1(1-h,h;\tfrac{3}{2};t)\,,
\end{equation}
where for the integrand we can use the
Gauss hypergeometric connection formula
\begin{equation}
    {}_2 F_1(1-h,h;\tfrac{3}{2};t) = \frac{\cos(h\pi)}{1-2h}{}_2 F_1 (1-h,h,\tfrac{1}{2},1-t)-(1-t)^\frac{1}{2} \sin(h \pi) {}_2 F_1 (\tfrac{1}{2}+h,\tfrac{3}{2}-h,\tfrac{3}{2},1-t)\,.
\end{equation}
Now we change variables in the integral, $t \mapsto 1-t$. This yields two integrals of ${}_3F_2$ type (i.e. of the form \eqref{intrep}). To identify them unambiguously as ${}_3F_2$ functions, we must specify their domains of convergence. Although the convergence regions differ slightly for the two integrals, both are valid in the strip $-\tfrac14 < \mathrm{Im}\,u < \tfrac34$. We therefore obtain the following relation,
\begin{equation}\label{eq:FLHP}
\begin{split}
    F(u) = -\frac{\cos(h\pi)}{2h-1}&{}_3 F_2 (1-h,h,\tfrac{1}{4}-i u;\tfrac{1}{2},1;1)\\
    &-\frac{2}{\sqrt{\pi}}\sin(h \pi)\frac{\Gamma(\tfrac{3}{4}-i u)}{\Gamma(\tfrac{1}{4}-i u)}{}_3 F_2 (\tfrac{1}{2}+h,\tfrac{3}{2}-h,\tfrac{3}{4}-i u;\tfrac{3}{2},\tfrac{3}{2};1)\,.
\end{split}
\end{equation}
which is convergent using the standard series representation for $\Im u<1/4$. Since \eqref{eq:FUHPapp} and \eqref{eq:FLHP} together cover the entire complex plane, we now have perfect control over $q^\downarrow(u)$.
Next in order to build the special set of solutions $q_i^\downarrow(u)$ we have to analyse the analytic structure of the solution we found above.

\paragraph{Poles and zeros.}

Let us proceed to use our exact expressions to deduce the poles and zeros of $q_{\rm seed}(u)$. As we remarked in section~\ref{sec:2DQSCDef}, in the case of no magnons an UHP-analytic function solving Baxter will in general have poles of order $J$ at $-\frac{3}{4}\ii-\ii n\,, n=0,1,\dots$. This is indeed true for $q_{\rm seed}(u)$ as can be seen from the explicit $\Gamma$-functions appearing in \eqref{eq:qSeed1App}. The residues at these points are correlated with the value of $q_{\rm seed}(u)$ in the upper-half-plane, and we find
\begin{equation}\label{resreln}
    \frac{q_{\rm seed}(u)}{q_{\rm seed}(-p)}=\frac{\sin(h\pi)}{\pi^2(u-p)^2}+\lO\left(\frac{1}{u-p}\right),\quad p=-\frac{3i}{4}-i n,\quad n=0,1,\dots \,.
\end{equation}
Note also that
\begin{equation}\label{zers}
    q_{\rm seed}(-\tfrac{i}{4}-i n)=0,\quad n=0,1,\dots\,,
\end{equation}
which are all simple zeroes.

\paragraph{Constructing UHPA and LHPA solutions.}

We now try to take linear combinations of $q_{\rm seed}(u)$ and $q_{\rm seed}(-u)$ with $i$-periodic coefficients such that the solutions are UHPA or LHPA and with polynomial asymptotics according to \eqref{eq:AsymptoticsL2M2}. Let us first discuss how to construct UHPA solutions. First, $q_{\rm seed}(-u)$ comes with a ladder of double poles at $\frac{3i}{4}+i n$, $n=0,1,\dots$. We can cancel these by introducing a factor $\tanh\left(\pi \left(u-\tfrac{3i}{4}\right)\right)$, leaving us with simple poles. These simple poles in the UHP can then be cancelled by $\coth(\pi \left(u-\tfrac{3i}{4}\right))q_{\rm seed}(u)$, thanks to the relation $\eqref{resreln}$. Note however, than $\tanh(\pi \left(u-\tfrac{3i}{4}\right))$ also introduces poles in the UHP at $\frac{i}{4}+i n$, $n=0,1,\dots$, but thankfully these are cancelled as a result of \eqref{zers}. The fact that the zeroes in \eqref{zers} are simple is the reason we cannot just cancel the double poles immediately with $\tanh(\pi \left(u-\tfrac{3i}{4}\right))^2$.

In summary, an UHPA solution will be of the form
\begin{equation}\label{eq:ParamQUpper}
   \left(x+y \coth(\pi \left(u-\tfrac{3i}{4}\right))\right)q_{\rm seed}(u)+z \tanh(\pi \left(u-\tfrac{3i}{4}\right)) q_{\rm seed}(-u)
\end{equation}
where $y=-z \sin(h \pi)$ in order for all the poles in the UHP to cancel and in the same way, we can construct a general LHPA solution as
\begin{equation}\label{eq:ParamQLower}
    x \tanh(\pi \left(u+\tfrac{3i}{4}\right))q_{\rm seed}(u) + (-x\sin(h\pi)\coth(\pi \left(u+\tfrac{3i}{4}\right))+z)q_{\rm seed}(-u)\,.
\end{equation}

\paragraph{Purifying solutions.} We now turn to the task of tuning the parameters in \eqref{eq:ParamQUpper} and \eqref{eq:ParamQLower} in order to produce what we refer to as \emph{pure solutions}. These are solutions of the form $q\simeq u^{\pm h\mp \frac{1}{2}}\left(1+\frac{*}{u}+\dots\right)$ where $\star$ is some $u$-independent quantity. To simplify notation we use
\begin{equation}
    \hat{M}_{i} = \{h-\tfrac{1}{2},-h+\tfrac{1}{2}\}\,
\end{equation}
and we denote the pure solutions as $q_{i}$.

The first step towards finding these solutions is to deduce the particular linear combination of $q_1$ and $q_2$ that makes $q_{\rm seed}(u)$. At $u\rightarrow \infty$ we find
\begin{equation}
\begin{split}
    q_{\rm seed}(u) \simeq
    \ii\frac{\sin (\pi  h)}{2 \pi  (2 h-1)}&\bigg(4^{\hat{M}_1}e^{-\frac{\ii \pi \hat{M}_1}{2}}\Gamma\left[\frac{1}{2}-\hat{M}_1\right]u^{\hat{M}_1}-\left(\hat{M}_1\leftrightarrow \hat{M}_2\right)\bigg)+\dots\,.
\end{split}
\end{equation}
Using these asymptotics we can readily construct
\begin{equation}
\begin{split}
    &q_i^\downarrow/p^{\downarrow}_{i} = \,\left[\mp_i \ii \cos(h\pi)- \sin(h \pi)\coth_{(-)}\right] q_{\rm seed}(u)+\tanh_{(-)}  q_{\rm seed}(-u) \\
    &q_i^\uparrow/p^{\uparrow}_{i}=\tanh_{(+)} q_{\rm seed}(u) + \left[\pm_i \ii \cos(h\pi)-\sin(h\pi)\coth_{(+)}\right] q_{\rm seed}(-u)
\end{split}
\end{equation}
where $\pm_i=-\mp_i=\{+1,-1\}$ and
\begin{equation}
    \tanh_{(\pm)} = \tanh{\pi(u\pm \ii \tfrac{3}{4})}\,,
    \quad
    \coth_{\pm} = \coth{\pi(u\pm \ii \tfrac{3}{4})}.
\end{equation}
Finally, the explicit form of the normalization factors are
\begin{equation}
    p_{i}^{\downarrow} = e^{\ii \pi\left(\hat{M}_i+\frac{1}{2}\right)}\,p^{\uparrow}_{i}= \frac{\pi  (2 h-1) 4^{\frac{1}{2}-\hat{M}_i} e^{\frac{i \pi  \hat{M}_i}{2}}}{\sin\left(2\pi h\right)\Gamma
   \left(\frac{1}{2}-\hat{M}_i\right)}\,.
\end{equation}

\paragraph{Finding $\Omega$.}

We recall that $\Omega$is an $\ii$-periodic matrix defined to satisfy
\begin{equation}\label{eq:Omegaqq}
    q^{\uparrow}_i = \Omega_{i}{}^{j} q_j^\downarrow\,.
\end{equation}
Now, in order to find this matrix in our current set-up we can leverage various symmetry considerations and consistency conditions. First of all, we note that the asymptotics of $\Omega$ is fixed from \eqref{eq:Omegaqq} to be
\begin{equation}\label{eq:AsymptoticsOmegaL2}
    \Omega \simeq_{u\rightarrow \infty} = 1_{2\times 2}\,,
    \quad
    \Omega \simeq_{u\rightarrow \infty} = \begin{pmatrix}
        e^{-2\pi \ii \hat{M}_1} & 0 \\
        0 & e^{-2\pi \ii \hat{M}_2}
    \end{pmatrix}\,.
\end{equation}
Secondly, we notice that since the Baxter equation \eqref{eq:L2Baxter} is parity symmetric, we must have $q^{\downarrow}_i(-u) \propto q^{\uparrow}_i(u)$, the exact proportionality is fixed from asymptotics and we find
\begin{equation}
    q^{\uparrow}_i(u) = \mathbb{g}_{i}{}^{j}q^{\downarrow}(-u)\,,
    \quad
    \mathbb{g} = \begin{pmatrix}
        e^{-\pi \ii \hat{M}_1} & 0 \\
        0 & e^{-\pi \ii \hat{M}_2}
    \end{pmatrix}\,.
\end{equation}
We can use this relation to derive a consistency equation, namely we find
\begin{equation}\label{eq:ParityOmegaL2}
    \Omega(u) \mathbb{g}^{-1} \Omega(-u) \mathbb{g}^{-1} = 1_{2\times 2}\,.
\end{equation}
Using together \eqref{eq:AsymptoticsOmegaL2} and \eqref{eq:ParityOmegaL2} fixes $\Omega$ up to 1 constant. To also find this constant we can evaluate \eqref{eq:Omegaqq} at $u\simeq -\frac{3\ii}{4}$. This fully fixes the matrix to be
\begin{equation}
    \Omega = \frac{1}{\left(e^{2 \pi  u}+i\right)^2}\left(
\begin{array}{cc}
 e^{4 \pi  u}+e^{-2 i \pi  h} & \ii \, e^{2 \pi  u}\, 2^{3-4 h}\sin (\pi  h)\, \frac{ \Gamma (h)}{\Gamma (1-h)} \\
 i\, e^{2 \pi  u} \,2^{4 h-1}\,\sin (\pi  h)\,\frac{  \Gamma (1-h)}{\Gamma (h)} & e^{4 \pi  u}+e^{2 i \pi  h} \\
\end{array}
\right)\,.
\end{equation}
The remainder of the details can be found in the main text. 

\section{Direct diagonalization of graph-building operator}\label{app:DirectDiag}

The spin operators can be explicitly diagonalised for $J=2$ for any choice of the local spins $s_1$ and $s_2$. We find the following eigenfunction

\begin{equation}
    \Psi(z_1,z_2)=\frac{1}{[z_1]^{\{h,\dot{h}\}+s_1-s_2}[z_2]^{\{h,\dot{h}\}-s_1+s_2}[z_1-z_2]^{-\{h,\dot{h}\}+s_1+s_2}}\,.
\end{equation}

We will now show that this function diagonalizes the graph-building operator $\hat{B}$ after appropriately tuning the local spins according to the values in Table \ref{tab:configs}. 

\paragraph{No magnons.}

For no magnons $s_1 = s_2 = \tfrac{1}{4}$ and the wave function reads
\begin{equation}
    \Psi(z_1,z_2) = \frac{1}{[z_1]^{\{h,\dot{h}\}} [z_2]^{\{h,\dot{h}\}} [z_1 - z_2]^{\frac{1}{2}-\{h,\dot{h}\}}}\,.
\end{equation}
and the graph-building operator is
\begin{equation}
    \hat{B}\Psi(z_1,z_2) = \frac{\xi^4}{\pi^2}\displaystyle \int {\rm d}w_1 {\rm d}w_2 \frac{1}{[z_1-w_1]^\frac{1}{2}[z_2-w_2]^\frac{1}{2}[w_1-w_2]}\Psi(w_1,w_2)\,.
\end{equation}
We can now perform the integral using the star-triangle relation \eqref{eqn:startriangle}, first performing the $w_2$ integral and then the $w_1$ integral. The final result yields
\begin{equation}
     \hat{B}\Psi(z_1,z_2) = \xi^4 a\left(\tfrac{1}{2},\tfrac{3}{2}-h,h\right)a\left(\tfrac{1}{2},1-h,h+\tfrac{1}{2}\right)\Psi(z_1,z_2)
\end{equation}
where the function $a(\alpha,\beta,\gamma)$ is defined in \eqref{eqn:afngamma}. Finally, requiring that the eigenvalue for $\hat{B}$ is equal to $1$ for physical states, and using the relations $\Delta=h+\bar{h}$ and $S=\bar{h}-h$ together with the expression \eqref{eqn:afngamma} for $a(\alpha,\beta,\gamma)$ in terms of $\Gamma$-functions we immediately obtain the relation
\begin{equation}
    \frac{1}{4}(1+S-\Delta)(-1+S+\Delta)=\xi^4\,,
\end{equation}
perfectly matching \cite{Kazakov:2018qbr}.

\paragraph{One magnon or one anti-magnon.}

Repeating as above we find the following relation between $\xi$ and $h$, valid for either one magnon or one anti-magnon:

\begin{equation}
    \xi^4 = \frac{\Gamma(\tfrac{5}{4}-h)^2}{\Gamma(\tfrac{3}{4}-\dot{h})^2}\frac{\Gamma(\tfrac{1}{4}+h)^2}{\Gamma(-\tfrac{1}{4}+\dot{h})^2}\,.
\end{equation}
Now we use $h=\frac{1}{2}(\Delta-S)$ and $\dot{h}=\frac{1}{2}(\Delta+S)$. For these states we expect $\Delta=\tfrac{3}{2}+\lO(\xi^2)$. Plugging in, we find
\begin{equation}
    \Delta=\frac{3}{2}-2\xi^2 -4\log(4)\, \xi^4 +\lO(\xi^6)\,,
\end{equation}
perfectly matching the ABA expression \cite{Basso:2019xay}.

\section{Finding ${\mathbb Q}_+(-\frac{3\ii}{4})$ and the cyclicity condition}\la{app:cyc}
In order to impose the cyclicity condition in section~\ref{ABACyc} we have to repeat the calculation of
${\mathbb Q}_+$ for another special point. Here we assume $M<J$, for simplicity.
In this appendix we compute the pole of ${\mathbb Q}_{+}(u)$ at $u=-\frac{3\ii}{4}$. The discussion here is essentially the same as in section~\ref{ABACyc}.

We start by introducing
\begin{equation}
    q_{i}^{\downarrow}\left(-\frac{3\ii}{4}+\epsilon\right) \simeq \frac{D_{i}}{\epsilon^{J-M}}\,,
    \quad
    \dot{q}_{i}^{\downarrow}\left(-\frac{3\ii}{4}+\epsilon\right) \simeq \frac{\dot{D}_i}{\epsilon^{J-M}}\,.
\end{equation}
We once again want to establish a relation among $D_{i}$. To more easily follow the analysis above, it turns out to be useful to consider $\Omega^{-1}$, we introduce
\begin{equation}
    F_{i}{}^{j} = \lim_{\epsilon\rightarrow 0} \epsilon^{J-M}\left(\Omega^{-1}\right)_{i}{}^{j}\left(-\frac{3\ii}{4}\right)\,.
\end{equation}
We once again have a number of relations among $F_{i}{}^{j}$, they are precisely the same as for $R$ and we are once again left with $3$ unknowns, $F_{2}{}^{1},\dot{F}_{2}{}^{1}$ and $F_{2}{}^2$. Repeating similar steps we have
\begin{equation}\label{eq:Qpatm34}
    \lim_{\epsilon\rightarrow 0} \epsilon^{J-M}{\mathbb Q}_{+}\left(-\frac{3\ii}{4}+\epsilon\right) = \frac{cD_2\dot{D}_{2}}{F_{2}{}^{2}}\,.
\end{equation}

We proceed to calculate $D_{2},\dot{D}_{2}$ and $F_{2}{}^{2}$ using the same methods as before. Using that $D_{1}$ is suppressed compared to $D_{2}$ we evaluate the Wronskian equation \eqref{eq:WronskianDown} to find
\begin{equation}
    D_2 = C^{\downarrow}_{W}\frac{(-2)^{M}}{\mathbf{q}_1^{\downarrow}\left(\frac{\ii}{4}\right)}\left(\frac{\ii}{\sqrt{\pi}}\right)^{J-M}\,,
    \quad
    \dot{D}_{2} = \dot{C}^{\downarrow}_{W}\frac{(-2)^{M}}{\dot{\mathbf{q}}_1^{\downarrow}\left(\frac{\ii}{4}\right)}\left(\frac{\ii}{\sqrt{\pi}}\right)^{J-M}\,.
\end{equation}
And from the explicit ABA form of $\Omega$ we obtain
\begin{equation}
\begin{split}
F_2{}^2 = \lim_{\epsilon\rightarrow0}\epsilon^{J-M}\frac{\mathbf{\Omega}_{1}{}^{1}\left(-\frac{3\ii}{4}\right)}{\mathbb{D}\left(-\frac{3\ii}{4}+\epsilon\right)} &= e^{-\frac{\ii \pi \gamma}{2}}\pi^{M-J} \prod_{i=1}^{M}\frac{1}{\cosh\pi\left(u_{i}-\frac{\ii}{4}\right)} \\
&=\frac{e^{-\frac{\ii}{2}\pi \gamma}}{\pi^{J}}\prod_{i=1}^{M}\Gamma\left(\ii u_{i}+\frac{3}{4}\right)\Gamma\left(-\ii u_{i}+\frac{1}{4}\right)\,.
\end{split}
\end{equation}
Plugging in these expressions into \eqref{eq:Qpatm34} we finally find
\begin{equation}\label{eq:QpRes111}
    \lim_{\epsilon\rightarrow 0} \epsilon^{J-M}{\mathbb Q}_{+}\left(-\frac{3\ii}{4}+\epsilon\right) \propto  4^{M}c\left(-1\right)^{J-M}\frac{C_{W}^{\downarrow}\dot{C}^{\downarrow}_{W}}{\betheQ\left(\frac{\ii}{4}\right)\dot{\betheQ}\left(\frac{\ii}{4}\right)}\prod_{i=1}^{M} \frac{\Gamma\left(\ii u_{i}+\frac{3}{4}\right)}{\Gamma\left(-\ii u_{i}+\frac{1}{4}\right)} \ .
 \end{equation}
This is the main result of this appendix.

\section{Generalization to anisotropic fishnets}
\subsection{Set-up and graph-building operator}

The 2d fishnet CFT considered in the main text can be generalised to an anisotropic fishnet CFT \cite{Kazakov:2018qbr} where the bare dimensions of the $\phi_1$ and $\phi_2$ fields are no longer equal to $1/4$ but are now given by $\delta/2$ and $\bar{\delta}/2$ subject to the relation $\bar{\delta}+\delta=1$, where $0<\delta<1$. The Lagrangian for the deformed theory is given by \cite{Kazakov:2018qbr}
\begin{equation}
    \mathcal{L} = N\, \mathrm{tr}\!\left[
\phi_1^\dagger (-\partial_\mu \partial^\mu)^{\delta} \phi_1
+ \phi_2^\dagger (-\partial_\mu \partial^\mu)^{\bar{\delta}} \phi_2
+ 4\pi \xi^2\, \phi_1^\dagger \phi_2^\dagger \phi_1 \phi_2
\right]
\end{equation}
and the TBA for the set-up with magnons was worked out in \cite{Basso:2019xay}.

We now generalise the QSC equations presented in the main text to the anisotropic set-up. As in the main text, we consider a CFT wave function with a vacuum of $\phi_1^\dagger$ fields with excitations of magnons, anti-magnons or magnon-anti-magnon pairs on top. The local spins and inhomogeneities for these configurations are given by

\begin{table}
\begin{equation}\nn
\begin{array}{c|c|c|c}
{I}_n & \text{fields} & s_n & \theta_n \\
\midrule
0 & \phi_1^\dagger(x_n) & \frac{\bar{\delta}}{2} & 0 \\
+1 &\phi_1^\dagger(x_n) \phi_2^\dagger(x_n) & \frac{\bar{\delta}}{2}+\frac{\delta}{2} & +\frac{i\delta}{4} \\
-1 & \phi_1^\dagger(x_n)\phi_2(x_n) & \frac{\bar{\delta}}{2}+\frac{\delta}{2} & -\frac{i\delta}{4} \\
\bar{0} & \phi_2^\dagger(x_n)\phi_1^\dagger(x_n)\phi_2(x_n) & \frac{\bar{\delta}}{2}+\delta & 0
\end{array}
\end{equation}
\caption{The table shows the relation between the fields sitting in the CFT wave function and the corresponding values of the local spins and inhomogeneities of the spin chain for the anisotropic 2D biscalar fishnet CFT. 
\label{tab:anisoparams}}
\end{table}
Note that since $\delta+\bar{\delta}=1$ we have $s_k=\tfrac{1}{2}$ for both magnons and anti-magnons.

\paragraph{Graph-building operator.}
As before, the graph-building operator $\hat{B}$ in the anisotropic case is specified by its integration kernel \cite{Kazakov:2018qbr}
\begin{equation}
    \hat{B}\circ f(z_1,\dots z_J) = \frac{\xi^{2J}}{\pi^J}\int \prod_{j=1}^J {\rm d}^2 w_j \, b_{I_j} (z_j,w_j,w_{j-1}) f(w_1,\dots,w_J)
\end{equation}
where $b_{I_j}$ now takes the form
\begin{equation}
    b_I(z,w,\tilde{w}) = \frac{1}{[z-w]^{1-\delta}} \times \left\{
      \begin{array}{ll}
        \dfrac{1}{[w-\tilde w]^{\delta}}, & \ \ I = 0 \quad\text{(no magnon)}, \\
        \dfrac{1}{[z-\tilde w]^{\delta}}, & \ \ I = 1 \quad\text{(magnon)}.
      \end{array}
    \right.
\end{equation}

As in the main text, we restrict our attention to configurations with only magnons on top of the vacuum.

We will now extract the graph-building operator from the Q-operator.

\subsection{Q-operator}

\paragraph{Q-operator.}
We will now construct the Q-operator and identify its properties. As the derivations are almost identical to those in the main text we only sketch the results. 

Like in the isotropic $\delta=\bar{\delta}=\frac{1}{2}$ case we write the Q-operator $\hat{\mathbb{Q}}_+(u,\dot{u})$ as an integral operator 
\begin{equation}
    [\hat{\mathbb{Q}}_+(u,\dot u)\Phi](z_1,\dots,z_J) = \int {\rm d}^{2J} w\, \mathcal{Q}_+(u, \dot u)(z_1,\dots,z_J|w_1,\dots,w_J)\,\Phi(w_1,\dots,w_J)\,,
\end{equation}
with the kernel $\mathcal{Q}_+(u,\dot{u})$ given by 
\begin{equation}
   \mathcal{Q}_+(u,\dot u)(z|w)=\displaystyle \prod_{k=1}^J  [w_{k}-z_k]^{\{\alpha^+_k,\dot\alpha^+_k\}}[w_{k-1}-z_k]^{\{\beta_k^+,\dot\beta_k^+\}} [w_{k}-w_{k-1}]^{\{\gamma^+_k,\dot\gamma^+_k\}}
\end{equation}
where now the exponents are given by
\begin{equation}
    \alpha^+_k = i u-\frac{\bar{\delta}}{2},\quad \beta_k^+=-iu-\frac{\bar{\delta}}{2}-\delta I_k,\quad \gamma^+_k = \delta(I_k-1)\,.
\end{equation}
Like in the isotropic case $I_k=1$ if a site contains a magnon and $I_k=0$ if it does not. 

\paragraph{Relation to graph-building operator.} The relation between the graph-building operator $\hat{B}$ and the Q-operator $\hat{\mathbb{Q}}_+$ is a simple generalization of the isotropic relation \eqref{QPtoB1}:
\begin{equation}
    \hat{\mathbb{Q}}_+\left(\frac{i\bar{\delta}}{2} \right)=  \frac{\pi^J}{\xi^{2J}}\hat{B}\,.
\end{equation}
As before for equal arguments $u=\dot{u}$ in the Q-operator we have only displayed one of them for brevity. Using the relation to the identity operator 
\begin{equation}
    \lim_{\epsilon\rightarrow 0}\epsilon^J \hat{\mathbb{Q}}_+\left(\frac{i}{2}+\frac{i\delta}{2}-i\epsilon \right) = \pi^J \times 1
\end{equation}
we then obtain the normalization-independent relation between the Q-operator eigenvalues $\mathbb{Q}_+$ and the coupling $\xi$
\begin{equation}\label{eqn:anisocoupling}
    \lim_{\epsilon\rightarrow 0}\epsilon^J \frac{\mathbb{Q}_+\left(\frac{i}{2}+\frac{i\delta}{2}-i\epsilon \right)}{\mathbb{Q}_+\left(\frac{i\bar{\delta}}{2}\right)} = \xi^{2J}\,.
\end{equation}

\paragraph{Relation to shift operator.} We also need the relation to the shift-operator $\hat{U}$ in order to select physical states corresponding to single-trace operators. For $M\neq J$ this relation is given by

\begin{equation}\label{eqn:anisoU}
    \hat U=
\frac{(-1)^J}{ (4\xi^2)^M}
\lim_{\epsilon\to 0}
\frac{1}{\epsilon^M}
\frac{\hat {\mathbb Q}_{+}\left(-\frac{i}{2}-\frac{i\delta}{2}-i\epsilon\right)}
{\hat {\mathbb Q}_{+}\left(+\frac{i}{2}+\frac{i\delta}{2}-i\epsilon\right)}
\end{equation}
while for $M=J$ the relation is given by 
\begin{equation}\label{eqn:anisoUJ}
    \hat U^{M=J}=
\(-1\)^J
\lim_{\epsilon\to0}
\frac{\hat{\mathbb Q}_+\left(-\tfrac{i}{2}+\tfrac{i\delta}{2}-i\epsilon\right)}{\hat{\mathbb{Q}}_+\left(+\tfrac{i}{2}+\tfrac{i\delta}{2}-i\epsilon\right)}\,.
\end{equation}
Like in the isotropic case we require the eigenvalue $U$ of $\hat{U}$ to be equal $1$ for single-trace operators.

\paragraph{Baxter equation.} The Q-operator $\hat{\mathbb{Q}}_+$ continues to satisfy the operatorial Baxter equation now given by
\begin{equation}\label{eqn:anisotropicbaxter}
    \left(u+\frac{i\bar{\delta}}{2} \right)^J \hat{\mathbb{Q}}(u+i,\dot{u}) -\hat{t}(u) \hat{\mathbb{Q}}(u+i,\dot{u}) + \left(u-\frac{i\delta}{2}-\frac{i}{2} \right)^M\left( u-\frac{i\bar{\delta}}{2} \right)^{J-M} \hat{\mathbb{Q}}(u-i,\dot{u})=0
\end{equation}
and similarly for the dotted sector with the transfer matrix $\hat{t}(u)$ given by \eqref{eqn:transfermatrix} with the appropriate selection of local spins and inhomogeneities as specified in \eqref{tab:anisoparams}.

\paragraph{Asymptotics.} We can straightforwardly repeat the derivation of the Q-operator asymptotics as carried out in the main text. For $u=v+\tfrac{i n}{2}$ and $\dot{u} = v-\tfrac{i n}{2}$ and $v\rightarrow \infty$ we find
\begin{equation}
    \mathbb{Q}_+(v+\tfrac{i n}{2},v-\tfrac{i n}{2}) \simeq u^{h+\dot{h}-(J\bar{\delta}+M\delta)}\times \lO(1) + u^{2-h-\dot{h}-(J\bar{\delta}+M\delta)}\times \lO(1)\,.
\end{equation}

\subsection{Quantum Spectral Curve}

We now have all the ingredients necessary to write down the QSC equations describing the exact spectrum of the anisotropic 2D fishnet CFT. We denote the UHPA and LHPA pure solutions of the undotted Baxter equation as $q_i^\downarrow$ and $q_i^\uparrow$ respectively, with 
\begin{equation}
    q_1 \simeq u^{h-(J\bar{\delta}+M\delta)/2},\quad q_2 \simeq u^{1-h-(J\bar{\delta}+M\delta)/2}
\end{equation}
which are related by the $i$-periodic matrix $\Omega_i^{\ j}$ with $q_i^\uparrow = \Omega_i^{\ j}q_j^\downarrow$, and similarly for the dotted sector. We then build $\mathbb{Q}_+(u,\dot{u})$ as a bilinear combination of these two families of solutions
\begin{equation}\label{eqn:anisoQP}
    \mathbb{Q}_+(u,\dot{u}) = \mathbb{c}\, \Gamma^{ij} q_i^\downarrow \dot{q}_j^\uparrow(\dot{u}) = \mathbb{c}\, \Gamma^{ij} q_i^\uparrow(u)\dot{q}^\downarrow(\dot{u})
\end{equation}
consistent with the asymptotics and the location of the poles as in \eqref{poles_locattions}. Again we choose to normalize $\Gamma^{ij}$ according to 
\begin{equation}
    \Gamma^{ij} = \left(\begin{array}{cc}
        1 & 0 \\
        0 & c
    \end{array} \right)\,.
\end{equation}
The consistency of \eqref{eqn:anisoQP} then leads to the condition 
\begin{equation}\label{eqn:anisoquant}
    \Gamma^{ik}\Omega_{k}^{\ j} = \Gamma^{jk}\dot{\Omega}_{k}^{\ i}
\end{equation}
as in the isotropic case. By solving the Baxter equation for the Q-functions $q_i^\uparrow$ and $q_i^\downarrow$, imposing the quantisation condition \eqref{eqn:anisoquant} as well as the relation \eqref{eqn:anisocoupling} to the coupling and the zero-momentum conditions \eqref{eqn:anisoU} for $M\neq J$ or \eqref{eqn:anisoUJ} for $M=J$, we deduce the spectrum of conformal dimensions $\Delta$ of single-trace operators of the anisotropic fishnet CFT as a function of the coupling $\xi$.

\section{Integral identities}

The following integral relations involving complex propagators are useful.

\paragraph{Chain relation}

\begin{equation}\label{eqn:chainreln}
    \int {\rm d}^{2}w\,
\frac{1}{[x-w]^{\{\beta,\dot{\beta}\}}\,[w-z]^{\{\alpha,\dot{\alpha}\}}}
=
\pi\,(-1)^{\gamma-\dot{\gamma}}\,
a(\alpha,\beta,\gamma)\,
\frac{1}{[x-z]^{\{\alpha,\dot{\alpha}\}+\{\beta,\dot{\beta}\}-1}}\,, 
\end{equation}
where
\beq\label{eqn:afngamma}
\gamma = 2-\alpha-\beta,\quad a(\alpha,\beta,\gamma)=\frac{\Gamma(1-\dot \alpha)\Gamma(1-\dot \beta)\Gamma(1-\dot \gamma)}{\Gamma(\alpha)\Gamma(\beta)\Gamma(\gamma)}\,.
\eeq
\paragraph{Star-triangle relation}

\begin{equation}\label{eqn:startriangle}
    \int d^2w \, \frac{1}{[z - w]^{\{\alpha,\dot{\alpha}\}} [x - w]^{\{\beta,\dot{\beta}\}} [y - w]^{\{\gamma,\dot{\gamma}\}}}
= \frac{\pi\, a(\alpha, \beta, \gamma)}{[x - z]^{1 - \{\gamma,\dot{\gamma}\}} [z - y]^{1 - \{\beta,\dot{\beta}\}} [y - x]^{1 - \{\alpha,\dot{\alpha}\}}}
\end{equation}
$\text{with } \alpha + \beta + \gamma = \dot{\alpha} + \dot{\beta} + \dot{\gamma} = 2$. As a special case,
\begin{equation}\label{eqn:chainsimp}
    \int d^2 w \, \frac{1}{[x - w]^{2 - \{\alpha,\dot{\alpha}\}} [w - z]^{\{\alpha,\dot{\alpha}\}}} = \pi^2 a(\alpha, 2 - \alpha) \, \delta^{(2)}(x - z)\,.
\end{equation}

\bibliographystyle{JHEP}
\bibliography{references}

\end{document}